\documentclass[journal,twocolumn]{IEEEtran}
\usepackage[cmex10]{amsmath}
\usepackage{amsfonts}
\usepackage{amssymb}
\usepackage{amsbsy}
\usepackage{graphicx}
\usepackage{rotating}
\usepackage{bm}
\usepackage{bbm}
\usepackage[table]{xcolor}
\usepackage{cite}
\usepackage{enumerate} 
\usepackage{dsfont}
\usepackage{multirow}
\usepackage{tikz}
 \usetikzlibrary{positioning,shapes,arrows, fit, patterns}
\usepackage{amsthm}

\allowdisplaybreaks

\newtheoremstyle{numbered}
  {3pt plus 1pt minus 2pt} % Space above
  {3pt plus 1pt minus 2pt} % Space below
  {} % Body font
  {1em} % Indent amount
  {\itshape} % Theorem head font\itshape\bfseries
  {:} % Punctuation after theorem head
  {.3em} % Space after theorem head
  {#1~#2} % Theorem head spec (can be left empty, meaning `normal')
\theoremstyle{numbered}
\newtheorem{theorem}{Theorem}

\newtheorem{pro}[theorem]{Proposition}
\newtheorem{cor}[theorem]{Corollary}
\newtheorem{lemma}[theorem]{Lemma}
\newtheorem{remark}{Remark}%[section]

\makeatletter
\renewcommand\section{\@startsection{section}{1}{\z@}{2.5ex plus 1.5ex minus 0.8ex}{1.5ex plus 1ex minus 0.5ex}{\normalfont\normalsize\centering\scshape}} %change spacing before and after sections
\renewcommand\subsection{\@startsection{subsection}{2}{\z@}{2.5ex plus 1.5ex minus 0.8ex}{1.5ex plus 1ex minus 0.5ex}{\normalfont\normalsize\itshape}}%change spacing before and after subsections
\renewcommand\@IEEEauthorblockconfadjspace{-0.1em} % spacing between title and authors
\renewcommand\@IEEEauthorblockNtopspace{0.0ex} 
\renewcommand\@IEEEauthorblockAtopspace{1ex}% spacing between authors and affiliations

\makeatother

\newcommand{\be}{\begin{equation}}
\newcommand{\ee}{\end{equation}}
\newcommand{\ist}{\hspace*{.3mm}}
\newcommand{\rmv}{\hspace*{-.3mm}}
\providecommand{\abs}[1]{\lvert#1\rvert}
\providecommand{\bigabs}[1]{\bigg\lvert#1\bigg\rvert}
\providecommand{\norm}[1]{\lVert#1\rVert}

\def\ba#1\ea{\begin{align}#1\end{align}}
\def\bas#1\eas{\begin{align*}#1\end{align*}}

\def\L{N}

\def\R{R}
\def\Q{Q}

\def\T{T}
\def\Tprop{\widetilde{\T}}

\def\nreq{\ell}

\def\0v{\boldsymbol{0}}
\def\Iv{\mathbf{I}}
\def\av{\boldsymbol{a}}

\def\sv{\boldsymbol{s}}
\def\svt{\tilde{\sv}}

\def\uv{\boldsymbol{u}}

\def\vv{\boldsymbol{v}}

\def\xv{\boldsymbol{x}}
\def\xvt{\tilde{\xv}}
\def\yv{\boldsymbol{y}}
\def\yvb{\bar{\yv} }  %%%%%%%%%%%%%%%%%%

\def\zv{\boldsymbol{z}}

\def\xiv{\boldsymbol{\xi}}

\def\betav{\boldsymbol{\beta}}

\def\Am{\boldsymbol{A}}

\def\Bm{\boldsymbol{B}}
\def\Cm{\boldsymbol{C}}

\def\Jm{\boldsymbol{J}}
\def\Mm{\boldsymbol{M}}
\def\Km{\boldsymbol{K}}
\def\Qm{\boldsymbol{Z}}

\def\Xm{\boldsymbol{X}}

\def\Zm{\boldsymbol{Z}}
\def\Sigm{\boldsymbol{\Sigma}}

\def\rs{\mathsf{s}}

\def\randk{\mathsf{k}}

\def\r0v{\boldsymbol{\mathsf{0}}}

\def\rcv{\boldsymbol{\mathsf{c}}}

\def\rhv{\boldsymbol{\mathsf{h}}}

\def\rnv{\boldsymbol{\mathsf{w}}}
\def\rnvt{\tilde{\rnv}}

\def\rsv{\boldsymbol{\mathsf{s}}}

\def\ruv{\boldsymbol{\mathsf{u}}}

\def\rvv{\boldsymbol{\mathsf{v}}}

\def\rxv{\boldsymbol{\mathsf{x}}}

\def\ryv{\boldsymbol{\mathsf{y}}}
\def\ryvb{\bar{\ryv} }  %%%%%%%%%%%%%%%%%%
\def\ryvt{\tilde{\ryv}}

\def\rAm{\boldsymbol{\mathsf{A}}}

\def\rBm{\boldsymbol{\mathsf{B}}}

\def\rXm{\boldsymbol{\mathsf{X}}}

\def\IN{\mathbb{N}}
\def\IC{\mathbb{C}}
\def\IR{\mathbb{R}}
\def\IZ{\mathbb{Z}}
\def\sA{\mathcal{A}}
\def\sB{\mathcal{B}}

\def\sD{\mathcal{D}}
\def\sDt{\widetilde{\sD}}
\def\sE{\mathcal{E}}
\def\sF{\mathcal{F}}
\def\sI{\mathcal{I}}
\def\sJ{\mathcal{J}}
\def\sG{\mathcal{G}}
\def\sM{\mathcal{M}}
\def\sN{\mathcal{N}}
\def\sL{\mathcal{L}}
\def\sLt{\widetilde{\sL}}

\def\sS{\mathcal{S}}

\def\sP{\mathcal{P}}
\def\sPt{\widetilde{\sP}}

\def\sU{\mathcal{U}}

\def\sW{\mathcal{W}}
\def\sZ{\mathcal{Z}}
\providecommand{\absdet}[1]{\lvert#1\rvert}
\def\diag{\operatorname{diag}}
\def\rank{\operatorname{rank}}
\def\lcm{\operatorname{lcm}}
\def\mymod{\operatorname{mod}^*}
\def\shortlcm{L}
\def\trans{{\operatorname{T}}}
\def\E{\mathbb{E}}

\begin{document}

\IEEEoverridecommandlockouts

\title{
Degrees of Freedom of Generic Block-Fading \\
MIMO Channels without A Priori  \\
Channel State Information
}

\author{
G\"unther~Koliander,~\IEEEmembership{Student~Member,~IEEE,} 
Erwin~Riegler,~\IEEEmembership{Member,~IEEE,}\\
Giuseppe~Durisi,~\IEEEmembership{Senior~Member,~IEEE,}
and~Franz~Hlawatsch,~\IEEEmembership{Fellow,~IEEE}%
%\IEEEauthorblockA{
%$^1$Institute of Telecommunications, Vienna University of Technology, 1040 Vienna, Austria\\
%$^2$Department of Signals and Systems, Chalmers University of Technology, 41296 Gothenburg, Sweden\\
%}
\thanks{This paper was presented in part at the Allerton Conference on Communication, Control, and Computing, Monticello, IL, Oct. 2012 
and at the IEEE International Symposium on Information Theory (ISIT),  Istanbul, Turkey, July 2013.
}%
\thanks{This work was supported by the WWTF under grant ICT10-066 (NOWIRE) and by the Swedish Research Council under grant 2012-4571.}
\thanks{G. Koliander and F. Hlawatsch are with the Institute of Telecommunications, Vienna University of Technology, 1040 Vienna, Austria (e-mail: guenther.koliander@nt.tuwien.ac.at, franz.hlawatsch@nt.tuwien.ac.at).}%
%\thanks{
\thanks{E.~Riegler is with the Department of Information Technology and Electrical Engineering, ETH Zurich, 8092 Zurich, Switzerland (e-mail: eriegler@nari.ee.ethz.ch).}%
\thanks{G. Durisi is with the Department of Signals and Systems, Chalmers University of Technology, 41296 Gothenburg, Sweden (e-mail: durisi@chalmers.se).}%
\thanks{Copyright (c) 2014 IEEE. Personal use of this material is permitted.  However, permission to use this material for any other purposes must be obtained from the IEEE by sending a request to pubs-permissions@ieee.org.}
}

\maketitle

%\renewcommand{\baselinestretch}{1}\small\normalsize
%\vspace{-10mm}

\begin{abstract}
We study
%% investigate 
the high-SNR capacity of \emph{generic} MIMO Rayleigh block-fading channels in the noncoherent setting where neither transmitter nor receiver has \emph{a priori} channel state information but both are aware of the channel statistics. 
In contrast to the well-established constant block-fading model, we allow the fading to vary within each block with a temporal correlation that is ``generic'' (in the sense used in the interference-alignment literature).
%%We show that when the number of receive antennas is sufficiently large, % and 
%% when 
%the temporal correlation within each block is ``generic'' , 
%%the number of degrees of freedom of the channel is given by $\T(1-1/\L)$ for $\T<\L$, where $\T$ denotes the number of transmit antennas and $\L$ denotes the block length.
We show that the number of degrees of freedom of a generic MIMO Rayleigh block-fading channel with $\T$ transmit antennas and block length $\L$ is given by $\T(1-1/\L)$ provided that $\T<\L$ and the number of receive antennas is at least $\T(\L-1)/(\L-\T)$.
%%  (provided that the receiver is equipped with a sufficient number of receive antennas). 
A comparison with the 
%% noncoherent capacity of the 
%widely used 
constant block-fading channel (where the fading 
%% coefficients are 
is constant within each block)
shows that, for large 
%% $N$,
block lengths, 
generic correlation increases the number of degrees of freedom by a factor of up to four.
\end{abstract}

%%%%%%%%%%%%%%%%%%%%%%%%%%%%%%%
\section{Introduction} \label{sec:intro}   
% \cha{and Summary of Contributions}
%%%%%%%%%%%%%%%%%%%%%%%%%%%%%%%
%\vspace{1mm}
The use of multiple antennas is a well-established method to increase data rates in wireless systems. 
A classic result in information theory states that the 
%degrees of freedom (i.e., the asymptotic ratio between capacity and the logarithm of the SNR as the SNR grows large, also referred to as \emph{capacity pre-log})
throughput achievable with 
%% multiuser 
%of
multiple-input multiple-out\-put (MIMO) wireless systems grows linearly in the number of antennas when  
 perfect channel state information (CSI) is available  at the receiver~\cite{te99}. 
In practice, though, 
the MIMO data rates are 
%% fundamentally 
limited by the need to acquire CSI~\cite{MaHo99, zhengtse02, live04, Schuster08, mo09, adca12}.
%%  that needs to be acquired in both downlink and uplink to release MIMO gains~\cite{adca12,hokode11}.
A fundamental way to assess the rate penalty 
%incurred by the need of estimating the channel
due to channel estimation
(relative to the unrealistic case where perfect CSI is available) 
is to study capacity in the 
%% so-called 
\emph{noncoherent setting} where neither the transmitter nor the receiver has \emph{a priori} CSI but both are aware of the channel statistics.

The model most commonly used to capture channel variations for capacity analyses in the noncoherent MIMO setting is the 
%% model in capacity analyses in the noncoherent setting is the 
Rayleigh-fading \emph{constant block-fading channel model} 
%% introduced in 
\cite{MaHo99},
according to which
the fading process takes on independent realizations across blocks of  $\L$ channel uses (``block-memoryless'' assumption), 
and within each block the fading coefficients stay constant. 
Thus,
%% Specifically, 
the $\L$-dimensional vector  describing 
the channel between antennas $t$ and 
%% antenna 
$r$ 
%% of length $\L$ 
(hereafter briefly termed ``$(t,r)$ channel'') within a block 
%% %\vspace{-1mm}
is 
%\vspace{-2mm}
\begin{equation}\label{eq:constant_block_memoryless}
  \rhv_{r,t}=\rs_{r,t} \ist \mathbf{1}_{\L\times 1} \,.
\end{equation}
Here, $\mathbf{1}_{\L\times 1}$ denotes the $\L$-dimensional all-one vector and $\rs_{r,t}$, $r \in \{1,\dots,R\}, t \in \{1,\dots,T\}$, are independent $\mathcal{CN}(0,1)$ random variables;
%% $r=1,\dots,R$, $t=1,\dots,T$, 
$T$ and $R$ denote the number of transmit and receive antennas, respectively.
%%  the number of receive antennas, 
%
Unfortunately, even for this simple channel model, a closed-form expression for the capacity in the noncoherent setting 
is unavailable.
%% difficult to characterize 
%% analytically 
%% Even 
%% %% in the case 
%% when full cooperation between users is assumed and, thus, the multiuser MIMO system reduces to a point-to-point MIMO channel
%% (which is the scenario studied in this paper),
%% %% \footnote{In the remainder of the paper, we shall focus on this scenario.}, 
%% closed-form expressions are not available.
However,
%% Nevertheless, 
an accurate characterization exists for
%% capacity can be characterized quite accurately at 
high signal-to-noise ratio (SNR).
Specifically, Zheng and Tse~\cite{zhengtse02}  proved that the number of degrees of freedom (i.e., the asymptotic ratio between capacity and the logarithm of the SNR as the SNR grows large, also referred to as \emph{capacity pre-log}) 
for the constant block-fading model is given by
\ba
\hspace*{-.5mm}\chi_{\text{const}}=\ist M \bigg(\rmv 1 \rmv-\frac{M}{\L} \bigg) \ist , \;\; \text{with} \;\, M \rmv = \ist \min\bigg\{\T, \R, \bigg\lfloor\frac{\L}{2}\bigg\rfloor\bigg\} \ist .%\label{eq:pre_log_constant}
 \label{eq:pre_log_constant}
\ea
%
%% where $M^*\triangleq \min\{\T, \R, \lfloor\L/2\rfloor\}$. 
For the case $\R+\T\leq \L$, they also provided a high-SNR capacity expansion 
 that is accurate up to a $o(1)$ term (i.e., a term that vanishes as the SNR grows).
This expansion was recently extended
%% generalized 
in~\cite{yaduri12} to the ``large-MIMO'' setting $\R+\T>\L$.

%% \newpage %%%%%%%%%
\subsection{Extending the Constant Block-fading Model}

One limitation
%% deficiency 
of the constant block-fading model is that it fails to describe 
%% accurately a special 
a specific setting where block-fading models are of interest, namely, cyclic-prefix orthogonal frequency division multiplexing (CP-OFDM) systems \cite{tsvi05}.
In such systems, the channel input-output relation is most conveniently described in the frequency domain: the 
%% In this domain, the 
vector of channel gains  $\rhv_{r,t}$ is equal to the Fourier transform of the discrete-time impulse response $\rcv_{r,t}$ of
%% associated to 
the $(t,r)$ channel.
The constant block-fading model here corresponds to the situation where the impulse response of each $(t,r)$ channel consists of a single tap, i.e., 
$\rcv_{r,t}=\sqrt{\L}\rs_{r,t}(1\; 0 \cdots 0)^{\trans}$,
a situation for which the use of OFDM is unnecessary.

In this paper, we focus on a channel model that allows for impulse responses with multiple taps.
Furthermore, we shall allow different $(t,r)$ channels to have different correlation structures. 
One way to achieve these goals is to model the channel gains as
\be\label{eq:generic_block_memorylessQ1}
\rhv_{r,t} = \rs_{r,t}\zv_{r,t}\,.
\ee
Here, the squared magnitude of the inverse Fourier transform of each deterministic vector $\zv_{r,t}$ is equal to the power-delay profile of the corresponding $(t,r)$ channel.
%Assuming that the vectors $\rcv_{r,t}$ change independently across blocks and antennas, we can try to use the constant block-fading model~\eqref{eq:constant_block_memoryless} to describe the channel gains $\rhv_{r,t}$. 
%This corresponds to a discrete-time impulse response $\rcv_{r,t}=\sqrt{\L}\rs_{r,t}(1\; 0 \dots 0)^{\trans}$
%which is to a very specific channel model.
%We want to consider a system model that allows for more complicated $\rcv_{r,t}$ and, in particular, different correlations for different $(t,r)$ channels.
%As a first step we model the channel gains as 
%\be\label{eq:generic_block_memorylessQ1}
%\rhv_{r,t} = \rs_{r,t}\zv_{r,t} 
%\ee
%introducing deterministic vectors $\zv_{r,t}$ whose squared inverse
%Fourier transform equals the power-delay profile of the $(t,r)$
%channel.
%The correlation in a block is still quite simple although we already introduced different correlations across the $(t,r)$ channels. 
To obtain an even more general system model,
we assume that in each block the correlation is described by $\Q \geq 1$ independent random variables according 
%\vspace{-3mm}
to
\begin{align}\label{eq:generic_block_memoryless}
  \rhv_{r,t} = \Zm_{r,t} \rsv_{r,t}
\end{align}
where $\Zm_{r,t}\in \IC^{\L\times \Q}$ with $\Q\leq\L$ is a deterministic matrix and $\rsv_{r,t}\in \IC^{\Q}$ contains independent $\mathcal{CN}(0,1)$ entries, which are also independent across
$r \in \{1,\dots,\R\}$ and $t \in \{1,\dots,\T\}$. 
A similar system model, with the simplifying assumption that all matrices $\Zm_{r,t}$ are equal, was analyzed in~\cite{live04}, where a lower bound on the number of degrees of freedom was derived.
This lower bound is tight only for the single-antenna case \cite{modubo10, rimodulistbo11, moridu13}. 

\subsection{Main Result} %% Contributions % (fold)
Building on our previous work in~\cite{koridumohl12}  and~\cite{koriduhl13},
we study
%% analyze 
the high-SNR capacity of 
%% the ???generic 
MIMO block-fading channels modeled according to
\eqref{eq:generic_block_memoryless} and show that when the deterministic matrices $\Zm_{r,t}$ are 
\emph{generic},
%A rigorous definition will be provided in Section~\ref{sec:syst}.
%% We will term the channel model~\eqref{eq:generic_block_memoryless}
%% with generic vectors $\{\zv_{r,t}\}$ the ``generic block-fading 
%% model.''}  %%%%%%%%
the number of degrees of freedom can be larger than %the degrees of freedom 
in the constant block-fading case as given in~\eqref{eq:pre_log_constant}.
Coarsely speaking, we can think of generic $\Zm_{r,t}$ as being generated from an underlying joint probability density function.\footnote{We %%%%%%%%
use the term ``generic''  in the same sense as it is used in the interference-alignment literature \cite{ja11}.}
We shall refer to~\eqref{eq:generic_block_memoryless} with generic $\Zm_{r,t}$ as \emph{generic block-fading model}.
Our specific contribution is as follows: we show that for all matrices $\Zm_{r,t}$ except for a set of Lebesgue measure zero, the number of degrees of freedom is given by
%
%% \footnote{\cha{Our results do not encompass the case where all $\Zm_{r,t}$ are equal. This specific case remains an open problem.}}
\be\label{eq:pre_log_generic}
  \chi_{\text{gen}}=T\bigg(1\rmv-\rmv \frac{1}{\L}\bigg) \ist
%\vspace{1mm}
\ee
%for the \emph{generic block-fading model} (i.e.,~\eqref{eq:pre_log_generic} holds for the block-fading model~\eqref{eq:generic_block_memoryless} with any matrices  $\{\Zm_{r,t}\}$, but a set of Lebesgue measure zero), 
provided that  $\T \!<\rmv \L/\Q$ and 
%the number of receive antennas is sufficiently large, namely  
$\R\geq \T(\L \rmv-\rmv 1) / (\L \rmv-\rmv  \T\Q)$.
We note that the set corresponding to the case where all matrices $\Zm_{r,t}$ are \emph{exactly} equal has Lebesgue measure zero,
and thus we do not know whether \eqref{eq:pre_log_generic} holds for equal $\Zm_{r,t}$.
%%  in that case. 
Therefore, this specific case remains an open problem.
%However, we conjecture that it will rarely occur in practical systems.}
%
%
We also provide an upper bound and a lower bound on $\chi_{\text{gen}}$ for the case $\R< \T(\L \rmv-\rmv 1) / (\L \rmv-\rmv  \T\Q)$.

\subsection{Comparison with the Constant Block-fading Model}

Let us compare the maximal values of $\chi_{\text{const}}$ and $\chi_{\text{gen}}$  for a fixed $\L$, which are obtained for optimal choices of $\T$ and $\R$.
For the constant block-fading model~\eqref{eq:constant_block_memoryless} with block length $\L$, it can be easily verified that the number of degrees of freedom $\chi_{\text{const}}$ given in~\eqref{eq:pre_log_constant} is maximized for $M=\lfloor\L/2\rfloor$. Setting $\T = \R = \lfloor\L/2\rfloor$ to obtain $M=\lfloor\L/2\rfloor$, we conclude that the maximal $\chi_{\text{const}}$ is given 
%\vspace{-1.5mm}
by
\[
\chi_{\text{const,max}}=
\bigg\lfloor\frac{\L}{2}\bigg\rfloor \bigg(\rmv 1 \rmv-\frac{\big\lfloor\frac{\L}{2}\big\rfloor}{\L} \bigg)\,.
%\vspace{1mm}
\]
This can be easily shown to be upper-bounded by $N/4$.
For the generic block-fading model
with $\Q=1$ and $\T \rmv<\rmv \L$, it follows from 
\eqref{eq:pre_log_generic} that the number of degrees of freedom is maximized for $\T=\L \rmv-\rmv 1$ and $\R=(\L \rmv-\rmv 1)^2$, which results in 
\[
\chi_{\text{gen,max}}=\frac{(\L \rmv- 1)^2}{\L}\,.
\]

\begin{figure}[t]
\centering
\includegraphics[width=\linewidth, trim=2cm 11.75cm 11cm 10cm, clip]{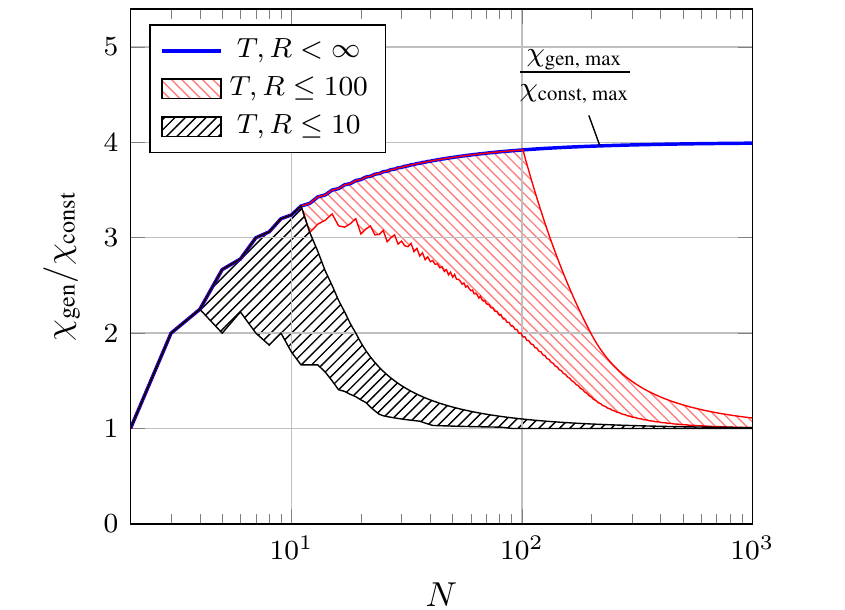} %lbrt
\caption{Ratio between the maximal value of $\chi_{\text{gen}}$ (for the case $\Q=1$) and the maximal value of $\chi_{\text{const}}$ as a function of $\L$, with and without a constraint on the maximal number of antennas. The shaded areas indicate the regions of $\chi_{\text{gen}}/\chi_{\text{const}}$ delimited by the upper bound \eqref{eq:upperbound} and lower bound \eqref{eq:lowerbound} on $\chi_{\text{gen}}$.}
\label{fig:optimized}
\end{figure}

Fig.~\ref{fig:optimized} shows the ratio between the maximal value of $\chi_{\text{gen}}$ (for $\Q=1$) and the maximal value of $\chi_{\text{const}}$
as a function of $\L$.
Because for the generic block-fading model the optimal number of receive antennas grows quadratically  with $\L$, which may yield an unreasonably large number of antennas for practically relevant values of $\L$ (e.g., $1000$ symbols or more), in Fig.~\ref{fig:optimized} we also show the ratio between the maximal values of $\chi_{\text{gen}}$ and $\chi_{\text{const}}$ under  a constraint on the maximal number of antennas.
For the case $\R< \T(\L-1)/(\L-\T)$, which is relevant in the constrained setting, our upper and lower bounds on $\chi_{\text{gen}}$ (see~\eqref{eq:upperbound} and~\eqref{eq:lowerbound} below) do not match. 
The degrees-of-freedom region delimited by the two bounds is represented in Fig.~\ref{fig:optimized} by shaded areas.
%%along with the optimal values of $\T$ and $\R$.
%%  achieving $\chi_{\text{const,max}}$ or $\chi_{\text{gen,max}}$. 
One can see from Fig.~\ref{fig:optimized} that $\chi_{\text{gen, max}}$
 is about four times 
%% the highest achievable 
$\chi_{\text{const, max}}$ when $\L$ grows large.
However, when the maximal number of transmit and receive antennas is constrained, the ratio $\chi_{\text{gen}}/\chi_{\text{const}}$ converges to $1$.
%the number of receive antennas guaranteeing the maximal $\chi_{\text{gen}}$, $\R=(\L-1)^2$, grows quadratically in the block length $\L$.
%Restricting the number of antennas to high, but realistic scenarios negates the degrees-of-freedom gain for large block lengths.
%Still, this setting might, e.g.,  be of interest %for
%This suggests that the degrees-of-freedom gains illustrated in Fig.~\ref{fig:optimized} may be leveraged 
%% may occur 
%in the uplink of massive-MIMO systems \cite{rupe13}, where the use of $100$ or more receive antennas is realistic.}
%\end{itemize}

We emphasize that the only difference between the channel models~\eqref{eq:generic_block_memorylessQ1} and~\eqref{eq:constant_block_memoryless} is that the generic (but deterministic) vectors $\zv_{r,t}$ of~\eqref{eq:generic_block_memorylessQ1} are replaced by the all-one vector in~\eqref{eq:constant_block_memoryless}. 
It is important to note that the generic vectors $\zv_{r,t}$ for which~\eqref{eq:pre_log_generic} holds include vectors that are arbitrarily close to the all-one vector. 
Hence, \emph{arbitrarily small perturbations of the constant block-fading model may result in a significant increase in the number of degrees of freedom}.
%Instead, when the $\{\zv_{r,t}\}$ are generic the number of degrees of freedom is given by~\eqref{eq:pre_log_generic}. 
As we will demonstrate, the potential increase in the number of the degrees of freedom obtained when going from~\eqref{eq:constant_block_memoryless} to~\eqref{eq:generic_block_memorylessQ1} is due to the fact that, under the generic block-fading model~\eqref{eq:generic_block_memorylessQ1},
the received signal vectors 
in the absence of noise span a subspace of higher dimension than under the constant block-fading model~\eqref{eq:constant_block_memoryless}. 
We conclude that the commonly used constant block-fading model results in largely pessimistic capacity estimates at high SNR.

\subsection{Proof Techniques}

To establish~\eqref{eq:pre_log_generic}, we derive upper and lower bounds on capacity that match asymptotically (i.e., in terms of degrees of freedom). 
A similar approach was recently used in \cite{rimodulistbo11} to establish the degrees of freedom for the single-input mul\-tiple-output (SIMO) case. 
However, the proof techniques in \cite{rimodulistbo11} cannot be directly applied 
to the MIMO setting. 
%The main inconvenience originating in the MIMO case is that in order to obtain a sharp lower bound on the the capacity pre-log, we have to deal with mappings which are one-to-one in the SIMO case but turn out to be much more complicated in the MIMO setting.
A key step in \cite{rimodulistbo11} to obtain a tight lower bound on the number of degrees of freedom for the SIMO setting is to perform a change of variables using specific one-to-one mappings that relate the channel gains, the input signals, and the noiseless output signals.
Unfortunately, the corresponding mappings for the MIMO case are not one-to-one,
and hence the change-of-variable argument used in~\cite{rimodulistbo11} cannot be applied.
To overcome this problem, we invoke B\'ezout's theorem in algebraic geometry \cite[Prop.~B.2.7]{VdE00} and show that these mappings are at least finite-to-one almost everywhere.
We also derive a bound on the change of differential entropy that occurs when a random variable undergoes a finite-to-one mapping. 
Finally, we use a property  of subharmonic functions \cite[Th.~2.6.2.1]{Azarin09} to establish that a term appearing in this change of differential entropy is finite.
% of the lower bound in~\cite{koridumohl12}.
%(For concreteness,
%%  and simplicity, 
%we present the proof for a special choice of $T$, $R$, and $\L$; however, the extension to the general case is straightforward.)
%To demonstrate the simplification, we illustrate the main steps in the proof of the lower bound through an example.
%As the proof in~\cite{koridumohl12} is rather involved, this serves also the purpose of making the remaining steps more accessible.

%% \newpage %%%%%%%

%% \subsection{Notation}\label{notation}
\subsection{Notation}
Sets are denoted by calligraphic letters (e.g., $\sI$), and $|\mathcal{I}|$ 
denotes the cardinality of the set $\mathcal{I}$. The indicator function of a set $\sI$ is denoted by $\mathbbmss{1}_{\sI}$. 
Sets of sets are denoted by fraktur letters (e.g., $\mathfrak{M}$).
The set of natural numbers (including zero) $\{0, 1, 2, \dots\}$ is denoted as $\IN$.
We use the notation $[M\!:\!N]$ to indicate the set
%\linebreak %%%%%%%
$\{n\in \IN: M\leq n \leq N\}$ for $M, N \!\in\! \IN$.
Boldface uppercase and lower\-case letters denote matrices and vectors, respectively. Sans serif letters denote random quantities,  e.g., $\rAm$ is a random matrix, 
$\rxv$ is a random vector, and $\rs$ is a random scalar ($\Am, \xv$, and $s$ denote the deterministic counterparts). 
%All conventions for algebraic manipulations that are introduced for deterministic quantities will also be used for random quantities.
The superscripts ${}^{\operatorname{T}}$ and ${}^{\operatorname{H}}$ stand for transposition and 
Hermitian transposition, respectively. 
The all-zero vector or matrix of appropriate size is written as $\0v$, and the $M \rmv\times\rmv M$ identity matrix as
$\Iv_{M}$. The entry
%% element 
in the $i$th row and $j$th column of a matrix $\Am$
is denoted by $[\Am]^{j}_{i}$, and the $i$th entry of a vector $\xv$ by $[\xv]^{}_i$.
For an $M\times N$ matrix $\Am$, we denote by ${[\Am]}_{\sI}^{\sJ}$, where $\sI\subseteq [1\!:\!M]$ and $\sJ\subseteq [1\!:\!N]$,  
the $|\sI|\times|\sJ|$ submatrix of $\Am$ containing the entries $[\Am]^{j}_{i}$ with $i \!\in\! \sI$ and $j \!\in\! \sJ$;
furthermore, we let
${[\Am]}_{\sI} \!\triangleq {[\Am]}_{\sI}^{[1:N]}$ and ${[\Am]}^{\sJ} \! \triangleq {[\Am]}_{[1:M]}^{\sJ}$. 
We denote by ${[\xv]}_\sI\in\IC^{|\sI|}$ the subvector of $\xv$ containing the entries
$[\xv]^{}_i$ with $i\in\sI$.
The diagonal matrix with the entries of $\xv$ in its main diagonal is denoted by $\diag(\xv)$.
We let
$\diag(\Am_1,\dots,\Am_K)$ be the block-diagonal matrix having the matrices $\Am_1,\dots,\Am_K$ on the main block diagonal.
By $\abs{\Am}$ we denote the modulus of the determinant of the square matrix $\Am$.
%The Kronecker product 
%is denoted by $\Am\otimes\Bm$. % with the con\-vention that 
%$\Am\otimes\Bm\Cm \triangleq \Am\otimes (\Bm\Cm)$. 
For $x \!\in\!\IR$, we define $\lfloor x\rfloor\triangleq \max\{m \!\in\! \IZ : m \!\leq\! x\}$ 
and $\lceil x\rceil\triangleq \min\{m \!\in\! \IZ : m \!\geq\! x\}$. 
We write $\E[\cdot]$ for the expectation operator,
and $\rxv\sim\mathcal{CN}(\0v,\Sigm)$ to indicate
%% denote 
that $\rxv$  is a circularly symmetric complex Gaussian random vector with covariance matrix $\Sigm$.
The Jacobian matrix of a differentiable function $\phi$ is written as $\Jm_{\phi}$.
For a function $\phi$ with domain $\sD$ and a subset $\sDt\subseteq \sD$, we denote by $\phi\big|_{\sDt}$ the restriction of $\phi$ to the domain $\sDt$.
We use the Landau notation $f(\rho)=\mathcal{O}(g(\rho))$ to indicate that there exist constants $c_1, c_2 > 0$ such that 
$\abs{f(\rho)}\leq c_1 \,\abs{g(\rho)}$ for $\rho>c_2$. Similarly, we use $f(\rho)=o(g(\rho))$ to indicate that for every $\varepsilon>0$ there exists a constant $c_3 > 0$ such that 
$\abs{f(\rho)}\leq \varepsilon \,\abs{g(\rho)}$ for $\rho>c_3$.
%\vspace{.5mm}

\subsection{Organization of the Paper}
The rest of this paper is organized as follows. 
The system model is formulated in  Section~\ref{sec:syst}. 
In Section~\ref{sec:q1}, we present and discuss our main result on the number of degrees of freedom of the generic block-fading MIMO channel. 
An underlying upper bound is stated and proved in Section~\ref{sec:upperbound}, and a corresponding lower bound is given in Section~\ref{sec:lowerbound}. 
In Section~\ref{sec:prooflowerbound} and in four appendices, we provide a proof of the lower bound.

%%%%%%%%%%%%%%%%%%%%%%%%%%%%%%%
\section{System Model} \label{sec:syst}   
%%%%%%%%%%%%%%%%%%%%%%%%%%%%%%%

%\vspace{1mm}

We consider a MIMO channel with $\T$ transmit and $\R$ receive antennas.
The discrete-time fading process associated with
%% between 
each transmit-receive antenna pair conforms to a
%% same temporally correlated 
block-fading model, %~\eqref{eq:generic_block_memoryless}, 
which results in the following channel input-output relations within a given block of $\L$ channel 
%\vspace{2mm}
uses:
\be\label{eq:model1}
\ryv_{r} \ist=\, \sqrt{\frac{\rho}{\T}} \!\sum_{t\in [1:\T]} \! \diag(\rhv_{r,t}) \, \rxv_{t} \ist+\, \rnv_{r} \,,\quad r\in [1\!:\!\R] \,.
%\vspace{.5mm}
\ee
Here,
$\rxv_{t}\in\IC^{\L}$ is the signal vector originating from the $t$th transmit antenna; 
$\ryv_{r}\in\IC^{\L}$ is the signal vector at the $r$th receive antenna; 
$\rhv_{r,t}\sim\mathcal{CN}(\0v,\Sigm_{r,t})$ is the vector of $\L$ channel coefficients between the $t$th transmit antenna and the $r$th receive antenna;
$\rnv_{r}\sim\mathcal{CN}(\0v,\Iv_{\L})$ is the 
%% additive 
noise vector at the $r$th receive antenna; 
and
$\rho\in \IR^+$ is the SNR. 
The vectors $\rhv_{r,t}$ and $\rnv_{r}$ are assumed to be mutually independent and independent across 
$r\in [1\!:\!\R]$ and $t\in [1\!:\!\T]$, and to change in an independent fashion from block to block (``block-memoryless'' assumption). 
The transmitted signal vectors $\rxv_{t}$ are assumed to be independent of the vectors $\rhv_{r,t}$ and $\rnv_{r}$.
We consider the noncoherent setting, where transmitter and receiver know the covariance matrix $\Sigm_{r,t}$ of $\rhv_{r,t}$ but have no \emph{a priori} knowledge of the realization of $\rhv_{r,t}$. 
% and to satisfy the average per-user power constraints 
%\be
%\label{eq:powerconstr}
%\sum_{t\in \Tind_u} \rmv\E[\|\xv_{t}\|^2] \ist\leq\ist \L\abs{\Tind_u} \,, \quad u\in [1\!:\!\U]\,, 
%\ee
%where $\Tind_u$ denotes the index set of the transmit antennas belonging to user $u\in [1\!:\!\U]$. 

Because the covariance matrix $\Sigm_{r,t}$ is positive-semi\-definite, it can be factorized as
\[
\Sigm_{r,t}=\Zm_{r,t}\Zm_{r,t}^{\operatorname{H}}
\]
with $\Zm_{r,t}\in\IC^{\L\times\Q}$ and $\Q=\rank (\Sigm_{r,t})=\rank (\Zm_{r,t})$.
We can then rewrite the channel coefficient vectors $\rhv_{r,t}$ in terms of $\Zm_{r,t}$ as in~\eqref{eq:generic_block_memoryless}, i.e., 
\vspace*{-.5mm}
\be\label{eq:hass}
\rhv_{r,t} \ist=\ist \Zm_{r,t}\rsv_{r,t}
\ee
where $\rsv_{r,t}\in \IC^{\Q}$, $\rsv_{r,t}\sim\mathcal{CN}(\0v,\Iv_{Q})$. 
%We shall refer to the vectors $\rsv_{r,t}$ as \emph{whitened channel vectors}.
Using~\eqref{eq:hass}, the $\R$ input-output relations~\eqref{eq:model1} can be rewritten 
as
%\vspace{2mm}
\be\label{eq:model2}
\ryv_{r} \ist=\, \sqrt{\frac{\rho}{\T}} \!\sum_{t\in [1:\T]} \! \diag(\Zm_{r,t}\rsv_{r,t}) \, \rxv_{t} \ist+\, \rnv_{r} \,,\quad r\in [1\!:\!\R]
%\ryv\,=\ist \sqrt{\frac{\rho}{\T}} \ist \ryvb \ist+\ist \rnv \,,
\vspace*{-.5mm}
\ee
%with
%\vspace*{2mm}
%\be\label{eq:ybar}
%\ryvb \ist\triangleq\! \sum_{t\in[1:\T]}
%\!\begin{pmatrix}
%\rXm_t\Zm_{1,t} \hspace*{-2mm} &\hspace*{-4mm}&\hspace*{-2mm} \\[-1.5mm]
%\hspace*{-2mm} & \hspace*{-2mm}\ddots\hspace*{-2mm} \\[-.3mm]
%\hspace*{-2mm} & \hspace*{-4mm} & \hspace*{-2mm} \rXm_t\Zm_{\R,t}
%\end{pmatrix}
%\rmv \rsv_t \rmv\in\IC^{\R\L} \ist,
%%%\vspace{-.7mm}
%\ee
%where we have defined 
%$\rXm_{t}\triangleq\diag(\rxv_{t})\in\IC^{\L\times\L}$
%and $\rsv_{t}\triangleq(\rsv_{1,t}^{\operatorname{T}},\dots,$ $\rsv_{\R,t}^{\operatorname{T}})^{\operatorname{T}}\rmv\in\IC^{\R\Q}$.
or in stacked form as
%\vspace{1mm}
\be\label{eq:modelsimp}
\ryv\,=\ist \sqrt{\frac{\rho}{\T}} \ist \ryvb \ist+\ist \rnv, \quad \text{ with }
\ryvb \triangleq \rBm\rsv
%\vspace{1.5mm}
\ee
where 
$\ryv\!\triangleq(\ryv_{1}^{\operatorname{T}} \cdots \ryv_{\R}^{\operatorname{T}})^{\operatorname{T}}\rmv\in\IC^{\R\L}\!$,
$\rnv\!\triangleq\rmv(\rnv_{1}^{\operatorname{T}} \cdots \rnv_{\R}^{\operatorname{T}})^{\operatorname{T}}\in\IC^{\R\L}$,
$\rsv\triangleq(\rsv_{1}^{\operatorname{T}} \cdots \rsv_{\R}^{\operatorname{T}})^{\operatorname{T}}\rmv\in\IC^{\R\T\Q}$ with
$\rsv_{r}\triangleq(\rsv_{r,1}^{\operatorname{T}} \cdots$ $\rsv_{r,\T}^{\operatorname{T}})^{\operatorname{T}}\rmv\in\IC^{\T\Q}$, and
\ba\label{eq:defBm}
& \rBm\triangleq 
\begin{pmatrix}
%\begin{matrix}
 \rBm_{1} \\[-1.5mm]
 & \hspace{-2.5mm}\ddots  \\[-1.5mm]
 &  &  \hspace{-1.5mm}\rBm_{\R}
\end{pmatrix} \in \IC^{\R\L\times \R\T\Q}, \notag\\
& \rule{8mm}{0mm} \text{ with } \rBm_r\triangleq (\rXm_{1}\Zm_{r,1} \cdots \rXm_{\T}\Zm_{r,\T}) \in \IC^{\L\times \T\Q}
\ea
where
$\rXm_{t}\triangleq\diag(\rxv_{t})\in\IC^{\L\times\L}$.
%$\rBm_r\triangleq (\rXm_{1}\Zm_{r,1} \cdots \rXm_{\T}\Zm_{r,\T}) \in \IC^{\L\times \T\Q}$ and
%$\rXm_{t}\triangleq\diag(\rxv_{t})\in\IC^{\L\times\L}$.
%For later use, we define 
For later use, we also define 
$\rxv\triangleq(\rxv_{1}^{\operatorname{T}} \cdots \rxv_{\T}^{\operatorname{T}})^{\operatorname{T}}\rmv\in\IC^{\T\L}$ and
\[
\Zm \ist\triangleq\ist \begin{pmatrix}
\Zm_{1,1} & \hspace*{-2mm}\cdots\hspace*{-2mm} & \Zm_{1,\T}\\[-.5mm]
\vdots & & \vdots \\[-.8mm]
\Zm_{\R,1} & \hspace*{-2mm}\cdots\hspace*{-2mm} & \Zm_{\R,\T}
\end{pmatrix}
\rmv\in \IC^{\R\L\times \T\Q}.
\]
The matrix $\Zm$ contains all information about the correlation of the channel coefficients $\rhv_{r,t}$ (recall that $\Sigm_{r,t}=\Zm_{r,t}\Zm_{r,t}^{\operatorname{H}}$).
We will refer to $\Zm$ as \emph{coloring matrix} and 
use the phrase ``for a generic coloring matrix $\Zm$'' to indicate
%% denote 
that a property holds for \emph{almost every} matrix $\Zm$. Here, ``almost every'' is understood in the precise mathematical sense that the set of all matrices $\Zm$ for which the 
property does \emph{not} hold has Lebesgue measure zero.

In the special (nongeneric) case where $\Q=1$ and each $\Zm_{r,t}\in \IC^{\L\times 1}$ is the all-one vector,~\eqref{eq:model2} reduces to the input-output relation 
of the constant block-fading model given by 
%\vspace{1.5mm}
(cf.\ \eqref{eq:constant_block_memoryless})
\be\label{eq:modelconst}
\ryv_{r} \ist=\, \sqrt{\frac{\rho}{\T}} \!\sum_{t\in [1:\T]} \! \rs_{r,t} \, \rxv_{t} \ist+\, \rnv_{r} \,,\quad r\in [1\!:\!\R] \,.
%\ryv\,=\ist \sqrt{\frac{\rho}{\T}} \ist \ryvb \ist+\ist \rnv \,,
\vspace*{-.5mm}
\ee

%\subsection{MIMO Cyclic Prefix OFDM}
%In a MIMO Cyclic Prefix OFDM setting the discrete system model is given by
%\be\label{eq:ofdm}
%\ryv_r =  \sqrt{\frac{\rho}{\T}} \!\sum_{t\in [1:\T]}\rHm_{r,t}\rxv_t+\rnv_r\,, \quad r\in[1\!:\!\R]
%\ee
%where $\rHm_{r,t}\in \IC^{\L\times \L}$ are circulant matrices with first columns $\rhv_{r,t}\triangleq ([\rh_{r,t}]^{}_1, [\rh_{r,t}]^{}_2, \dots, [\rh_{r,t}]^{}_{\tilde{\Q}}, 0, \dots, 0)^{\trans}$. A well known result for circulant matrices is that the eigenvectors are the columns of the DFT matrix $\Fm$. Thus we can rewrite the system~\eqref{eq:ofdm} in the frequency domain as
%\[
%\tilde{\ryv}_r =  \sqrt{\frac{\rho}{\T}} \!\sum_{t\in [1:\T]}\rLamm_{r,t}\tilde{\rxv}_t+\tilde{\rnv}_r\,, \quad r\in[1\!:\!\R]
%\]
%where $\tilde{\ryv}_r = \Fm^{\operatorname{H}}\ryv_r$, $\Fm\tilde{\rxv}_t = \rxv_t$, $\tilde{\rnv}_r = \Fm^{\operatorname{H}}\rnv_r$, and $\rHm_{r,t}=\Fm\rLamm_{r,t}\Fm^{\operatorname{H}}$.
% The nonzero entries of each diagonal matrix $\rLamm_{r,t}$ correspond to the Fourier transformed power delay profile of the antenna pair $t$-$r$ and, hence, are complex Gaussian distributed with a covariance matrix of a rank $\Q$ that is less than or equal to the delay spread $\tilde{\Q}$ of the channel. Since this is always assumed to be less than the block length $\L$ this system model coincides with the one in~\eqref{eq:model1}. Thus, our results can also be interpreted as the limit of reliable communication over a cyclic prefix OFDM system without a-priori channel state information. 

%\vspace{1.5mm}

%%%%%%%%%%%%%%%%%%%%%%%%%%%%%%%
\section{Characterization of the Number of Degrees of Freedom} \label{sec:q1}   
%%%%%%%%%%%%%%%%%%%%%%%%%%%%%%%

%\vspace{1mm}

\subsection{Main Result} % (fold)
\label{sec:main_result}
%%%%%%%%%%%%%%%%%%%%%%%%%%%%%%%

Because of the block-memoryless assumption, the coding theorem in \cite[Section 7.3]{Gallager68} implies that the capacity of the channel~\eqref{eq:model2} is given by 
\be\label{eq:capacity}
C(\rho) \ist=\ist \frac{1}{\L}\sup  I(\rxv \ist;\ryv) \,.
%\vspace{1mm}
\ee
Here,
 $I(\cdot \ist;\cdot)$ denotes mutual information \cite[p.\,251]{Cover91} and 
the supremum is taken over all probability distributions of $\rxv$
that satisfy the average-power constraint
\be
\label{eq:powerconstr}
\E[\|\rxv\|^2] \ist\leq\ist \T\L \,.
\ee
%Because of~\eqref{eq:powerconstr} and the noise variance normalization, we can think of $\rho$ as the SNR.
The number of degrees of freedom  is defined as  
\be\label{eq:prelog1}
\chi \ist \triangleq\lim_{\rho\to\infty}\frac{C(\rho)}{\log \rho }
\ee
which corresponds to the expansion
\be\label{eq:capacityexpansion}
C(\rho) = \chi \log \rho + o(\log \rho)\,.
\ee

Our main result is stated in the following 
%\vspace{1.5mm}
theorem.

\begin{theorem}\label{THmaintheoremq1}
Let $\T \!<\rmv \L/\Q$ and $\R\geq \T(\L \rmv-\rmv 1)/(\L \rmv-\rmv \T\Q)$. For a channel conforming to the generic block-fading model, i.e., the channel~\eqref{eq:model2}  with generic coloring matrix $\Zm$, the number of degrees of freedom is 
given 
%\vspace{-1mm}
by 
\be\label{eq:maintheoremq1}
\chi_{\text{gen}}=\ist
\T\bigg(1\rmv-\rmv \frac{1}{\L}\bigg)\,.
\ee
%\vspace{-1.5mm}
\end{theorem} 

\begin{IEEEproof}[\hspace{-1em}Proof]
%% To establish~\eqref{eq:maintheoremq1}, we derive in 
In Section~\ref{sec:upperbound}, we will 
%% derive a capacity upper bound with pre-log equal to
%% coinciding with the right-hand-side of~
show that $\chi_{\text{gen}}$ is upper-bounded by $\T\ist (1\rmv-\rmv 1/\L)$ for all choices of $\T, \R, \L, \Q,$ and $\Zm$.
%%~\eqref{eq:maintheoremq1}.
%% The proof is concluded by noting that $\T(1-{1}/{\L})$ 
%For $\T \!<\rmv \L/\Q$, $\R\geq \T(\L \rmv-\rmv 1)/(\L \rmv-\rmv \T\Q)$, and a generic correlation,
In Section~\ref{sec:lowerbound}, we will show that this upper bound is achievable when $\T \!<\rmv \L/\Q$, $\R\geq \T(\L \rmv-\rmv 1)/(\L \rmv-\rmv \T\Q)$, and $\Zm$ is generic (see Corollary~\ref{cor:lowerbound}).
%% In Section~\ref{sec:lowerbound}, we will illustrate the main steps of the proof of \cite[Th.~1]{koridumohl12} for the simple case 
%% $\T \!=\rmv 2, \R \rmv=\rmv 3, \L \rmv=\rmv 4$. 
%% We also present a novel bound on the change in differential entropy when a random variable undergoes a finite-to-one mapping, 
%% which allows us to drastically simplify the proof compared to the one reported in~\cite{koridumohl12}.
%\vspace{1.5mm} 
\end{IEEEproof}

%%\vspace{1mm}

\subsection{Degrees of Freedom Gain}
%% Discussion %% Rationale Behind the Pre-Log Gain} % (fold)
\label{sec:rationale_behind_the_pre_log_gain}
As discussed in Section~\ref{sec:intro},~\eqref{eq:maintheoremq1} implies that the maximal achieveable number of degrees of freedom in the generic block-fading model can be about four times as large as the number of degrees of freedom in the constant block-fading model~\eqref{eq:pre_log_constant}.
We will now provide some intuition regarding
%% behind 
this gain. For concreteness, we 
%% will focus on 
consider the case $\T \rmv=\rmv 2, \R \rmv=\rmv 3, \Q \rmv=\rmv 1, \L \rmv=\rmv 4$.
In this case,~\eqref{eq:pre_log_constant} and~\eqref{eq:maintheoremq1} give $\chi_{\text{const}}=1$ and $\chi_{\text{gen}}=3/2$, respectively.
%(hence, $\Lt=\L$).
%%  (this case will also be used
%% we shall use this example 
%% to illustrate the proof of the lower bound in Section~\ref{sec:lowerbound}).

The number of degrees of freedom characterizes the channel capacity in a regime where the noise can ``effectively'' be ignored. 
Thus, according to the intuitive argumentation in~\cite[Section~III]{moridu13}, the number of degrees of freedom should be equal to the
number of entries of $\rxv \rmv\in\rmv \IC^{8}$ that can be deduced from the corresponding received vector $\ryv \rmv\in\rmv \IC^{12}$ %\linebreak %%%%% 
in the absence of noise, 
divided by the block length $\L \!=\rmv 4$.

In the constant block-fading model~\eqref{eq:modelconst}, the %\nolinebreak %%%%%%
noise\-less %\nolinebreak %%%%%% 
received vectors $\ryvb_r \rmv= \rs_{r,1}\rxv_1 \rmv+ \rs_{r,2} \ist \rxv_2$, $r \rmv=\rmv 1,2,3$ 
%% all 
belong to the two-dimensional subspace spanned by $\{\rxv_1, \rxv_2\}$. 
Hence, the received vectors $\ryvb_1, \ryvb_2, \ryvb_3$ are linearly dependent, and two of them 
%% already 
contain all the information available about $\rxv$. 
From two of the received vectors, we obtain $2\cdot 4$ scalar
%% $8$ 
equations in $8 + 4$ scalar
variables ($\rxv,
%% \rxv_1, \rxv_2, 
\rs_{1,1}, \rs_{1,2}, \rs_{2,1}, \rs_{2,2}$). Since we do not have control 
of the variables $\rs_{r,t}$, one way to reconstruct $\rxv$ is to fix
%% freeze 
four
%% $4$ 
of its entries (or, equivalently, to transmit four
%% $4$ 
pilot symbols) to obtain
%% end up with 
eight
%% $8$ 
equations in eight
%% $8$ 
variables. By solving this system of equations, we obtain the remaining four
%% $4$ 
entries of $\rxv$. Hence, we can deduce four entries of $\rxv$ from $\ryvb$. We conclude that the
%% yields 
 number of degrees of freedom is $4/4=\rmv 1$, which is in agreement with~\eqref{eq:pre_log_constant}.
%% In contrast, 

In the generic block-fading model~\eqref{eq:model2}, on the other hand, the received vectors without noise 
\[
\ryvb_r = \diag(\Zm_{r,1}\rs_{r,1}) \ist \rxv_1 + \diag(\Zm_{r,2}\rs_{r,2}) \ist \rxv_2, \quad r \rmv= 1,2,3
\] 
span a three-dimensional subspace almost surely. 
Hence, we obtain a system of $3\cdot 4$ equations in $8 + 6$ variables ($\rxv,
%% \rxv_1, \rxv_2, 
\rs_{1,1}, \rs_{1,2}, \rs_{2,1}, \rs_{2,2}, \rs_{3,1}, \rs_{3,2}$). 
Fixing
%% Freezing 
two
%% $2$ 
entries of $\rxv$, we are able to recover the remaining six
%% $6$ 
entries.
Hence, the number of degrees of freedom is $6/4=3/2$, which is in agreement with~\eqref{eq:maintheoremq1}.

This argument suggests that the reason why the generic block-fading
%% memoryless 
model yields a larger number of degrees of freedom than the constant block-fading model is that 
%% under the generic model, 
the noiseless received vectors span a subspace of $\IC^{\L}$ of higher dimension.
%%  than in the constant model case.

%%%%%%%%%%%%%%%%%%%%%%%%%%%%%%%
\section{Upper Bound} \label{sec:upperbound}   
%%%%%%%%%%%%%%%%%%%%%%%%%%%%%%%
%\vspace{1mm}

The following upper bound on the number of degrees of freedom of the 
%% generic block-fading 
channel~\eqref{eq:model2} 
%% In this section, we will prove 
%% is stated by the following result, which 
holds for every
%% values of 
$\T$, $\R$, $\Q$, $\L$, and  
%\vspace{1.5mm}
$\Zm$. The assumption of a generic coloring matrix $\Zm$ is not required.

\begin{theorem}\label{THupperbound}
The number of degrees of freedom of the channel~\eqref{eq:model2} satisfies 
\be\label{eq:upperbound}
\chi_{\text{gen}} \ist\leq\ist \T\bigg(\rmv 1 \rmv-\rmv \frac{1}{\L}\rmv \bigg)\,. %\vspace{1.5mm}.
\ee
\end{theorem}

\begin{IEEEproof}[\hspace{-1em}Proof]
We will show that the 
%% capacity 
number of degrees of freedom is upper-bounded by $\T$
times the number of degrees of freedom of a constant block-fading SIMO channel; the 
%% desired 
result then follows from~\eqref{eq:pre_log_constant}.
To this end, we will rewrite each output vector $\ryv_r$ as the sum of the output vectors of $\T$ SIMO systems with $\R\Q$ receive antennas each. This will be achieved by splitting the additive noise variables 
%\pagebreak %%%%%%%
appropriately.

From~\eqref{eq:model2},
%%  we have that 
the $i$th entry of the received vector $\ryv_r$ is given 
%\vspace{1mm}
by
\ba
& [\ryv_r]^{}_i =\ist \sqrt{\frac{\rho}{\T}} \!\rmv \sum_{t\in[1:\T]} \sum_{q\in[1:\Q]}\!\!  [\Zm_{r,t}]^{q}_{i} \ist [\rsv_{r,t}]^{}_{q} \ist   [\rxv_t]^{}_i \ist+\ist [\rnv_r]^{}_i 
%\notag \\[-2mm]
%& \rule{55mm}{0mm}  r \!\in\! [1\!:\!\R]\,.
\label{eq:singleinout}
%\vspace{1mm}
\ea
for $r \in [1\!:\!\R]$.
%To introduce virtual constant block-fading SIMO channels, 
We first decompose the noise variables according 
%\vspace{.5mm}
to
\be\label{eq:noisedecomposition}
[\rnv_r]^{}_i = \sum_{t\in [1:\T]}\sum_{q\in[1:\Q]}\frac{[\Qm_{r,t}]^{q}_{i}\ist }{\sqrt{K\T}} [\rnvt_{q,r,t}]^{}_i+ [\rnv'_r]^{}_i\,.
%\vspace{1mm}
\ee
Here, all $[\rnvt_{q,r,t}]^{}_i$ and $[\rnv'_r]^{}_i$ are mutually independent and independent of all $\rxv_t$ and $\rsv_{r,t}$.
Furthermore, $[\rnvt_{q,r,t}]^{}_i  \sim \mathcal{CN}(0,1)$,
\ba
% \notag \\
[\rnv'_r]^{}_i & \sim \mathcal{CN} \bigg( 0, 1 - \sum_{t\in[1:\T]}\sum_{q\in[1:\Q]} \frac{\abs{[\Qm_{r,t}]^{q}_{i}}^2}{K\T} \bigg)\,,\notag 
\ea
 and $K$ is a finite constant satisfying\footnote{This condition on $K$ is required to ensure that the variance of all random variables $[\rnv'_r]^{}_i$ is positive.}
\[
K >\rmv \max_{r\in [1:\R],\ist i\in[1:\L]}\sum_{t\in[1:\T]}\sum_{q\in[1:\Q]}\abs{[\Zm_{r,t}]^{q}_{i}}^2\,.
\]
We next define $\T$ ``virtual'' constant block-fading SIMO channels with 
%SNR $K\rho$ and 
$\R\Q$ receive antennas each:
\ba\label{eq:SIMO}
& [\ryvt_{q,r,t}]^{}_i =\ist \sqrt{K\rho} \,\ist [\rsv_{r,t}]^{}_{q} \ist [\rxv_t]^{}_i +\ist [\rnvt_{q,r,t}]^{}_i \,, \notag \\
& \rule{25mm}{0mm}    i \!\in\! [1\!:\!\L], r \!\in\! [1\!:\!\R], q\in[1\!:\!\Q]
\ea
for $t \!\in\! [1\!:\!\T]$.
%% We write $\ryvt_t$ for the stacked output vector of the $t$th SIMO channel.
Inserting~\eqref{eq:noisedecomposition} into~\eqref{eq:singleinout} and using~\eqref{eq:SIMO}, it can be verified that~\eqref{eq:singleinout} can be rewritten 
%using~\eqref{eq:SIMO} and~\eqref{eq:noisedecomposition} 
as
\be\label{eq:MIMO_SIMO}
[\ryv_r]^{}_i =\ist \frac{1}{\sqrt{K\T}} \!\sum_{t\in[1:\T]} \sum_{q\in[1:\Q]}\! [\Zm_{r,t}]^{q}_{i} \ist [\ryvt_{q,r,t}]^{}_i +\ist [\rnv'_r]^{}_i \,.
%\vspace{.5mm}
\ee
Let $\ryvt_t \triangleq (\ryvt_{1,1,t}^{\operatorname{T}} \cdots\ist \ryvt_{\Q,\R,t}^{\operatorname{T}})^{\operatorname{T}}\!\in\rmv\IC^{\Q\R\L}\rmv$.
By~\eqref{eq:MIMO_SIMO}, the random variable $\ryv$ depends on $\rxv$ only via the random variables $\{\ryvt_t\}_{t \in [1:\T]}$.
Hence, the data-processing inequality~\cite[eq.~(2.3.19)]{Gallager68} %applied to~\eqref{eq:MIMO_SIMO} 
%% %\vspace{-2mm}
yields
\be\label{eq:dataproc1}
I(\rxv\ist ;\ryv) \ist\leq\ist I(\rxv\ist ;\ryvt_1,\dots, \ryvt_{\T})\,.
\ee
%with $\ryvt_t \triangleq (\ryvt_{1,1,t}^{\operatorname{T}} \cdots\ist \ryvt_{\Q,\R,t}^{\operatorname{T}})^{\operatorname{T}}\!\in\rmv\IC^{\Q\R\L}\rmv$.
The right-hand side of~\eqref{eq:dataproc1} can be upper-bounded as follows:
\begin{align}
I(\rxv\ist ;\ryvt_1,\dots, \ryvt_{\T}) & =  h(\ryvt_1,\dots, \ryvt_{\T}) \ist-\ist h(\ryvt_1,\dots, \ryvt_{\T} |\ist \rxv) \notag \\[.5mm]
& \stackrel{\hidewidth (a) \hidewidth}= h(\ryvt_1,\dots, \ryvt_{\T}) \ist-\! \sum_{t\in[1:\T]} \!h(\ryvt_t | \ist \rxv_t) \notag \\[-.3mm]
& \stackrel{\hidewidth (b) \hidewidth}\leq \!\sum_{t\in[1:\T]} \!\! \big[ h(\ryvt_t) - h(\ryvt_t | \ist \rxv_t) \big] \notag \\[.5mm]
& =\sum_{t\in[1:\T]} \!\! I(\rxv_t\ist ;\ryvt_t) \,.
%& \stackrel{\hidewidth (b) \hidewidth}\leq \T\L \ist C_{\text{const}}(K\rho) \notag \\[0mm]
\label{eq:upperbound1}%\\[-8.5mm]
%\notag
\end{align}
Here, $h(\cdot)$ denotes differential entropy \cite[Ch.~8]{Cover91},
$(a)$ holds because $\ryvt_1,\dots,\ryvt_{\T}$ are conditionally independent given $\rxv$, and $(b)$ follows from the chain rule for differential entropy \cite[Th.~8.6.2]{Cover91} and because conditioning does not increase differential entropy. 
Since (by assumption) the input vector $\rxv$ satisfies the power constraint~\eqref{eq:powerconstr}, we conclude that, trivially, also each subvector $\rxv_t$ satisfies the individual power constraint $\E[\norm{\rxv_t}^2]\leq \T\L$. 
Thus, the SNR (i.e., the expected power of the noiseless received signal divided by the noise power) of each ``virtual'' constant block-fading SIMO channel~\eqref{eq:SIMO} is given 
%\vspace{1mm}
by 
\bas
\frac{\E[\norm{\sqrt{K \rho}\,[\rsv_{r,t}]^{}_{q}\rxv_t}^2]}{\E[\norm{\rnvt_{q,r,t}}^2]}
& = \frac{K \rho \,\E[\abs{[\rsv_{r,t}]^{}_{q}}^2]\,\E[\norm{\rxv_t}^2]}{\E[\norm{\rnvt_{q,r,t}}^2]} \\
& \leq \frac{K \rho \, \T \L}{\L} \\
& = \T K \rho \,.
%\vspace{1mm}
\eas
%Hence, we can upper-bound each mutual information $I(\rxv_t\ist ;\ryvt_t)$ by the capacity of a SIMO constant block-fading channel of SNR $\T K \rho$. 
By~\eqref{eq:pre_log_constant} and~\eqref{eq:capacityexpansion}, the capacity of a constant block-fading SIMO channel of SNR $\T K \rho$ is of the form\footnote{Since the number of transmit antennas is one for a SIMO channel, we have $M=1$ in~\eqref{eq:pre_log_constant}.} $(1-1/\L) \log(\T K \rho)+ o(\log \rho )$. 
Since, by~\eqref{eq:capacity}, the capacity is the supremum of the mutual information divided by the block length, we can upper-bound each mutual information $I(\rxv_t\ist ;\ryvt_t)$, $t\in[1\!:\!\T]$ by $\L$ times the capacity. This results 
%\vspace{-2.5mm}
in 
\bas
I(\rxv_t\ist ;\ryvt_t) & \leq \L \bigg(\Big(1 - \frac{1}{\L}\Big) \log (\T K \rho) + o(\log \rho )\bigg) \\
& =(\L - 1)\log (\T K \rho) + o(\log \rho )\,.
\eas
Hence, continuing~\eqref{eq:dataproc1} and~\eqref{eq:upperbound1}, we obtain
\begin{align}\label{eq:boundconstSIMO}
I(\rxv\ist ;\ryv) & \leq 
\sum_{t\in[1:\T]} \!\! I(\rxv_t\ist ;\ryvt_t) \notag \\
%& \ist\stackrel{(b)}\leq \T\L \ist C_{\text{const}}(K\rho) \notag \\[0mm]
& \leq \, \T(\L \rmv-\rmv 1)\log(\T K \rho) \ist+\ist o(\log \rho ) \notag \\[.5mm]
& \stackrel{\hidewidth(a)\hidewidth}=\, \T(\L \rmv-\rmv 1)\log \rho  \ist+\ist o(\log \rho )
\end{align} 
where $(a)$ holds because $\log(\T K \rho) = \log \rho+\log(\T K)$.
%from~\eqref{eq:pre_log_constant} with $M=1$, 
%after noting that $I(\rxv_t\ist ;\ryvt_t)$ is upper-bounded by the capacity of a constant block-fading SIMO channel with $\R\Q$ receive antennas,  block length $\L$, and SNR equal to $K\rho$.
%% , since here $C_{\text{const}}(K\rho)$ refers to the capacity of constant block-fading SIMO channels.
Thus, the mutual information $I(\rxv;\ryv)$ with $\rxv$ satisfying the power constraint~\eqref{eq:powerconstr} is upper-bounded by~\eqref{eq:boundconstSIMO}. 
Inserting~\eqref{eq:boundconstSIMO} into~\eqref{eq:capacity} yields
\begin{align*}
C(\rho) & 
\leq \T\ist\frac{\L \!-\! 1}{\L}\log \rho  \ist+\ist o(\log \rho )
\end{align*}
from which~\eqref{eq:upperbound} follows via~\eqref{eq:prelog1}.
%% %\vspace{1.5mm} 
\end{IEEEproof}

\section{Lower Bound} \label{sec:lowerbound}   
%%%%%%%%%%%%%%%%%%%%%%%%%%%%%%%

%\vspace{1mm}

We first derive a lower bound on $\chi_{\text{gen}}$ assuming that $\Tprop\leq \min\{\T,\R\}$ transmit antennas are effectively used (i.e., $\rxv_{\Tprop+1}, \dots, \rxv_{\T}$ are set to zero).
Then we maximize the lower bound by identifying the optimal number $\Tprop$ of transmit antennas to use.%\vspace{1.5mm} ~
%We will obtain the main result of this section, which is stated in Theorem~\ref{th:lowerbound} below, by maximizing with respect to $\T$ the lower bound given in the following proposition. 

\begin{pro}\label{pro:prelim}
%For $\T \!\leq\!\R$ and ,
The number of degrees of freedom of the chan\-nel~\eqref{eq:model2} for a generic coloring matrix $\Zm$
is lower-bounded 
%\vspace{-1mm}
by
\ba
\chi_{\text{gen}} \,\geq\, \chi_{\text{low}}(\Tprop) \,\triangleq\, \min\rmv\bigg\{\Tprop\bigg(1 \rmv-\rmv \frac{1}{\L}\bigg), \R\bigg(1 \rmv-\rmv \frac{\Tprop\Q}{\L}\bigg)\bigg\} %\notag \\
\label{eq:prelim} %\\[-10mm] \notag
\ea
for all 
%\pagebreak %%%%%%%
$\Tprop\leq \min\{\T,\R\}$.%\vspace{1.5mm} 
\end{pro}
\begin{IEEEproof}[\hspace{-1em}Proof]
See Section~\ref{sec:prooflowerbound}.%\vspace{1.5mm} 
\end{IEEEproof}

%In Section~\ref{sec:rationale_behind_the_pre_log_gain} we compared the number of equations we obtain by the noiseless receive vectors and the number of variables we have to derive from those equations. 
The minimum in~\eqref{eq:prelim} is given by $\chi_{\text{low}}(\Tprop)=\Tprop(1-1/\L)$ when the number $\R$ of receive antennas is large enough (i.e., $\R\geq \Tprop(\L-1)/(\L-\Tprop\Q)$).
% to allow the inference of all unknown variables 
%(see the discussion in Section~\ref{sec:rationale_behind_the_pre_log_gain}); 
%we will show in Section~\ref{sec:dimensioncounting} that this amounts to the condition $\R\geq \Tprop(\L-1)/(\L-\Tprop\Q)$.
%it is possible to obtain all available information about the input signals due to (cf.\ Section~\ref{}). 
In contrast, $\chi_{\text{low}}(\Tprop)=\R(1-\Tprop\Q/\L)$ when the number of degrees of freedom is constrained by the limited number of receive antennas (i.e., $\R< \Tprop(\L-1)/(\L-\Tprop\Q)$).

The main result of this section is stated in the following 
%\vspace{1.5mm}
theorem. 

\begin{theorem}\label{th:lowerbound}
The number of degrees of freedom of the channel~\eqref{eq:model2} for a generic coloring matrix $\Zm$
is lower-bounded 
%\vspace{-2mm}
by
\ba\label{eq:lowerbound}
\chi_{\text{gen}}  \geq \chi^*_{\text{low}}  & \triangleq
\max_{\Tprop\leq \min\{\T,\R\}}\chi_{\text{low}}(\Tprop) \notag \\
& = \begin{cases}
\T\bigg(1 \rmv-\rmv \dfrac{1}{\L}\bigg), & \text{ if } \T\leq\T_{\text{opt}} \\
\eta, & \text{ if } \T>\T_{\text{opt}}
\end{cases}
%%\vspace{-2mm}
\ea
where
%\vspace{.5mm}\vspace{1mm}
\be\label{eq:topt}
\T_{\text{opt}} \,\triangleq\, \frac{\R\L}{\L+\R\Q-1}
%\,\leq\, \min\{\R, \L/\Q\} \,.
%\vspace{2mm}
\ee
and
\be\label{eq:eta}
\eta \,\triangleq\, \max\bigg\{ \R\bigg(1-\frac{\lceil \T_{\text{opt}}\rceil\Q}{\L}\bigg), \lfloor\T_{\text{opt}}\rfloor\bigg(1\rmv-\rmv \frac{1}{\L}\bigg) \bigg\} \,.
%\vspace{-.2mm}
\ee
\end{theorem} 
\begin{IEEEproof}[\hspace{-1em}Proof]
The idea behind the bound $\chi^*_{\text{low}}$ in~\eqref{eq:lowerbound} is to obtain the tightest (i.e., largest) of the lower bounds $\chi_{\text{low}}(\Tprop)$ in~\eqref{eq:prelim} for $\T$ transmit antennas by maximizing $\chi_{\text{low}}(\Tprop)$ 
with respect to the number of effectively used transmit antennas $\Tprop\leq \min\{\T,\R\}$.
%(note that Proposition~\ref{pro:prelim} holds only for $\Tprop\leq \R$).
%(note that we can always switch off some antennas).
%Thus, we will take the maximum of $\chi_{\text{low}}(\Tprop)$ in~\eqref{eq:prelim} with respect to $\Tprop\leq \min\{\T,\R\}$. Here, $\Tprop$ is also restricted by $\Tprop\leq\R$ because Proposition~\ref{pro:prelim} holds only for $\Tprop\leq \R$.
According to~\eqref{eq:prelim}, $\chi_{\text{low}}(\Tprop)$ is the minimum of two quantities where the first, $\Tprop(1-1/\L)$, is monotonically increasing in $\Tprop$ 
and the second, $\R(1-\Tprop\Q/\L)$, is monotonically decreasing in $\Tprop\rmv$. 
Hence, $\chi_{\text{low}}(\Tprop)$ attains its maximum at the intersection point $\T_{\text{opt}}$ defined in~\eqref{eq:topt}.
If $\T\leq\T_{\text{opt}}$, we are for all $\Tprop\leq \min\{\T, \R\}$ in the regime where $\chi_{\text{low}}(\Tprop)$ is monotonically increasing, and thus the best choice is to use $\Tprop=\T$ transmit antennas 
(note that because $\T\leq \T_{\text{opt}} \stackrel{\hidewidth \eqref{eq:topt}\hidewidth }\leq \R\L/\L = \R$, the choice $\Tprop=\T$ in Proposition~\ref{pro:prelim} is possible). 
Thus, in this case we have 
$
\chi^*_{\text{low}}=\chi_{\text{low}}(\T) = \T(1-1/\L)
$,
which yields the first case in~\eqref{eq:lowerbound}.
If $\T>\T_{\text{opt}}$, we would like to use $\T_{\text{opt}}$ transmit antennas, but we have to take into account that $\T_{\text{opt}}$ may  be noninteger. 
Thus, we take the maximum of the bounds $\chi_{\text{low}}(\Tprop)$ resulting from the closest integers, $\chi_{\text{low}}(\lfloor\T_{\text{opt}}\rfloor)$ and $\chi_{\text{low}}(\lceil \T_{\text{opt}}\rceil)$, which yields $\eta$ in~\eqref{eq:eta}. 
This concludes the proof.%\vspace{1.5mm} 
\end{IEEEproof}

%\begin{remark}
%The set $\sQ$ will be specified in Definition~\ref{def:qind} in Section~\ref{sec:boundh}. 
%\end{remark}
%
%\begin{remark}
%For a fixed $\R$, the maximum value of $\chi^*_{\text{low}}$ in~\eqref{eq:lowerbound} is obtained by using either $\lfloor \T_{\text{opt}}\rfloor$ or $\lceil \T_{\text{opt}}\rceil$ transmit antennas. This implies that the optimal number of transmit antennas is 
%% always 
%upper-bounded by $\lceil \T_{\text{opt}}\rceil$.
%\end{remark}

%\begin{remark}
%For $\L=1$ we have that $\chi^*_{\text{low}} =0$.
%\end{remark}

\begin{remark}
For $\L\geq 2$, the optimal number of transmit antennas $\T_{\text{opt}}$ is upper-bounded as follows:
\be\label{toptbound}
\T_{\text{opt}} < \frac{\L}{\Q}\,.
\ee
In fact, $\T_{\text{opt}}= \R\L/(\L+\R\Q-1) < \R\L/(\R\Q)=\L/\Q$.
%\vspace{1.5mm}
\end{remark}

\begin{remark}
For $\L \rmv=\rmv \Q \geq 2$, we have by~\eqref{toptbound} that $\T_{\text{opt}}<1$. Hence, $\T>\T_{\text{opt}}$ and thus, by~\eqref{eq:lowerbound} and~\eqref{eq:eta}, $\chi^*_{\text{low}}= \eta = \max\big\{ \R(1- \Q/\L), 0 \big\}=0$. 
Similarly, we obtain for $\L=1$ that $\chi_{\text{low}}(\Tprop)\leq 0$ for all $\Tprop$, which yields $\chi^*_{\text{low}}  \leq 0$. 
Hence, our lower bound $\chi^*_{\text{low}}$ is trivial. 
In these scenarios, the capacity grows double-logarithmically in the SNR $\rho$ \cite{lamo03, dubo11}.%\vspace{1.5mm} 
\end{remark}

\begin{remark} \label{re:min}
The lower bound $\chi^*_{\text{low}}$ in~\eqref{eq:lowerbound} can be equivalently expressed as
\[
\chi^*_{\text{low}} \ist=\, \min\bigg\{\T\bigg(1 \rmv-\rmv \frac{1}{\L}\bigg), \eta\bigg\}\,.\vspace{1.5mm}
\]
\end{remark}

\begin{cor}\label{cor:lowerbound}
Let $\L\geq 2$. For the lower bound $\chi_{\text{low}}^*$ in Theorem~\ref{th:lowerbound}, the following properties hold:
\begin{enumerate}
\renewcommand{\theenumi}{(\roman{enumi})}
\renewcommand{\labelenumi}{(\roman{enumi})}
\item For $\T\geq \L/\Q$, we have $\T>\T_{\text{opt}}$ and $\chi_{\text{low}}^*=\eta$. \label{en:lowerbound1}
\item For $\T < \L/\Q$ and $\R\geq \T(\L-1)/(\L-\T\Q)$, we have $\T\leq\T_{\text{opt}}$ and $\chi_{\text{low}}^*=\T(1-1/\L)$.\label{en:lowerbound2}
\item For $\T < \L/\Q$ and $\R < \T(\L-1)/(\L-\T\Q)$, we have $\T>\T_{\text{opt}}$ and $\chi_{\text{low}}^*=\eta$.\label{en:lowerbound3}
\item For fixed $\L$ and $\Q$, $\chi_{\text{low}}^*$ attains its maximal value  
for $\T=\lfloor(\L-1)/\Q\rfloor$ transmit antennas and $\R=\lceil(\L-1)^2/\Q\rceil$ receive antennas; this maximal value of $\chi^*_{\text{low}}$ equals $\lfloor(\L-1)/\Q\rfloor(1-1/\L)$. %\vspace{1.5mm} 
\label{en:lowerbound4}
\end{enumerate}
\end{cor}
\begin{IEEEproof}[\hspace{-1em}Proof]
By~\eqref{toptbound}, the inequality $\T\geq \L/\Q$ implies $\T>\T_{\text{opt}}$, from which Property~\ref{en:lowerbound1} follows by~\eqref{eq:lowerbound}. 
For $\T<\L/\Q$, the following equivalence holds:
\begin{align*}
\T\leq\T_{\text{opt}} \stackrel{ \eqref{eq:topt} }= \frac{\R\L}{\L+\R\Q-1}\quad & %\stackrel{\hidewidth \eqref{eq:topt} \hidewidth}\Leftrightarrow \quad \T \leq \frac{\R\L}{\L+\R\Q-1} \notag \\
%& \Leftrightarrow \quad \T (\L+\R\Q-1) \leq \R\L \notag \\
%& \Leftrightarrow \quad \T(\L -1) \leq \R(\L-\T\Q) \notag \\
%& 
\Leftrightarrow \quad \T\frac{\L -1}{\L-\T\Q} \leq \R \,.
%\label{eq:equivalencetr}
\end{align*}
Thus, the conditions in Properties~\ref{en:lowerbound2} and~\ref{en:lowerbound3} imply $\T\leq\T_{\text{opt}}$ and $\T>\T_{\text{opt}}$, respectively, and the expressions of $\chi_{\text{low}}^*$ given in Properties~\ref{en:lowerbound2} and~\ref{en:lowerbound3} follow immediately from the case distinction in~\eqref{eq:lowerbound}. 

To prove Property~\ref{en:lowerbound4}, we first show that $\chi_{\text{low}}^*\leq \lfloor(\L-1)/\Q\rfloor(1-1/\L)$ for arbitrary $\T$ and $\R$. Subsequently, we will show that this upper bound is achievable for the proposed number of antennas. 
We first note that for each $\Tprop\leq \L/\Q$, the lower bound $\chi_{\text{low}}(\Tprop)$ in~\eqref{eq:prelim} is monotonically nondecreasing in $\R$. 
Furthermore, for $\Tprop> \L/\Q$, $\chi_{\text{low}}(\Tprop)$ is negative and can be ignored in the maximization process, i.e., we have $\chi_{\text{low}}^*=\max_{\Tprop\leq \min\{\T,\R, \L/\Q\}}\chi_{\text{low}}(\Tprop)$.
This implies that $\chi_{\text{low}}^*$ is---as a maximum of nondecreasing functions---also monotonically  nondecreasing in $\R$. 
Hence, to obtain an upper bound on $\chi_{\text{low}}^*$, we can assume $\R$ arbitrarily large without loss of generality. We choose $\R >(\L-1)^2/\Q$. 
%With this choice, we obtain
Simple algebraic manipulations yield the equivalence
\begin{align}
\R >\frac{(\L-1)^2}{\Q} \quad & \Leftrightarrow 
%\quad \R\Q >\L^2-2\L+1 \notag \\
%& \Leftrightarrow \quad \R\Q >\L^2-2\L+1+\R\L\Q-\R\L\Q \notag \\
%& \Leftrightarrow \quad \R\L\Q >\L^2+\R\L\Q-\L-\L-\R\Q+1 \notag \\
%& \Leftrightarrow \quad \R\L\Q > (\L-1)(\L+\R\Q-1) \notag \\
%& \Leftrightarrow 
\quad \T_{\text{opt}}=\frac{\R\L}{\L+\R\Q-1} > \frac{\L-1}{\Q} %\notag \\
%& \Rightarrow \quad \lceil\T_{\text{opt}}\rceil \geq \frac{\L}{\Q}
 \label{eq:equivrlarge}\,.
\end{align}
This implies $\lceil\T_{\text{opt}}\rceil\Q>\L-1$ and further, because both sides of this strict inequality are integers, that $\lceil\T_{\text{opt}}\rceil\Q\geq\L$.
Thus, the first argument of the maximum defining $\eta$ in~\eqref{eq:eta} satisfies
\[
\R\bigg(1-\frac{\lceil \T_{\text{opt}}\rceil\Q}{\L}\bigg) \leq \R(1-1)=0
\]
and, hence, $\eta$ reduces to $\eta =\lfloor\T_{\text{opt}}\rfloor(1-1/\L)$. 
By~\eqref{eq:lowerbound}, we have that $\chi_{\text{low}}^*$ is either equal to $\T(1-1/\L)$ (for $\T\leq \T_{\text{opt}}$) or equal to  $\eta= \lfloor\T_{\text{opt}}\rfloor(1-1/\L)$ (for $\T> \T_{\text{opt}}$). 
In both cases we have $\chi_{\text{low}}^*\leq \lfloor\T_{\text{opt}}\rfloor(1-1/\L)$.
Since $\lfloor\T_{\text{opt}}\rfloor \leq \lfloor(\L-1)/\Q\rfloor$ 
by\footnote{By~\eqref{toptbound}, $\lfloor\T_{\text{opt}}\rfloor<\L/\Q$ and thus $\Q\lfloor\T_{\text{opt}}\rfloor<\L$. 
Since both sides of this strict inequality are integers, we have $\Q\lfloor\T_{\text{opt}}\rfloor\leq\L-1$ and hence $\lfloor\T_{\text{opt}}\rfloor \leq(\L-1)/\Q$, which in turn implies $\lfloor\T_{\text{opt}}\rfloor \leq \lfloor(\L-1)/\Q\rfloor$.}~\eqref{toptbound}, this implies $\chi_{\text{low}}^*\leq \lfloor(\L-1)/\Q\rfloor(1-1/\L)$. 

It remains to be shown that this upper bound is achievable.
For $\R=\lceil(\L-1)^2/\Q\rceil\geq(\L-1)^2/\Q$, we obtain (see~\eqref{eq:equivrlarge} with ``$>$'' replaced by ``$\geq$'') that $\T_{\text{opt}}\geq (\L-1)/\Q$.
Hence, for $\T=\lfloor(\L-1)/\Q\rfloor\leq \T_{\text{opt}}$, the lower bound~\eqref{eq:lowerbound} simplifies to $\chi_{\text{low}}^*= \T(1-1/\L) =\lfloor(\L-1)/\Q\rfloor(1-1/\L)$.
Thus, we have shown that $\chi_{\text{low}}^*$ is maximized for $\T=\lfloor (\L-1)/\Q\rfloor$ and $\R=\lceil(\L-1)^2/\Q\rceil$ and its maximum equals $\lfloor(\L-1)/\Q\rfloor(1-1/\L)$.
%\vspace{1.5mm} 
\end{IEEEproof}

\begin{remark}
Property~\ref{en:lowerbound2} in Corollary~\ref{cor:lowerbound} shows that for a fixed $\T \!<\rmv \L/\Q$, we can achieve $\chi^*_{\text{low}} \!= \T(1 \rmv-\rmv 1/\L)$ by using a sufficiently large number of receive antennas $\R$. 
This coincides with the upper bound presented in Section~\ref{sec:upperbound}.
Thus, in this regime, the number of degrees of freedom grows linearly in the number of transmit antennas.
%Property~\ref{en:lowerbound4} in Corollary~\ref{cor:lowerbound} shows that for $\T=\lfloor(\L-1)/\Q\rfloor$, we obtain the best possible $\chi^*_{\text{low}}$.
%as far as it is suggested by our lower bound
\end{remark}

\section{Proof of Proposition~\ref{pro:prelim}} \label{sec:prooflowerbound}
%%%%%%%%%%%%%%%%%%%%%%%%%%%%%%%
%\vspace{1mm}

In this section, we establish the lower bound~\eqref{eq:prelim}.
For $\L\leq\Tprop\Q$, the inequality in~\eqref{eq:prelim} is trivially true, because in this case $\R(1-\Tprop\Q/\L)\leq 0$ and hence  $\chi_{\text{low}}\leq 0$. 
Therefore, we focus on the case 
\[%\label{eq:assumption1}
\L>\Tprop\Q
\]
 which will thus be assumed in the remainder of this section. 
Furthermore, recall that we assumed  in Proposition~\ref{pro:prelim} that $\Tprop\leq \min\{\T, \R\}$. 
Thus, setting $\rxv_{\Tprop+1}, \dots, \rxv_{\T}$ to zero, we can replace $\T$ by $\Tprop$ in the input-output relation~\eqref{eq:modelsimp} and the power constraint~\eqref{eq:powerconstr}.
Finally, we shall assume that  
%\vspace{-1mm}
%We start with Proposition~\ref{pro:prelim} and conclude with maximizing the result to the general lower bound Theorem~\ref{th:lowerbound}.
%It turns out convenient to restrict without loss of generality to the case 
\[%\label{eq:assumption2}
\R\leq \bigg\lceil\frac{\Tprop(\L-1)}{\L-\Tprop\Q}\bigg\rceil \,.
%\vspace{1mm}
\] 
If more receive antennas are available, we simply  turn them  off. % or ignore some receive antennas. 
%An intuitive explanation why this is reasonable is the comparison between the numbers of unknown variables and equations. 
The following dimension counting argument provides some intuition on why the use of more than $\lceil\Tprop(\L-1)/(\L-\Tprop\Q)\rceil$ receive antennas is not beneficial.

\subsection{Dimension Counting}\label{sec:dimensioncounting} 
The noiseless received vector $\ryvb=\rBm\rsv\in \IC^{\R\L}$ in~\eqref{eq:modelsimp} corresponds to $\R\L$ polynomial equations. 
The unknown variables of these equations are the entries of the vectors $\rsv_{r,t}\in \IC^{\Q}$, $r\in [1\!:\!\R]$, $t\in [1\!:\!\Tprop]$ ($\R\Tprop\Q$ unknown variables) and of the transmitted signal vectors $\rxv_t\in \IC^{\L}$, $t\in [1\!:\!\Tprop]$ ($\Tprop\L$ unknown variables).
Consider now a pair $(\rxv_{t},\rsv_{r,t})$, consisting of a transmitted signal vector  $\rxv_{t}$ and a fading vector $\rsv_{r,t}$ that is a solution of $\ryvb=\rBm\rsv$.
Then the pair $(c_t \rxv_t,\rsv_{r,t}/c_t)$,
where  $c_t$ is an arbitrary nonzero constant, is also a solution of $\ryvb=\rBm\rsv$.
This implies that each $\rxv_t$ can be recovered from $\ryvb$ only up to a scaling factor. 
To resolve this ambiguity, we fix one entry in each $\rxv_{t}$. 
Hence, the total number of unknown variables becomes $\R\Tprop\Q+\Tprop\L-\Tprop$. 
As long as the number of equations is larger than or equal to the number of unknown variables, i.e., $\R\L\geq\R\Tprop\Q+\Tprop\L-\Tprop$, we are able to recover\footnote{Strictly speaking, this argument is true for linear equations. In our case, because we have polynomial rather than linear equations, we obtain in general a finite number of solutions for the variables $\rxv$ and not a unique solution, as will be discussed further in Section~\ref{sec:boundh}.} the $\L-1$ unknown entries of each  $\rxv_{t}$. 
%Thus as long as $\R\L\geq\R\Tprop\Q+\Tprop\L-\Tprop$ we can omit additional equations. 
The above condition is equivalent to $\R\geq \Tprop(\L-1)/(\L-\Tprop\Q)$.
Hence, it is reasonable to consider only the case $\R\leq \lceil\Tprop(\L-1)/(\L-\Tprop\Q)\rceil$, as the received vectors resulting from the use of additional receive antennas would not help us gain more information about the transmit vectors $\{\rxv_{t}\}_{t\in [1:\Tprop]}$.

%Many parts of the proof are hard to follow due to the heavy notation. To make those steps easier accessible we present them first for the specific setting $\L=4, \R=3, \Tprop=2,$ and $\Q=1$ and prove them rigorously afterwards. 

\subsection{Bounding $I(\rxv \ist;\ryv)$}\label{sec:boundixy} 

By~\eqref{eq:capacity}, the capacity $C(\rho)$ and, hence, $\chi_{\text{gen}}$ (cf.~\eqref{eq:prelog1}) can be lower-bounded by evaluating $I(\rxv \ist;\ryv)$ for any specific input distribution that satisfies the power constraint~\eqref{eq:powerconstr}. 
%$C(\rho) \ist\geq\ist (1/\L) I(\rxv \ist;\ryv)$ with the specific input distribution
In particular, in what follows, we will assume $\rxv \sim \mathcal{CN}(\0v,\Iv_{\Tprop\L})$. 
Thus, 
%Inserting this lower bound into~\eqref{eq:prelog1} then gives
\be \label{eq:capacitygauss}
C(\rho) \geq \frac{1}{\L}\, I(\rxv \ist;\ryv) \,.
\ee
%This implies that \cite[Lemma 6.7]{lamo03}
%\[
%% \label{eq:propx}
%\E\ist[\ist \log(| x_{t,i} |)] \ist>\ist -\infty \,,\quad t\in [1\!:\!\Tprop] \, ,\; i\in[1\!:\!\L] \,, 
%\]
%where $x_{t,i}$ denotes the $i$th element of $\rxv_{t}$.
%In what follows, we thus assume that $\rxv \sim \mathcal{CN}(\0v,\Iv_{\Tprop\L})$.
As
%\vspace{-1mm}
\be\label{eq:decomposemutinf}
I(\rxv \ist;\ryv)=h(\ryv)-h(\ryv\ist|\ist\rxv)
\ee
 we can lower-bound $I(\rxv \ist;\ryv)$ by 
%% first finding an 
upper-bounding $h(\ryv\ist |\ist\rxv)$ and lower-bounding $h(\ryv)$.

%\vspace{-2mm}

%%%%%%%%%%%%%%%%%%%%%%%%%%%%
\subsubsection{Upper Bound on $h(\ryv|\rxv)$}
%%%%%%%%%%%%%%%%%%%%%%%%%%%%
It follows from~\eqref{eq:modelsimp} and~\eqref{eq:defBm} together with $\rsv_{r,t}\sim\mathcal{CN}(\0v,\Iv_{Q})$ and $\rnv_{r}\sim\mathcal{CN}(\0v,\Iv_{\L})$ that
%%  \sim \mathcal{CN}(\0v,\Iv_{8})$, 
%% the vector 
$\ryv$ is conditionally Gaussian  given $\rxv$, with 
conditional covariance matrix $(\rho/\Tprop) \ist \rBm\rBm^{\operatorname{H}} \rmv+\Iv_{\R\L}$ (note that $\rBm=\rBm(\rxv)$). 
Hence, 
%% $h(\ryv\ist |\ist\rxv)$ is given by
\[
h(\ryv\ist |\ist\rxv) 
%% \ist=\, \E_{\rxv'}[h(\ryv\ist |\ist\rxv=\rxv')] 
= \E_{\rxv}\bigg[ \rmv\log\rmv\bigg((\pi e)^{\R\L}\,\bigg\lvert\frac{\rho}{\Tprop} \ist \rBm\rBm^{\operatorname{H}}+\rmv \Iv_{\R\L}\bigg\rvert \ist \bigg)\bigg]
\]
 according to~\cite[Th.~2]{nema93}.
%\]
By \cite[Th.~1.3.20]{hojo85}, 
%% there is 
$\big\lvert(\rho/\Tprop) \ist \rBm\rBm^{\operatorname{H}} \rmv+\Iv_{\R\L}\big\rvert 
= \big\lvert(\rho/\Tprop) \ist \rBm^{\operatorname{H}}\rBm +\rmv \Iv_{\R\Tprop\Q}\big\rvert$. Furthermore, assuming without loss of generality  
that $\rho \rmv>\! 1$ (note that we are only interested in the asymptotic regime $\rho\rightarrow \infty$), we have
%% obtain 
$\big\lvert(\rho/\Tprop) \ist \rBm^{\operatorname{H}}\rBm +\rmv \Iv_{\R\Tprop\Q}\big\rvert \leq \big\lvert\rho \big((1/\Tprop) \ist \rBm^{\operatorname{H}}\rBm +\rmv \Iv_{\R\Tprop\Q}\big)\big\rvert = \rho^{\R\Tprop\Q}\big\lvert(1/\Tprop) \ist \rBm^{\operatorname{H}}\rBm +\rmv \Iv_{\R\Tprop\Q}\big\rvert$.
%\pagebreak %%%%%%%%
Thus,
\begin{align}
h(\ryv\ist |\ist\rxv) & \leq \E_{\rxv}\bigg[\log\rmv\bigg((\pi e)^{\R\L}\rho^{\R\Tprop\Q} \,\bigg\lvert\frac{1}{\Tprop} \ist \rBm^{\operatorname{H}}\rBm +\rmv \Iv_{{\R\Tprop\Q}}\bigg\rvert \ist \bigg)\bigg] \notag \\[.5mm]
& = \R\Tprop\Q\log \rho + \E_{\rxv}\bigg[\log\bigg\lvert\frac{1}{\Tprop}  \rBm^{\operatorname{H}}\rBm  
%\notag \\ & \rule{38mm}{0mm} 
+ \Iv_{{\R\Tprop\Q}}\bigg\rvert  \bigg]  %\\& \rule{45mm}{0mm}
+ \mathcal{O}(1).\label{eq:hygivenxbound}
\end{align}
By using Jensen's inequality for the concave function $\log(\cdot)$, we obtain
%% the average-power constraint 
%%~\eqref{eq:powerconstr} and Jensen's inequality (i.e., 
\begin{align}\label{eq:showconst}
& \E_{\rxv}\bigg[\log\ist\bigg\lvert\frac{1}{\Tprop}  \ist \rBm^{\operatorname{H}}\rBm \ist+ \Iv_{{\R\Tprop\Q}}\bigg\rvert\bigg] 
%\notag \\ & \rule{25mm}{0mm}
\leq \log \E_{\rxv}\bigg[\bigg\lvert\frac{1}{\Tprop}  \ist \rBm^{\operatorname{H}}\rBm+\, \Iv_{{\R\Tprop\Q}}\bigg\rvert\bigg]\,. %\stackrel{(b)}= \mathcal{O}(1)
\end{align} 
The right-hand side in~\eqref{eq:showconst} is independent of $\rho$ and the determinant $\big\lvert(1/\Tprop) \ist \rBm^{\operatorname{H}}\rBm+\, \Iv_{{\R\Tprop\Q}}\big\rvert$ is some polynomial in the entries of $\rxv$ and $\rxv^{\operatorname{H}}$ (cf.~\eqref{eq:defBm}). 
Since $\rxv \sim \mathcal{CN}(\0v,\Iv_{\Tprop\L})$, all moments of $\rxv$, and, hence, the expectation $\E_{\rxv}\big[\big\lvert(1/\Tprop) \ist \rBm^{\operatorname{H}}\rBm+\, \Iv_{{\R\Tprop\Q}}\big\rvert\big]$, are finite. 
Therefore, 
the right-hand side in~\eqref{eq:showconst} is a finite constant with respect to $\rho$.
% and $(b)$ holds because $\rBm^{\operatorname{H}}\rBm$ is a positive semidefinite matrix in which all entries are a multiple of the product of two entries of $\rxv \sim \mathcal{CN}(\0v,\Iv_{\Tprop\L})$.
Hence,~\eqref{eq:hygivenxbound} together with~\eqref{eq:showconst} 
%\vspace{-2mm}
implies
\be
h(\ryv\ist |\ist\rxv)  \ist\leq\ist \R\Tprop\Q \log \rho +\ist \mathcal{O}(1)\,. \label{eq:boundhygivenx}
%\vspace{-1mm}
\ee
%\end{IEEEproof}
%%\vspace{-2mm}

%%%%%%%%%%%%%%%%%%%%%%%%%%%
\subsubsection{Lower Bound on $h(\ryv)$}
%%%%%%%%%%%%%%%%%%%%%%%%%%%%
The dimension counting argument provided in Section~\ref{sec:dimensioncounting} suggests that $\R\leq \lceil\Tprop(\L-1)/(\L-\Tprop\Q)\rceil$ receive antennas are sufficient to identify all unknown input parameters.
%At the beginning of this section we motivated that more equations than unknown variables are not necessary; but so far we only reduced the number of equations by reducing the number of receive antennas. 
%To bound $h(\ryv)$ we first have to reduce the number of equations even further.
By comparing more carefully the number of equations $\R\L$ and the number of variables $\R\Tprop\Q+\Tprop\L-\Tprop$, we see that we can get rid of 
\be\label{eq:defnreq}
\nreq\triangleq\max\{0, \R\L-(\R\Tprop\Q+\Tprop\L-\Tprop)\}
\ee
equations.
%(obviously we are not able to add equations, hence we can reduce the  number of equations only by a nonnegative integer). 
Since we assumed that $\R\leq\lceil\Tprop(\L-1)/(\L-\Tprop\Q)\rceil$ and $\L>\Tprop\Q$, we have that
\begin{align}
\nreq & = \max\{0, \R(\L-\Tprop\Q)-\Tprop(\L-1)\} \notag \\
& \leq \max\bigg\{0, \bigg\lceil\frac{\Tprop(\L-1)}{\L-\Tprop\Q}\bigg\rceil(\L-\Tprop\Q)-\Tprop(\L-1)\bigg\} \notag \\
& = \max\bigg\{0, \underbrace{\bigg(\bigg\lceil\frac{\Tprop(\L-1)}{\L-\Tprop\Q}\bigg\rceil-\frac{\Tprop(\L-1)}{\L-\Tprop\Q}\bigg)}_{\geq 0}\underbrace{(\L-\Tprop\Q)}_{>0}\!\bigg\} \notag \\
& = \underbrace{\bigg(\bigg\lceil\frac{\Tprop(\L-1)}{\L-\Tprop\Q}\bigg\rceil-\frac{\Tprop(\L-1)}{\L-\Tprop\Q}\bigg)}_{<1}(\L-\Tprop\Q) \notag \\
& < \L-\Tprop\Q\,. \label{eq:boundnreq}
\end{align}
Thus, we can make the number of equations equal to the number of unknown variables by removing at most $\L-\Tprop\Q-1$ equations. 
To do so, it is convenient to separate the $\R\L$ received variables into a ``useful'' part, which we denote by $[\ryv]^{}_{\sI}$ with\footnote{It is convenient to choose $\sI$ this way, but other choices may also be possible.}
\be\label{eq:defi}
\sI\triangleq [1:\R\L-\nreq]
\ee
 and a ``redundant'' part $[\ryv]^{}_{\sJ}$ with $\sJ\triangleq [1:\R\L]\setminus \sI = [\R\L-\nreq+1:\R\L]$. 
Note that in the case $\nreq=0$, i.e., when the number of equations does not exceed the number of unknown variables, we have $[\ryv]^{}_{\sI}=\ryv$ and the redundant part $[\ryv]^{}_{\sJ}$ is empty.

We can now lower-bound $h(\ryv)$ as follows:
\begin{align}
h(\ryv) &=h([\ryv]^{}_{\sI}, [\ryv]^{}_{\sJ})\nonumber\\
&\stackrel{\hidewidth (a) \hidewidth}=h([\ryv]^{}_{\sI})+h\big([\ryv]^{}_{\sJ}\ist \big|\ist[\ryv]^{}_{\sI}\big)\nonumber\\
&\stackrel{\hidewidth (b) \hidewidth}\geq h\bigg(\sqrt{\frac{\rho}{\Tprop}}[\ryvb]^{}_{\sI}+[\rnv]^{}_{\sI}\ist \bigg|\ist[\rnv]^{}_{\sI}\bigg)+h\big([\ryv]^{}_{\sJ}\ist \big|\ist\rsv,\rxv,[\ryv]^{}_{\sI}\big)\nonumber\\
& \stackrel{\hidewidth (c) \hidewidth}= h\bigg(\sqrt{\frac{\rho}{\Tprop}}[\ryvb]^{}_{\sI}\bigg)+\mathcal{O}(1)\nonumber\\
& \stackrel{\hidewidth (d) \hidewidth}= \log\bigg(\sqrt{\frac{\rho}{\Tprop}}\bigg)^{2 (\R\L-\nreq)} + h([\ryvb]^{}_{\sI})+\mathcal{O}(1) \notag \\
& = (\R\L-\nreq) \log \rho 
 + h ([\ryvb]^{}_{\sI})+\mathcal{O}(1)\,. \label{eq:boundhy}
\end{align}
%where $\ryvb$ was defined in~\eqref{eq:ybarsimp}.
%\be \label{eq:Pm}
%\Pm \ist\triangleq\, \diag \rmv\big( {[\Iv_{\L}]}_{\sI_1}  ,\dots,{[\Iv_{\L}]}_{\sI_{\R}} \big) \in \IC^{\sum_{r\in[1:\R]} \!|\sI_{r}| \times \R\L}\ist,
%%\vspace{.7mm}
%\ee
%the
%%% for sets 
%$\sI_r\subseteq [1\!:\!\L]$ for $r\in[1\!:\!\R]$ are certain subsets that will be specified later, and
%$c$ is a finite constant. 
Here, $(a)$ follows from the chain rule for differential entropy, in $(b)$ we used~\eqref{eq:modelsimp} and the fact that conditioning reduces differential entropy, $(c)$ holds since $h\big([\ryv]^{}_{\sJ}\ist \big|\ist\rsv,\rxv,[\ryv]^{}_{\sI}\big)=h\big([\rnv]^{}_{\sJ}\big)$ is a finite constant, and $(d)$ holds by the transformation property of differential entropy\cite[eq.~(8.71)]{Cover91}.
Using~\eqref{eq:boundhygivenx} and~\eqref{eq:boundhy} in~\eqref{eq:decomposemutinf},
%$I(\rxv \ist;\ryv)=h(\ryv)-h(\ryv\ist|\ist\rxv)$
 we obtain
\begin{align}
I(\rxv \ist;\ryv) & \geq \,(\R\L-\nreq-\R\Tprop\Q)\log \rho + h ([\ryvb]^{}_{\sI})  \ist+\ist \mathcal{O}(1)\notag\\ 
& \stackrel{\hidewidth \eqref{eq:defnreq} \hidewidth}= \,\Big(\R\L-\max\{0, \R\L-(\R\Tprop\Q+\Tprop\L-\Tprop)\} \notag \\ 
& \rule{25mm}{0mm} -\R\Tprop\Q\Big)\log \rho + h ([\ryvb]^{}_{\sI})  \ist+\ist \mathcal{O}(1) \notag \\
& =\,\min\{\R\L-\R\Tprop\Q, \Tprop\L-\Tprop\}\log \rho \notag \\
& \rule{35mm}{0mm}+ h ([\ryvb]^{}_{\sI})  \ist+\ist \mathcal{O}(1)\,.
\label{EQmutual}
\end{align}
The degrees of freedom lower bound~\eqref{eq:prelim} follows by inserting~\eqref{EQmutual} 
%into~\eqref{eq:preloggauss} 
%\be\label{eq:preloggauss2}
%\chi \geq \frac{1}{\L}\Bigg(\Tprop\L-\! \sum_{t\in [1:\Tprop]}\! |\sP_t| + \lim_{\rho\to\infty}\frac{h \big(\Pm\ryvb\ist | \ist {[\rxv_1]}_{\sP_1}, \dots, {[\rxv_{\Tprop}]}_{\sP_{\Tprop}} \big)}{\log \rho }  \Bigg).
%\ee
%With
%and choosing 
%the sets $\{\sI_r\}_{r\in[1:\R]}$ such that
%\be
%\label{EQalpha}
%\sum_{r\in [1:\R]} \!\rmv |\sI_r|
%\,=\, \min\ist\{ \Tprop\L-\Tprop+\R\Tprop\Q, \R\L\} \,,
%%% %\vspace{-.7mm}
%\ee
%\vspace{1mm}
into~\eqref{eq:capacitygauss}:
\bas
C(\rho) & \geq \frac{1}{\L}I(\rxv;\ryv)  \notag \\ 
& \geq \min\rmv\bigg\{\R\bigg(1 \rmv-\rmv \frac{\Tprop\Q}{\L}\bigg), \Tprop\bigg(1 \rmv-\rmv \frac{1}{\L}\bigg) \bigg\}\log \rho  \notag \\ 
& \rule{35mm}{0mm} + \frac{1}{\L} \,h ([\ryvb]^{}_{\sI})  \ist+\ist \mathcal{O}(1)
%\vspace{1mm}
\eas
whence, by~\eqref{eq:prelog1} and because $h ([\ryvb]^{}_{\sI})$ does not depend on $\rho$,
\bas
\chi_{\text{gen}}& \geq \! \lim_{\rho\to\infty}\!\frac{\min\rmv\Big\{\!\R\Big(1 \rmv-\rmv \frac{\Tprop\Q}{\L}\Big), \Tprop\Big(1 \rmv-\rmv \frac{1}{\L}\Big) \!\Big\}\log \rho + \mathcal{O}(1)}{\log \rho} \\
& =
\min\rmv\bigg\{\R\bigg(1 \rmv-\rmv \frac{\Tprop\Q}{\L}\bigg), \Tprop\bigg(1 \rmv-\rmv \frac{1}{\L}\bigg) \bigg\}
\eas
provided that $h ([\ryvb]^{}_{\sI}) \rmv >\rmv -\infty$.
To conclude the proof, we will next show that $h ([\ryvb]^{}_{\sI}) \rmv >\rmv -\infty$ for a generic coloring matrix $\Zm$.
This is the most technical part of the proof.

%%%%%%%%%%%%%%%%%%%%%%%%%%%%%%%
%%%%%%%%%%%%%%%%%%%%%%%%%%%%%%%
\subsection{Proof that $\,h ([\ryvb]^{}_{\sI}) \rmv >\rmv -\infty$} \label{sec:boundh}   
%%%%%%%%%%%%%%%%%%%%%%%%%%%%%%%
%According to~\eqref{eq:ybarsimp},
%% the noiseless output vector 

As $[\ryvb]^{}_{\sI}$ is a function of $\rsv$ and $\rxv$ (see~\eqref{eq:modelsimp} and~\eqref{eq:defBm}), 
the idea behind our proof is to relate 
%% the differential entropy of $\ryvb$ to the differential entropy of the pair $(\rxv, \rsv)$. 
$h([\ryvb]^{}_{\sI})$, which we are not able to calculate directly, to $h(\rsv, \rxv)$, which can be calculated trivially.
The underlying intuition 
%% behind the fact that $h ([\ryvb]^{}_{\sI}) \rmv >\rmv -\infty$ 
is that the image of a random variable of finite differential entropy, such as $(\rsv, \rxv)$,
%% (???e.g., a Gaussian random variable) 
under a ``well-behaved'' mapping, such as $(\rsv, \rxv) \mapsto [\ryvb]^{}_{\sI}$, cannot have an infinite differential entropy.
At the heart of the proof is the bounding of differential entropy under finite-to-one mappings, to be established in Lemma~\ref{LEMchangeh} below.

We first need to characterize the mapping between 
%% the random variables 
$(\rsv, \rxv)$ and $[\ryvb]^{}_{\sI}$.
%and analyze in detail the mapping.
%% Note that $\ryvb\in \IC^{12}$ and $(\rxv^{\trans} \, \rsv^{\trans})^{\trans}\!\in \IC^{14}$. 
To equalize the dimensions---note that $[\ryvb]^{}_{\sI}\in \IC^{\abs{\sI}}$ and $( \rsv^{\trans} \, \rxv^{\trans})^{\trans}\!\in \IC^{\R\Tprop\Q+\Tprop\L}$---we 
condition on $\R\Tprop\Q+\Tprop\L-\abs{\sI}$ entries of $\rxv$, which we denote by $[\rxv]^{}_{\sP}$ (hence, $\abs{\sP}=\R\Tprop\Q+\Tprop\L-\abs{\sI}$).
This results in 
\be\label{eq:boundhycond}
h([\ryvb]^{}_{\sI})\geq h([\ryvb]^{}_{\sI}\ist |\ist [\rxv]^{}_{\sP})\,.
\ee
We shall denote by $[\rxv]^{}_{\sD}$ the remaining entries of $\rxv$, i.e., $\sD\triangleq [1:\Tprop\L]\setminus \sP$. 
Note that $\abs{\sD}+\abs{\sP}=\Tprop\L$ and 
%\vspace{-2mm}
thus
\be\label{eq:sizeIfromD}
\abs{\sI}=\R\Tprop\Q + \abs{\sD}\,.
\ee
One can think of $[\rxv]^{}_{\sP}$ as pilot symbols and of $[\rxv]^{}_{\sD}$ as data symbols. 
The set $\sP$ will be defined in Appendix~\ref{sub:part1}. 
At this point, we are only concerned with its  size, which is equal to
\be
\abs{\sP} = \R\Tprop\Q+\Tprop\L-\abs{\sI} \,.
%\notag \\
%& = \R\Tprop\Q+\Tprop\L-\R\L+\nreq \notag \\
%& = \R\Tprop\Q+\Tprop\L-\R\L+\max\{0, \R\L-(\R\Tprop\Q+\Tprop\L-\Tprop)\} \notag \\
%& = \max\{\Tprop, \R\Tprop\Q-(\R \rmv-\rmv \Tprop)\L\}\,.
\label{eq:indp}
\ee
%At a later point we will further specify the set $\sP$.
%For easier notation, we set
%%% will use the abbreviations 
%%% $[\xv]^{}_{\sP}\triangleq ([\xv_1]^{}_1, [\xv_2]^{}_2)$ and $[\xv]^{}_{\sD}\triangleq ([\xv_1]^{}_2$, $[\xv_1]^{}_3$, $[\xv_1]^{}_4$, $[\xv_2]^{}_1$, $[\xv_2]^{}_3$, $[\xv_2]^{}_4)$. 
%$[\rxv]^{}_{\sP}\triangleq ([\rxv_1]^{}_1 \,\ist [\rxv_2]^{}_2)^\trans$ and 
%$[\rxv]^{}_{\sD}\triangleq ([\rxv_1]^{}_2 \,\ist [\rxv_1]^{}_3\,\ist [\rxv_1]^{}_4$\linebreak %%%%%%
%$[\rxv_2]^{}_1 \,\ist [\rxv_2]^{}_3 \,\ist [\rxv_2]^{}_4)^\trans\rmv$.

Because of~\eqref{eq:boundhycond}, it suffices to show that 
\[
h([\ryvb]^{}_{\sI}\ist \big|\ist[\rxv]^{}_{\sP})>-\infty\,.
\]
This will be done by relating $h([\ryvb]^{}_{\sI}\ist \big|\ist[\rxv]^{}_{\sP})$ to $h(\rsv, [\rxv]^{}_{\sD})$. 
Before doing so, we have to understand the connection between the variables $[\ryvb]^{}_{\sI}$ and $(\rsv, [\rxv]^{}_{\sD})$. 
%With this knowledge we are able to use a novel result on the change in differential entropy when a random variable undergoes a finite-to-one mapping.
This leads us to the following program:
\begin{enumerate}
\renewcommand{\theenumi}{\roman{enumi}}
\renewcommand{\labelenumi}{(\roman{enumi})}
\renewcommand{\theenumii}{-\alph{enumii}}
\renewcommand{\labelenumii}{\alph{enumii})}
\item Define the polynomial mapping $\phi_{[\xv]^{}_{\sP}}$ relating $(\rsv, [\rxv]^{}_{\sD})$ and 
%% the noiseless observation 
$[\ryvb]^{}_{\sI}$. 
\label{step1}%This results in polynomial mappings $\phi_{[\xv]^{}_{\sP}}$. %(which depends on $[\xv]^{}_{\sP}$);
\item Prove that $\phi_{[\xv]^{}_{\sP}}$ satisfies the following two properties: \label{step2}
\begin{enumerate}
	\item Its Jacobian matrix %$\Jm_{\phi_{[\xv]^{}_{\sP}}}\!(\sv, 	[\xv]^{}_{\sD})\!$ 
	%of $\phi_{[\xv]^{}_{\sP}}$ 
	is nonsingular almost everywhere (a.e.)\ for almost all (a.a.)\ $[\xv]^{}_{\sP}$.\label{step2a}
	\item %The mappings $\phi_{[\xv]^{}_{\sP}}\rmv$ are 
	It is finite-to-one\footnote{A mapping is called finite-to-one if every element in the codomain has a preimage of finite cardinality.} a.e.\ for a.a.\ 	$[\xv]^{}_{\sP}$.\label{step2b}
	\end{enumerate}
\item Apply a novel result on the change in differential entropy that occurs
 when a random variable undergoes 
 a finite-to-one mapping to relate $h([\ryvb]^{}_{\sI}\ist \big|\ist [\rxv]^{}_{\sP})$ to $h(\rsv, [\rxv]^{}_{\sD})$.\label{step3}
\item Bound the terms resulting
%% emerging 
from this change in differential entropy.\label{step4}
\end{enumerate}
%Let us split the vector $\rxv$ into the vectors $[\rxv]^{}_{\sP}\triangleq \big({[\rxv_1]}_{\sP_1}^{\operatorname{T}}, \dots, {[\rxv_{\Tprop}]}_{\sP_{\Tprop}}^{\operatorname{T}}\big)^{\operatorname{T}}\rmv$ and $[\rxv]^{}_{\sD}\triangleq \big({[\rxv_1]}_{\sD_1}^{\operatorname{T}},\dots,{[\rxv_{\Tprop}]}_{\sD_{\Tprop}}^{\operatorname{T}}\big)^{\operatorname{T}}\rmv$, 
%where $\sP_t\subseteq[1\!:\!\L]$ and $\sD_t\triangleq[1\!:\!\L]\setminus\sP_t$ for $t\in [1\!:\!\Tprop]$.
%Because $h (\Pm\ryvb\ist | \ist[\rxv]^{}_{\sP})\leq h (\Pm\ryvb)$, it is sufficient to show that $h (\Pm\ryvb\ist | \ist[\rxv]^{}_{\sP})>-\infty$.
%We can think of the sets $\sP_t$ as positions of pilots where the conditioning reflects that they are known. Although, in fact they are only a mathematical trick to make the proof easier. Similarly, the sets $\sD_t$ can be interpreted as data sets.
%As in \cite{rimodulistbo11}, we wish to relate 
%%% the differential entropy 
%$h (\Pm\ryvb\ist | \ist[\rxv]^{}_{\sP})$ to the simpler quantity 
%$h(\rsv,[\rxv]^{}_{\sD}) \,=\, h(\rsv) \ist+ h([\rxv]^{}_{\sD})$. 
%Since the dimensions of $\Pm\ryvb$ and $(\rsv,\rxv)$ do not coincide 
%We now can relate the differential entropies $h (\Pm\ryvb\ist | \ist[\rxv]^{}_{\sP})$ and $h(\rsv,[\rxv]^{}_{\sD})$. 
%\vspace{-2mm}
\subsection*{Step~(\ref{step1}):} We consider the $[\xv]^{}_{\sP}$-parametrized mapping
\be\label{eq:phi}
\phi_{[\xv]^{}_{\sP}}\colon \IC^{\R\Tprop\Q+\abs{\sD}}\to \IC^{\abs{\sI}};\; (\sv,[\xv]^{}_{\sD}) \mapsto\, 
%\Pm \!
%% \Bigg(
%\sum_{t\in [1:\Tprop]} \!(\Iv_{\R}\otimes\rXm_{t}\Qm) \ist\rsv_{t}
%% \Bigg)
%\ist=\ist 
[\yvb]^{}_{\sI}
\ee
in which $\yvb$ is defined in~\eqref{eq:modelsimp} and~\eqref{eq:defBm}, i.e.,
\be\label{eq:ybardisc}
\yvb \,= \Bm\sv,
 \quad \text{ with }
\Bm=\begin{pmatrix}
\Bm_{1} \\[-1.5mm]
 & \hspace{-2.5mm}\ddots \hspace{-1mm} \\[-1.5mm]
 &  & \hspace{-1.5mm} \Bm_{\R} 
\end{pmatrix}
%\vspace{-.5mm}
\ee
where
%\vspace{-1mm}
%\be\label{eq:defXi}
%$\Bm_{r}=(\Xm_{1}\Zm_{r,1} \cdots \Xm_{\Tprop}\Zm_{r,\Tprop})$
\be\label{eq:ybardisc2}
\Bm_{r}\!=\rmv(\Xm_{1}\Zm_{r,1} \cdots \Xm_{\Tprop}\Zm_{r,\Tprop}),\quad\text{ with } \Xm_{t} \!=\rmv \diag(\xv_t).
\ee
We see from~\eqref{eq:ybardisc} and~\eqref{eq:ybardisc2} that the components of the vector-valued mapping $\phi_{[\xv]^{}_{\sP}}$ are multivariate polynomials of degree 2 in the entries of $\sv$ and $[\xv]^{}_{\sP}$.
The Jacobian matrix $\Jm_{\phi_{[\xv]^{}_{\sP}}}$ of $\phi_{[\xv]^{}_{\sP}}\rmv$
%\[
%\Jm_{\phi_{[\xv]^{}_{\sP}}}(\sv,[\xv]^{}_{\sD}) \ist\triangleq\ist \frac{\partial \Pm\yvb}{\partial (\sv,[\xv]^{}_{\sD}) } \,,
%\in \IC^{\sum_{r\in[1:\R]} \!|\I_{r}| \times\Tprop\ist (\Q\R\rmv+\rmv\L) - \sum_{t\in [1:\Tprop]}\! |\sP_t|} 
%\]
is equal to
%\vspace{-1mm}
\ba
& \Jm_{\phi_{[\xv]^{}_{\sP}}}\!(\sv,[\xv]^{}_{\sD}) \ist  \ist=\ist \left[ \left( 
\Bm \;\,
[\Am]^{\sD}
\right) \right]^{}_{\sI}
\in \IC^{\abs{\sI} \times \abs{\sI}},  \notag \\
& \rule{10mm}{0mm} \text{ with } 
\Am =\begin{pmatrix}
\Am_{1,1} & \hspace*{-2.5mm}\cdots\hspace*{-2.5mm}  & \Am_{1,\Tprop}  \\[-.8mm]
 \vdots    &   \hspace*{-2.5mm}\hspace*{-2.5mm}     & \vdots     \\[-1mm]
\Am_{\R,1} & \hspace*{-2.5mm}\cdots\hspace*{-2.5mm} & \Am_{\R,\Tprop}
\end{pmatrix} \in \IC^{\R\L\times \Tprop\L}
\label{eq:Jacobian1}
\ea
where 
\vspace*{-1.5mm}
\ba
%gin{multline}\label{EQArt}
& \Am_{r,t} \ist\triangleq\ist
\diag (\av_{r,t})
\rmv, \quad\!\!
t\in[1\!:\!\Tprop] \ist, \,r\in [1\!:\!\R]\,, \notag \\
& \rule{40mm}{0mm} \text{ with} \;\; \av_{r,t} \triangleq\ist \Zm_{r,t}\sv_{r,t} 
\label{EQArt}
%\vspace{.5mm}
\ea
and where in~\eqref{eq:Jacobian1} we used that $\abs{\sI}=\R\Tprop\Q + \abs{\sD}$ (see~\eqref{eq:sizeIfromD}).
Note that we did not take derivatives with respect to $[\xv]^{}_{\sP}$, since the entries of $[\xv]^{}_{\sP}$ are treated as fixed 
parameters.

%\vspace{-2mm}
\subsection*{Step~(\ref{step2a}):} 
We have to show that 
%% the matrix 
$\Jm_{\phi_{[\xv]^{}_{\sP}}}\!$ is nonsingular (i.e., $\absdet{\Jm_{\phi_{[\xv]^{}_{\sP}}}\rmv} \!\not=\! 0$) a.e.\ for a.a.\ $[\xv]^{}_{\sP}$ and a generic coloring matrix $\Zm$.
The determinant of $\Jm_{\phi_{[\xv]^{}_{\sP}}}\!$ is
%% can be seen as 
a polynomial $p(\Zm, \sv, \xv)$ (i.e., a polynomial in all the entries of $\Zm$, $\sv$, $[\xv]^{}_{\sD}$, and $[\xv]^{}_{\sP}$).
We will show 
%% that this polynomial is 
that $p(\Zm, \sv, \xv)$ does not vanish at a specific point $(\tilde{\Zm}, \svt, \xvt)$, i.e., $p(\tilde{\Zm}, \svt, \xvt)\neq 0$. 
This implies
%By fixing $\svt$ and $\xvt$, we can then conclude
%% deduce 
that $p(\Zm, \svt, \xvt)$ (as a function of $\Zm$, for fixed $\svt$ and $\xvt$) is not identically zero.
Since a polynomial vanishes either identically or on a set of measure 
zero\!\! %%%%%%%%% 
\cite[Cor.~10]{GR65},  %~\cite[]{bla}
 we conclude that $p(\Zm, \svt, \xvt)\neq 0$ for $\Zm\in \sZ$, where $\sZ$ is a set with a complement of measure zero. 
Using the same argument, we find that, for a fixed $\Zm\in \sZ$, the function $p(\Zm, \sv, \xv)$ does not vanish a.e.\ (as a function of $(\sv, \xv)$).
Hence, $\absdet{\Jm_{\phi_{[\xv]^{}_{\sP}}}(\sv,[\xv]^{}_{\sD})} \!\not=\! 0$ 
for a.a.\ $(\sv, [\xv]^{}_{\sD}, [\xv]^{}_{\sP})$ and all $\Zm\in \sZ$. In other words, for a generic coloring matrix $\Zm$, the matrix $\Jm_{\phi_{[\xv]^{}_{\sP}}}$ is nonsingular a.e.\ for a.a.\ 
%\pagebreak %%%%%%%%
$[\xv]^{}_{\sP}$.

It remains to find the point $(\tilde{\Zm}, \svt, \xvt)$, i.e., a specific point  $(\tilde{\Zm}, \svt, \xvt)$ such that $p(\tilde{\Zm}, \svt, \xvt)\neq 0$. This, in turn, requires to find a specific set $\sP$. This is done in the proof of the following %\vspace{1.5mm} 
lemma.

\begin{lemma}\label{LEMnotvanish}
Let $\R\geq \Tprop$, $\L>\Tprop\Q$, and $\R\leq \lceil\Tprop(\L-1)/(\L-\Tprop\Q)\rceil$.
Then there exists a triple $(\Zm, \sv, \xv)$ and a choice of $\sP$ for which %$p(\Zm, \sv, \xv)$ 
the determinant of  the Jacobian matrix $\Jm_{\phi_{[\xv]^{}_{\sP}}}$ in~\eqref{eq:Jacobian1} is  %\vspace{1.5mm} 
nonzero. 
\end{lemma}
\begin{IEEEproof}[\hspace{-1em}Proof]
See Appendix~\ref{app:proofnotvanish}.%\vspace{1.5mm} 
%\vspace{-3mm}
\end{IEEEproof}

\subsection*{Step~(\ref{step2b}):} 
We will invoke B\'ezout's theorem \cite[Prop.~B.2.7]{VdE00} to show that the mapping $\phi_{[\xv]^{}_{\sP}}$ is 
finite-to-one 
a.e.\ for a.a.\ $[\xv]^{}_{\sP}$. 
In what follows, note that for a given $[\yvb]^{}_{\sI}$ in the codomain of $\phi_{[\xv]^{}_{\sP}}$, the quantity $\phi_{[\xv]^{}_{\sP}}^{-1}([\yvb]^{}_{\sI})$ is the preimage $\phi_{[\xv]^{}_{\sP}}^{-1}([\yvb]^{}_{\sI})= \{(\sv,[\xv]^{}_{\sD}): \phi_{[\xv]^{}_{\sP}}(\sv,[\xv]^{}_{\sD})= [\yvb]^{}_{\sI}\}$ and not the function value of the inverse function (which does not even exist in most  cases). 
Furthermore, for a given $[\xv]^{}_{\sP}$, we denote by $\widetilde{\sM}\subseteq \IC^{\abs{\sI}}$ the set of all $(\sv,[\xv]^{}_{\sD})$ for which $\Jm_{\phi_{[\xv]^{}_{\sP}}}\!(\sv,[\xv]^{}_{\sD})$ is nonsingular, i.e.,
\[ %\label{eq:defmtilde}
\widetilde{\sM}\triangleq \big\{(\sv,[\xv]^{}_{\sD}) \in \IC^{\R\Tprop\Q \,+\, \abs{\sD}} \!: \absdet{\Jm_{\phi_{[\xv]^{}_{\sP}}}\!(\sv,[\xv]^{}_{\sD})}\neq 0\big\}\,.
\]

\begin{lemma}\label{LEMinjective}
For a given $[\xv]^{}_{\sP}$, let $\widetilde{\sM}$ be defined as above. Then for all $[\yvb]^{}_{\sI}\in \phi_{[\xv]^{}_{\sP}}(\widetilde{\sM})$,
\[ %\label{eq:finite}
\big|\phi_{[\xv]^{}_{\sP}}^{-1}([\yvb]^{}_{\sI})\cap \widetilde{\sM}\big|\,\leq\, \widetilde{m}\,\triangleq \,2^{\R\Tprop\Q \,+\, \abs{\sD}}\,.
%\vspace{1.5mm}
\]
\end{lemma}

\begin{IEEEproof}[\hspace{-1em}Proof]
Let $[\yvb]^{}_{\sI} \rmv\in\rmv \phi_{[\xv]^{}_{\sP}}\rmv(\widetilde{\sM})$. 
The set $\phi_{[\xv]^{}_{\sP}}^{-1}([\yvb]^{}_{\sI})$ contains all points $(\sv,[\xv]^{}_{\sD})$ such that $\phi_{[\xv]^{}_{\sP}}(\sv,[\xv]^{}_{\sD})=[\yvb]^{}_{\sI}$. 
Thus, these points are the zeros of the vector-valued mapping
\be \label{eq:vecmap}
(\sv,[\xv]^{}_{\sD})\mapsto \phi_{[\xv]^{}_{\sP}}(\sv,[\xv]^{}_{\sD})-[\yvb]^{}_{\sI}\,.
\ee
It follows from~\eqref{eq:phi}--\eqref{eq:ybardisc2} that each component of the vector-valued mapping~\eqref{eq:vecmap} is a polynomial of degree 2. 
Hence, the zeros of the mapping~\eqref{eq:vecmap}
are the common zeros of $\abs{\sI}=\R\Tprop\Q+\abs{\sD}$ polynomials of degree 2.
By a weak version of B\'ezout's theorem \cite[Prop.~B.2.7]{VdE00}, the number of isolated zeros (i.e., 
%% zeros where there are 
with no other zeros in some neighborhood) cannot exceed $\widetilde{m}=2^{\R\Tprop\Q \,+\, \abs{\sD}}$. 
Since $\Jm_{\phi_{[\xv]^{}_{\sP}}}\!$ 
is nonsingular on $\widetilde{\sM}$, the function $\phi_{[\xv]^{}_{\sP}}$ restricted to $\widetilde{\sM}$ is locally one-to-one \cite[Th.~9.24]{ru79} and, hence, each zero 
of $\phi_{[\xv]^{}_{\sP}} \!\rmv-\rmv[\yvb]^{}_{\sI}$ on $\widetilde{\sM}$ has to be an isolated zero. 
Therefore, the number of points $(\sv,[\xv]^{}_{\sD})\in \widetilde{\sM}$ such that $\phi_{[\xv]^{}_{\sP}}(\sv,[\xv]^{}_{\sD})=[\yvb]^{}_{\sI}$ cannot exceed $\widetilde{m}$. 
%Since these points are exactly those points in the inverse image $\phi_{[\xv]^{}_{\sP}}^{-1}(\yvb)$ that also belong to $\widetilde{\sM}$, we conclude that~\eqref{eq:finite} holds.
%\vspace{1.5mm} 
\end{IEEEproof}

By Lemma~\ref{LEMinjective}, the function $\phi_{[\xv]^{}_{\sP}}$ for a given $[\xv]^{}_{\sP}$ is finite-to-one on the set $\widetilde{\sM}$. 
Because by \mbox{Step~(\ref{step2a})} the matrix $\Jm_{\phi_{[\xv]^{}_{\sP}}}\!(\sv,[\xv]^{}_{\sD})$ is nonsingular a.e.\ for a.a.\ $[\xv]^{}_{\sP}$,
and because $\widetilde{\sM}\subseteq \IC^{\abs{\sI}}$ is the set of all $(\sv,[\xv]^{}_{\sD})$ for which $\Jm_{\phi_{[\xv]^{}_{\sP}}}\!(\sv,[\xv]^{}_{\sD})$ is nonsingular,
 we conclude that $\phi_{[\xv]^{}_{\sP}}$ is finite-to-one a.e.\ for a.a.\ $[\xv]^{}_{\sP}$.

%\vspace{-2mm}
\subsection*{Step~(\ref{step3}):}
We will use the following novel 
result bounding the change in differential entropy that occurs
when a random variable undergoes  a finite-to-one %\vspace{1.5mm}  
mapping.
%A proof is provided in the 
%%\vspace{1.5mm}
%appendix.

\begin{lemma}\label{LEMchangeh}
Let $\ruv\in \IC^n$ be a random vector with probability density function $f_{\ruv}$. 
Consider a 
continuously differentiable mapping $\kappa\colon \IC^n \!\rightarrow \IC^n$ with Jacobian matrix $\Jm_{\kappa}$. 
Assume that  $\Jm_{\kappa}$ is nonsingular a.e.\ and let
$\sM\triangleq \{\uv \rmv\in\rmv \IC^n \!: \absdet{\Jm_{\kappa}(\uv)}\neq 0\}$ (thus, $\IC^{n}\setminus \sM$ has Lebesgue measure zero).
Furthermore,
let $\rvv \triangleq \kappa(\ruv)$, and assume that for all $\vv\in \IC^n$, the cardinality of the set $\kappa^{-1}(\vv)\cap \sM$ satisfies
$\abs{\kappa^{-1}(\vv)\cap \sM}\leq m < \infty$, for some 
%% constant 
$m\in \IN$ (i.e., $\kappa\big|_{\sM}$ is finite-to-one). Then:

(I) There exist disjoint measurable sets $\{\sU_k\}_{k\in [1:m]}$ such that $\kappa\big|_{\sU_k}\!$ is one-to-one for each $k \rmv\in\rmv [1 \!:\rmv m]$ and 
%\vspace{.4mm}
$\bigcup_{k\in [1:m]}\sU_k$ covers almost all of $\sM$. 
%% \label{en:existence}

(II) For every choice of such sets %\vspace{-1mm} 
$\{\sU_k\}_{k\in [1:m]}$,
%\vspace{1.5mm}
\be
h(\rvv) \ist\geq\ist h(\ruv) +\rmv \int_{\IC^n} \!\rmv f_{\ruv}(\uv)\log (\absdet{\Jm_{\kappa}(\uv)}^2) \,d\uv -H(\randk)
\label{eq:bounddiffe}
%\vspace{0mm}
\ee
where $\randk$ is a discrete random variable that takes on the value $k$ when $\ruv\in\sU_k$ and $H$ denotes entropy.
%% \label{en:inequality}
%\vspace{1.5mm}

\end{lemma}
\begin{IEEEproof}[\hspace{-1em}Proof]
See Appendix~\ref{app:proofpartition}.
%\vspace{1.5mm} 
\end{IEEEproof}

Since by Step~(\ref{step2b}) the mapping $\phi_{[\xv]^{}_{\sP}}\big|_{\widetilde{\sM}}$ is finite-to-one for a.a.\ $[\xv]^{}_{\sP}$, 
we can use Lemma~\ref{LEMchangeh} with $\ruv = (\rsv,[\rxv]^{}_{\sD})$, $\kappa = \phi_{[\xv]^{}_{\sP}}$, $n=\R\Tprop\Q+\abs{\sD}$, $m=\tilde{m}$, and $\sM=\widetilde{\sM}$ and
%\vspace{1mm}
obtain
\ba
& h\big(\phi_{[\xv]^{}_{\sP}}(\rsv,[\rxv]^{}_{\sD})\big) \notag \\
& \quad\geq h(\rsv,[\rxv]^{}_{\sD})   + \int_{\IC^{\R\Tprop\Q+\abs{\sD}}}   f_{\rsv,[\rxv]^{}_{\sD}}(\sv,[\xv]^{}_{\sD}) \notag \\
& \quad\quad \times \log \rmv\big(\absdet{\Jm_{\phi_{[\xv]^{}_{\sP}}}\!(\sv,[\xv]^{}_{\sD})}^2\big) \ist d(\sv,[\xv]^{}_{\sD})  -H(\randk_{[\xv]^{}_{\sP}})
\label{eq:applylemma8}
\ea
where $\randk_{[\xv]^{}_{\sP}}$ corresponds to the random variable $\randk$ from Lemma~\ref{LEMchangeh} (since $\kappa = \phi_{[\xv]^{}_{\sP}}$, we have a different $\randk$ for each $[\xv]^{}_{\sP}$).
Because of $[\ryvb]^{}_{\sI}=\phi_{[\rxv]^{}_{\sP}}(\rsv,[\rxv]^{}_{\sD})$, we have $h\big([\ryvb]^{}_{\sI}\ist \big|\ist [\rxv]^{}_{\sP}\!=\![\xv]^{}_{\sP}\big)= h\big(\phi_{[\xv]^{}_{\sP}}(\rsv,[\rxv]^{}_{\sD})\big)$.
%Since $h(\rsv,[\rxv]^{}_{\sD})$ does not depend on $[\xv]^{}_{\sP}$ 
Thus,~\eqref{eq:applylemma8} entails
\begin{align}
& h\big([\ryvb]^{}_{\sI}\ist \big|\ist [\rxv]^{}_{\sP}\big) \notag \\
& \rule{1mm}{0mm}= \, \E_{[\rxv]^{}_{\sP}} \big[h\big(\phi_{[\rxv]^{}_{\sP}}(\rsv,[\rxv]^{}_{\sD})\big)\big] \notag \\
&\rule{1mm}{0mm}\geq\, h(\rsv,[\rxv]^{}_{\sD})   +\ist \E_{[\rxv]^{}_{\sP}}\! \bigg[\rmv \int_{\IC^{\R\Tprop\Q+\abs{\sD}}} \! f_{\rsv,[\rxv]^{}_{\sD}}(\sv,[\xv]^{}_{\sD}) \notag \\
&\rule{5mm}{0mm} \times \log \rmv\big(\big\lvert\Jm_{\phi_{[\rxv]^{}_{\sP}}}\!(\sv,[\xv]^{}_{\sD})\big\rvert^2\big) \ist d(\sv,[\xv]^{}_{\sD})  -H\big(\randk_{[\rxv]^{}_{\sP}}\big) \bigg]\ist. \label{eq:boundhybar} \\[-10mm] \notag
\end{align}

%\vspace{-5mm}

\subsection*{Step~(\ref{step4}):}
We show now that the right-hand side of~\eqref{eq:boundhybar} is low\-er-bound\-ed by a finite constant.
The differential entropy $h(\rsv,[\rxv]^{}_{\sD})$ 
%% does not depend on $[\xv]^{}_{\sP}$ and 
is the differential entropy of a standard multivariate Gaussian random vector and thus a finite constant. 
The entropy $H\big(\randk_{[\xv]^{}_{\sP}}\big)$ for a.a.\ $[\xv]^{}_{\sP}$
does not exceed $\log(\tilde{m})$, where $\tilde{m} = 2^{\R\Tprop\Q+\abs{\sD}}$.
%can be upper-bounded 
%by the entropy of a uniformly distributed discrete random variable.
%% of size $m$ 
%% Hence, its expectation is also finite. 
Hence, it remains to lower-bound
\begin{align}
&\hspace{-1.5mm}\E_{[\rxv]^{}_{\sP}} \!\bigg[\int_{\IC^{\R\Tprop\Q+\abs{\sD}}} \! f_{\rsv,[\rxv]^{}_{\sD}}(\sv,[\xv]^{}_{\sD}) \notag\\
&\rule{23mm}{0mm} \times \log \rmv\big(\absdet{\Jm_{\phi_{[\rxv]^{}_{\sP}}}\!(\sv,[\xv]^{}_{\sD})}^2\big) 
  \ist d(\sv,[\xv]^{}_{\sD})\bigg] \notag\\
& =
\int_{\IC^{\abs{\sP}}}\int_{\IC^{\R\Tprop\Q+\abs{\sD}}} \! f_{[\rxv]^{}_{\sP}}([\xv]^{}_{\sP})\,f_{\rsv,[\rxv]^{}_{\sD}}(\sv,[\xv]^{}_{\sD}) \notag\\
&\rule{23mm}{0mm} \times \log \rmv\big(\absdet{\Jm_{\phi_{[\xv]^{}_{\sP}}}\!(\sv,[\xv]^{}_{\sD})}^2\big) \ist d(\sv,[\xv]^{}_{\sD})\, d[\xv]^{}_{\sP}\notag\\[1mm]
& \stackrel{\hidewidth (a) \hidewidth}=
\int_{\IC^{\R\Tprop\Q+\Tprop\L}} \! f_{\rsv,\rxv}(\sv,\xv) \ist\log \rmv\big(\absdet{\Jm_{\phi_{[\xv]^{}_{\sP}}}\!(\sv,[\xv]^{}_{\sD})}^2\big) \ist d(\sv,\xv)
\label{eq:finlogdet}
\end{align}
where $(a)$ holds because $(\rsv,[\rxv]^{}_{\sD})$ and $[\rxv]^{}_{\sP}$ are independent.
A similar problem was recently solved in \cite{moridu13} using Hironaka's theorem on the resolution of singularities. 
Here, we take a much simpler approach, which relies on the fact that $\det\big(\Jm_{\phi_{[\xv]^{}_{\sP}}}\big)$ in~\eqref{eq:finlogdet} is an analytic function \cite[Ch.~10]{ru87} that does not vanish identically, 
and on a property of subharmonic functions%
\footnote{See~\cite[Ch.~2.6]{Azarin09} for a definition of subharmonic functions.}
  %\vspace{1.5mm} 
(see~\cite[Th.~2.6.2.1]{Azarin09}).

\begin{lemma}\label{LEMboundanalytic}
Let $f$ be an analytic function on $\IC^n$ that is not identically zero. Then
\be\label{eq:expec}
I_1 \ist\triangleq
\int_{\IC^n} \!\exp(-\norm{\xiv}^2)\log(\abs{f(\xiv)})\,d\xiv \ist> -\infty \,.
%\vspace{1.5mm}
\ee
\end{lemma}

\begin{IEEEproof}[\hspace{-1em}Proof]
See Appendix~\ref{app:proofboundanalytic}.
%\vspace{1.5mm} 
\end{IEEEproof}

The function $f_{\rsv,\rxv}$ is the probability density function of a standard multivariate Gaussian random vector.  
Furthermore, since the function $\det (\Jm_{\phi_{[\xv]^{}_{\sP}}}\!(\sv,[\xv]^{}_{\sD}))$ is a complex polynomial that is nonzero a.e.\ (see Step~(\ref{step2a})), it is an analytic function that is not identically zero. 
Hence, by Lemma~\ref{LEMboundanalytic}, the integral in~\eqref{eq:finlogdet} is finite. 
Thus, with~\eqref{eq:boundhybar}, we obtain $h\big([\ryvb]^{}_{\sI}\big|[\rxv]^{}_{\sP}\big)>-\infty$ and, because of~\eqref{eq:boundhycond}, that $h([\ryvb]^{}_{\sI})>-\infty$. 
This concludes the proof. 
% that $h ([\ryvb]^{}_{\sI}) \rmv >\rmv -\infty$.
%% of Proposition~\ref{pro:prelim}.

%%%%%%%%%%%%%%%%%%%%%%%%%%%%%%%%%%%%%%%%%%%%%
\section{Conclusion}
%%%%%%%%%%%%%%%%%%%%%%%%%%%%%%%%%%%%%%%%%%%
%\vspace{1mm}

\begin{figure*}[!t]
% ensure that we have normalsize text
\normalsize
% Store the current equation number.
\newcounter{MYtempeqncnt}
\setcounter{MYtempeqncnt}{\value{equation}}
% Set the equation number to one less than the one
% desired for the first equation here.
% The value here will have to changed if equations
% are added or removed prior to the place these
% equations are referenced in the main text.
\setcounter{equation}{54}
\be\label{eq:examplejacobian1}
\left(
\begin{smallmatrix}
[\Zm_{1,1}]^{}_1 & [\Zm_{1,2}]^{}_1 &&&&& 0 &&&                  [\Zm_{1,2}]^{}_1 s_{1,2} &&\\[.8mm]
[\Zm_{1,1}]^{}_2 & [\Zm_{1,2}]^{}_2 &&&&& [\Zm_{1,1}]^{}_2 s_{1,1} &&&& 0 &\\[.8mm]
[\Zm_{1,1}]^{}_3 & [\Zm_{1,2}]^{}_3 &&&&&& [\Zm_{1,1}]^{}_3 s_{1,1} &&& [\Zm_{1,2}]^{}_3 s_{1,2}&\\[.8mm]
[\Zm_{1,1}]^{}_4 & [\Zm_{1,2}]^{}_4 &&&&&&& [\Zm_{1,1}]^{}_4 s_{1,1} &   & & [\Zm_{1,2}]^{}_4 s_{1,2} 
\\[.8mm]  % end of R=1
&&[\Zm_{2,1}]^{}_1 & [\Zm_{2,2}]^{}_1 &&& 0 &&& [\Zm_{2,2}]^{}_1 s_{2,2}\\[.8mm]
&&[\Zm_{2,1}]^{}_2 & [\Zm_{2,2}]^{}_2 &&& [\Zm_{2,1}]^{}_2 s_{2,1}  &&&& 0 \\[.8mm]
&&[\Zm_{2,1}]^{}_3 & [\Zm_{2,2}]^{}_3 &&&& [\Zm_{2,1}]^{}_3 s_{2,1} &&& [\Zm_{2,2}]^{}_3 s_{2,2} \\[.8mm]
&&[\Zm_{2,1}]^{}_4 & [\Zm_{2,2}]^{}_4 &&&&& [\Zm_{2,1}]^{}_4 s_{2,1} &&& [\Zm_{2,2}]^{}_4 s_{2,2}
 \\[.8mm] % end of R=2
&&&&[\Zm_{3,1}]^{}_1 & \text{\setlength{\fboxsep}{0pt}\colorbox{lightgray}{$[\Zm_{3,2}]^{}_1$}} & 0 &&& \text{\setlength{\fboxsep}{0pt}\colorbox{lightgray}{$[\Zm_{3,2}]^{}_1 s_{3,2}$}}\\[.8mm]
&&&&\text{\setlength{\fboxsep}{0pt}\colorbox{lightgray}{$[\Zm_{3,1}]^{}_2$}} & [\Zm_{3,2}]^{}_2 & \text{\setlength{\fboxsep}{0pt}\colorbox{lightgray}{$[\Zm_{3,1}]^{}_2 s_{3,1}$}} &&&& 0 \\[.8mm]
&&&&[\Zm_{3,1}]^{}_3 & \text{\setlength{\fboxsep}{0pt}\colorbox{lightgray}{$[\Zm_{3,2}]^{}_3$}} && [\Zm_{3,1}]^{}_3 s_{3,1} &&& \text{\setlength{\fboxsep}{0pt}\colorbox{lightgray}{$[\Zm_{3,2}]^{}_3 s_{3,2}$}} \\[.8mm]
&&&&\text{\setlength{\fboxsep}{0pt}\colorbox{lightgray}{$[\Zm_{3,1}]^{}_4$}} & [\Zm_{3,2}]^{}_4 &&& \text{\setlength{\fboxsep}{0pt}\colorbox{lightgray}{$[\Zm_{3,1}]^{}_4 s_{3,1}$}} &&& [\Zm_{3,2}]^{}_4 s_{3,2}
\end{smallmatrix}
\right)
\ee
% Restore the current equation number.
\setcounter{equation}{\value{MYtempeqncnt}}
% IEEE uses as a separator
\hrulefill
% The spacer can be tweaked to stop underfull vboxes.
\vspace*{4pt}
\end{figure*}

%We calculated upper and lower bounds on the number of degrees of freedom of generic block-fading channels. 
%In our main result Theorem~\ref{THmaintheoremq1} we showed that these bounds coincide in many cases and gave an exact characterization of the number of degrees of freedom. 
We characterized the number of degrees of freedom for generic block-fading MIMO channels in the noncoherent setting. 
Although the generic block-fading model seems to be just a minor variation of the classically used constant block-fading model, our result shows that the assumption of generic correlation may strongly affect the number of degrees of freedom. 
In fact, we showed that the (potentially small) perturbation in the channel model that results from making the coloring matrix $\Zm$ generic may yield a significant increase in the number of degrees of freedom.
%lead to a number of degrees of freedom that is up to four times as large as for the constant block-fading model.
%Furthermore, the number of antennas needed to achieve the maximal number of degrees of freedom 
%Furthermore, in the special case $\Q=1$, the number of degrees of freedom increases with the number of transmit antennas up to the block length $$
This suggests once more (see also~\cite{lamo03,dumobo12})
that care must be exercised in using this asymptotic quantity as a performance measure.

The highest gain in terms of the number of degrees of freedom is obtained for a sufficiently large number of receive antennas.
In this case, the number of degrees of freedom is equal to $\T$ times the number of degrees of freedom in the SIMO case, as long as the number $\T$ of transmit antennas satisfies  $\T<\L/\Q$.
This may be of interest for the uplink of massive-MIMO systems \cite{rupe13}.

From a practical point of view, the generic block-fading model is of particular interest for CP-OFDM systems. 
These systems cannot be described appropriately by the constant block-fading model, which corresponds to an impulse response of each $(t,r)$ channel that consists of a single tap.
By contrast, the generic block-fading model allows for impulse responses with multiple taps.
%Our result also suggests that the use of a large number of receive antennas (i.e, $\R\geq \Tprop(\L-1)/(\L-\Tprop\Q)$) is beneficial as it maximizes the number of degrees of freedom. 
%This may be of interest for the uplink of large-MIMO systems \cite{rupe13}.
%Furthermore, the generic block-fading model is of particular interest for CP-OFDM systems, which cannot be described appropriately by the constant block-fading model.
%
%We showed that a small perturbation in the channel model, such as the one introduced by making the matrix $\Zm$ generic, yields a significant increase in the number of degrees of freedom.
%This suggests once more (see also~\cite{lamo03,dumobo12})
%that care must be exercised in using this asymptotic quantity as performance metric.

For CP-OFDM systems with colocated antennas, it may appear
%% is perhaps 
questionable to assume that all coloring matrices $\Zm_{r,t}$ are different---an assumption that is needed for our result to hold 
(although MIMO channel matrices with nonidentical distributions arise, e.g., when pattern diversity is used \cite{tulove05}). 
The case where all matrices $\Zm_{r,t}$ are \emph{exactly} equal is still an open problem. 
However, it should be noted that
%% Note though that 
any nonzero
%% infinitesimal 
perturbation of the model with exactly equal $\Zm_{r,t}$---be it arbitrarily small---yields the generic model considered in this paper.
One may then argue that the assumption of exactly equal $\Zm_{r,t}$ is an idealization that may be convenient in theoretical analyses 
but will not be satisfied in practical systems.
An important conclusion to be drawn from our analysis is the fact that, as far as the number of degrees of freedom is concerned, the model with exactly equal $\Zm_{r,t}$ 
is highly nonrobust, since arbitrarily small perturbations yield a potentially large change in the number of degrees of freedom.

The proof of Proposition~\ref{pro:prelim} in Section~\ref{sec:prooflowerbound} does not provide a characterization
%% description 
of the class of coloring matrices $\Zm$ for which Theorem~\ref{th:lowerbound}  does not hold.
However, the only part of the proof where a generic 
%% coloring matrix 
$\Zm$ is needed is in the statement that $\big\lvert\Jm_{\phi_{[\xv]^{}_{\sP}}}\!(\sv,[\xv]^{}_{\sD})\big\rvert\neq 0$ a.e.\ (see Step~(\ref{step2a}) in Section~\ref{sec:boundh}).
%%  on p.~\pageref{LEMnotvanish}).
If a specific $\Zm$
%% coloring matrix 
is given, one can search for two vectors $\sv$ and $\xv$ and a set $\sP$ for which $\big\lvert\Jm_{\phi_{[\xv]^{}_{\sP}}}\!(\sv,[\xv]^{}_{\sD})\big\rvert\neq 0$. 
If the search is successful, then Theorem~\ref{th:lowerbound} holds for this $\Zm$.
%% coloring matrix.
Note that the converse is not necessarily true: if $\big\lvert\Jm_{\phi_{[\xv]^{}_{\sP}}}\!(\sv,[\xv]^{}_{\sD})\big\rvert$ vanishes for all choices of $\sv$, $\xv$, and $\sP$, one cannot conclude that Theorem~\ref{th:lowerbound} does not hold.

An open problem is a characterization of the capacity of generic block-fading MIMO channels beyond the number of degrees of freedom. 
Such a characterization would help understand whether the sensitivity of the number of degrees of freedom discussed above is an indication of a similar sensitivity of the capacity that occurs already at moderate SNR, or merely an asymptotic peculiarity. 
Furthermore, a capacity characterization that is nonasymptotic in the SNR can be analyzed for asymptotic block length, which would enable a capacity analysis of, e.g., stationary channel models.

\appendices

%%%%%%%%%%%%%%%%%%%%%%%%%%%%%%%
\section{Proof of Lemma~\ref{LEMnotvanish}} \label{app:proofnotvanish}
%%%%%%%%%%%%%%%%%%%%%%%%%%%%%%%
\renewcommand*\thesubsectiondis{\Alph{subsection}.}
\renewcommand*\thesubsection{\thesection.\Alph{subsection}}

\begin{figure*}[!t]
% ensure that we have normalsize text
\normalsize
% Store the current equation number.
%\newcounter{MYtempeqncnt}
\setcounter{MYtempeqncnt}{\value{equation}}
% Set the equation number to one less than the one
% desired for the first equation here.
% The value here will have to changed if equations
% are added or removed prior to the place these
% equations are referenced in the main text.
\setcounter{equation}{55}
\be\label{eq:examplejacobian2}
\left(
\begin{smallmatrix}
[\Zm_{1,1}]^{}_1 & [\Zm_{1,2}]^{}_1 &&& 0 && \text{\setlength{\fboxsep}{0pt}\colorbox{lightgray}{$[\Zm_{1,2}]^{}_1 s_{1,2}$}} \\[.8mm]
[\Zm_{1,1}]^{}_2 & [\Zm_{1,2}]^{}_2  &&& [\Zm_{1,1}]^{}_2 s_{1,1} &&& 0\\[.8mm]
[\Zm_{1,1}]^{}_3 & [\Zm_{1,2}]^{}_3 &&&& 0 && \text{\setlength{\fboxsep}{0pt}\colorbox{lightgray}{$[\Zm_{1,2}]^{}_3 s_{1,2}$}}\\[.8mm]
[\Zm_{1,1}]^{}_4 & [\Zm_{1,2}]^{}_4 &&&& [\Zm_{1,1}]^{}_4 s_{1,1} && 0
 \\[.8mm]  % end of R=1
&&[\Zm_{2,1}]^{}_1 & [\Zm_{2,2}]^{}_1 & 0 && [\Zm_{2,2}]^{}_1 s_{2,2}\\[.8mm]
&&[\Zm_{2,1}]^{}_2 & [\Zm_{2,2}]^{}_2 & \text{\setlength{\fboxsep}{0pt}\colorbox{lightgray}{$[\Zm_{2,1}]^{}_2 s_{2,1}$}} &&& 0 \\[.8mm]
&&[\Zm_{2,1}]^{}_3 & [\Zm_{2,2}]^{}_3 && 0 && [\Zm_{2,2}]^{}_3 s_{2,2} \\[.8mm]
&&[\Zm_{2,1}]^{}_4 & [\Zm_{2,2}]^{}_4 && \text{\setlength{\fboxsep}{0pt}\colorbox{lightgray}{$[\Zm_{2,1}]^{}_4 s_{2,1}$}} && 0
\end{smallmatrix}
\right)
\ee
% Restore the current equation number.
\setcounter{equation}{\value{MYtempeqncnt}}
% IEEE uses as a separator
\hrulefill
% The spacer can be tweaked to stop underfull vboxes.
\vspace*{4pt}
\end{figure*}

%\vspace{1mm}

Since the proof of Lemma~\ref{LEMnotvanish} is quite technical, we shall first (in Section~\ref{sub:example}) illustrate its key steps by focusing on the special case 
%present the proof for the simplified setting 
$\Tprop=2, \R=3, \L=4$, and $\Q=1$. 
The proof for arbitrary $\Tprop, \R, \L,$ and $\Q$ will be provided in Section~\ref{sub:part1}.

\subsection{Special Case $\Tprop=2, \R=3, \L=4$, $\Q=1$}\label{sub:example}
By~\eqref{eq:defnreq}, we have $\nreq=0$ and, thus,  $\sI=[1\!:\! 12]$. 
Furthermore, by~\eqref{eq:indp}, we have $\abs{\sP}=2$.
We choose $\sP=\{1, 6\}$. 
Hence, recalling that $\xv=(\xv_1^{\trans} \; \xv_2^{\trans})^{\trans}\in \IC^8$, we have 
\[
[\xv]^{}_{\sP}=([\xv_1]^{}_{1} \; [\xv_2]^{}_{2})^{\trans}
\]
 and  
\[
[\xv]^{}_{\sD}=([\xv_1]^{}_{2} \; [\xv_1]^{}_{3} \;[\xv_1]^{}_{4} \;[\xv_2]^{}_{1} \;[\xv_2]^{}_{3} \;[\xv_2]^{}_{4})^{\trans}\,.
\]
We also choose $\xv$ as the all-one vector. % Setting all the entries of $\xv$ equal to $1$ we obtain that the 
For these choices, the Jacobian $\Jm_{\phi_{[\xv]^{}_{\sP}}}$ in~\eqref{eq:Jacobian1} is equal to \eqref{eq:examplejacobian1} \addtocounter{equation}{1}at the top of this page.
We have to find $\Zm$ and $\sv$ such that the determinant of this matrix is nonzero.
Setting $[\Zm_{3,2}]^{}_1 \!\rmv=\! [\Zm_{3,1}]^{}_2\!\rmv=\![\Zm_{3,2}]^{}_3
 \!=\! [\Zm_{3,1}]^{}_4 \!=\! 0$, 
the entries highlighted in gray in~\eqref{eq:examplejacobian1} become zero.
Furthermore, choosing nonzero $[\Zm_{3,1}]^{}_1$, $\rmv[\Zm_{3,1}]^{}_3$, $\rmv[\Zm_{3,2}]^{}_2$, $\rmv[\Zm_{3,2}]^{}_4$, 
$\rmv s_{3,1}$, and $\rmv s_{3,2}$ 
and operating a Laplace expansion on the last four rows in~\eqref{eq:examplejacobian1}, it is seen that the determinant of the matrix in~\eqref{eq:examplejacobian1} is nonzero if the determinant of the matrix in \eqref{eq:examplejacobian2} \addtocounter{equation}{1}at the top of the next page is nonzero.
%\be\label{eq:examplejacobian2}
%\left(
%\begin{smallmatrix}
%[\Zm_{1,1}]^{}_1 & [\Zm_{1,2}]^{}_1 &&& 0 && \text{\setlength{\fboxsep}{0pt}\colorbox{lightgray}{$[\Zm_{1,2}]^{}_1 s_{1,2}$}} \\[.8mm]
%[\Zm_{1,1}]^{}_2 & [\Zm_{1,2}]^{}_2  &&& [\Zm_{1,1}]^{}_2 s_{1,1} &&& 0\\[.8mm]
%[\Zm_{1,1}]^{}_3 & [\Zm_{1,2}]^{}_3 &&&& 0 && \text{\setlength{\fboxsep}{0pt}\colorbox{lightgray}{$[\Zm_{1,2}]^{}_3 s_{1,2}$}}\\[.8mm]
%[\Zm_{1,1}]^{}_4 & [\Zm_{1,2}]^{}_4 &&&& [\Zm_{1,1}]^{}_4 s_{1,1} && 0
 %\\[.8mm]  % end of R=1
%&&[\Zm_{2,1}]^{}_1 & [\Zm_{2,2}]^{}_1 & 0 && [\Zm_{2,2}]^{}_1 s_{2,2}\\[.8mm]
%&&[\Zm_{2,1}]^{}_2 & [\Zm_{2,2}]^{}_2 & \text{\setlength{\fboxsep}{0pt}\colorbox{lightgray}{$[\Zm_{2,1}]^{}_2 s_{2,1}$}} &&& 0 \\[.8mm]
%&&[\Zm_{2,1}]^{}_3 & [\Zm_{2,2}]^{}_3 && 0 && [\Zm_{2,2}]^{}_3 s_{2,2} \\[.8mm]
%&&[\Zm_{2,1}]^{}_4 & [\Zm_{2,2}]^{}_4 && \text{\setlength{\fboxsep}{0pt}\colorbox{lightgray}{$[\Zm_{2,1}]^{}_4 s_{2,1}$}} && 0
%\end{smallmatrix}
%\right)\,.
%\ee
This is the Jacobian matrix corresponding to the case $\Tprop=2, \R=2, \L=4$, and $\Q=1$. In other words, by performing the matrix manipulations just described, we reduced the case $\R=3$ to the case $\R=2$. 
A similar idea will be used in the proof for the general case provided in Section~\ref{sub:part1}, where we will reduce $\R$ inductively until $\R=\Tprop$.
Setting $s_{1,2} \!=\! s_{2,1} \!=\! 0$, the entries
highlighted in gray in~\eqref{eq:examplejacobian2} become zero. 
By choosing nonzero $[\Zm_{1,1}]^{}_2$, $\rmv[\Zm_{1,1}]^{}_4$, $\rmv[\Zm_{2,2}]^{}_1$, $\rmv[\Zm_{2,2}]^{}_3$, $s_{1,1}$,
and $s_{2,2}$ 
 and 
oper\-ating 
a Laplace expansion on the last 
four columns, it is seen that  it is sufficient to show that the determinant of the following matrix is nonzero:
\[ %\label{eq:examplejacobian3}
\left(
\begin{matrix}
[\Zm_{1,1}]^{}_1 & [\Zm_{1,2}]^{}_1 \\[.8mm]
[\Zm_{1,1}]^{}_3 & [\Zm_{1,2}]^{}_3 
 \\[.8mm]  % end of R=1
&&[\Zm_{2,1}]^{}_2 & [\Zm_{2,2}]^{}_2 \\[.8mm]
&&[\Zm_{2,1}]^{}_4 & [\Zm_{2,2}]^{}_4 
\end{matrix}
\right)\,.
\]
This can 
be achieved, e.g.,
by setting all off-diagonal entries 
(i.e., $[\Zm_{1,1}]^{}_3$, $[\Zm_{1,2}]^{}_1$, $[\Zm_{2,1}]^{}_4$, and $[\Zm_{2,2}]^{}_2$) to zero
and choosing all diagonal entries  
(i.e., $[\Zm_{1,1}]^{}_1$, $\rmv[\Zm_{1,2}]^{}_3$, 
$\rmv[\Zm_{2,1}]^{}_2$, and $\rmv[\Zm_{2,2}]^{}_4$) nonzero.

\subsection{Proof for the General Case}\label{sub:part1}
%%%%%%%%%%%%%%%%%%%%%%
%
%
We have to find $\Zm$, $\sv$, $\xv$, and $\sP$ such that 
\[
\big\lvert\Jm_{\phi_{[\xv]^{}_{\sP}}}\!(\sv,[\xv]^{}_{\sD})\big\rvert\neq 0\,.
\]
%\vspace{-2mm}
%
\subsubsection{Construction of $\sP$}
%%%%%%%%%%%
We start by constructing the set $\sP$. 
Recall that $\sP$ specifies the indices of the pilot symbols in the vector $\xv = (\xv_1^{\trans} \cdots \xv_{\Tprop}^{\trans})^{\trans}$. 
It will turn out convenient to use the expression 
\be\label{eq:pilots}
\sP=\{i+(t-1)\L: i\in \sP_t, t\in [1\!:\!\Tprop]\}
\ee
 where $\sP_t \subseteq[1\!:\!\L]$  specifies the indices of the pilot symbols in the vector $\xv_t$, $t\in [1\!:\!\Tprop]$.
%We first construct sets $\sP_t\subseteq[1\!:\!\L]$ that specify the position of the pilots for each $\xv_t$ for $t\in [1\!:\!\Tprop]$ separately. 
%Afterwards we obtain the set $\sP$ by $\sP\triangleq\{i+(t-1)\L: i\in \sP_t\}$.
%To this end we 
%With this idea we obtain the connection $\sP\triangleq\{i+(t-1)\L: i\in \sP_t\}$.
%introduce sets $\sP_t\subseteq[1\!:\!\L]$  that specify the pilots in each $\xv_t$ (recall that $\xv = (\xv_1^{\trans} \dots \xv_{\Tprop}^{\trans})^{\trans}$) for $t\in [1\!:\!\Tprop]$, i.e., $\sP_t$ specifies the entries of $\sP\cap[(t-1)\L+1:t\L]$. With this setting we obtain the set $\sP$ as $\sP\triangleq\{i+(t-1)\L: i\in \sP_t\}$. 
%\begin{example}
%In the simplified setting we have $\sP\subseteq [1\!:\!8]$ where, e.g., $1\in \sP_1$ impies $1\in \sP$ and $2\in \sP$ implies $4+2=6\in \sP$.
%\end{example}
The sets $\{\sP_t\}_{t\in[1:\Tprop]}$ have to satisfy
\begin{align}
\sum_{t\in[1:\Tprop]}\abs{\sP_t} & =
\abs{\sP} \notag \\[-4mm]
& \stackrel{\hidewidth \eqref{eq:indp} \hidewidth}= \,\R\Tprop\Q+\Tprop\L-\abs{\sI} \notag \\
& \stackrel{\hidewidth \eqref{eq:defi} \hidewidth}= \,\R\Tprop\Q+\Tprop\L-\R\L+\nreq \notag \\
& \stackrel{\hidewidth \eqref{eq:defnreq} \hidewidth}= \, \R\Tprop\Q+\Tprop\L-\R\L \notag \\
& \rule{10mm}{0mm}
+\max\{0, \R\L-(\R\Tprop\Q+\Tprop\L-\Tprop)\} \notag \\
& = \,\max\{\Tprop, \R\Tprop\Q-(\R \rmv-\rmv \Tprop)\L\}  \notag \\
& \triangleq\, \vartheta_{\R}\,.
\label{eq:sizep1}
\end{align}
%\be
%\sum_{t\in[1:\Tprop]}\abs{\sP_t} \,=\, \abs{\sP} =  \max\{\Tprop, \R\Tprop\Q-(\R \rmv-\rmv \Tprop)\L\} \,\triangleq\, \vartheta_{\R} \,. 
%\label{eq:sizep1}
%\ee
(We use the subscript $\R$ in $\vartheta_{\R}$ because the dependence on $\R$ will be important later.)
To provide intuition about our choice of the sets $\sP_t$, we use a card game metaphor.
Consider a deck of $\Tprop\L$ cards showing numbers from $1$ to $\L$ sorted as follows: $1, 2, \dots, \L,  \dots, 1, 2, \dots, \L$ (i.e., the sequence $1,2,\dots,\L$ repeated $\Tprop$ times).
The idea is to choose the $\vartheta_{\R}$ positions of the pilot symbols by assigning the indices $i\in [1\!:\!\L]$ to the sets $\sP_t$ in the same way as the first $\vartheta_{\R}$ cards are distributed to $\Tprop$ players (in Fig.~\ref{fig:cards}, we give an example of the algorithm for $\vartheta_{\R}=14$, $\L=6$, and $\Tprop=4$): 
The first card shows $1$ and goes to $\sP_1$, i.e., $1\in \sP_1$, and in the same way we proceed with $2\in \sP_2$, $\dots$, $\Tprop\in \sP_{\Tprop}$ (this corresponds to the $1$st to $4$th card in Fig.~\ref{fig:cards}). 
When we run out of sets (players), we start with the first set (player) again: $\Tprop+1\in \sP_1$, $\Tprop+2\in \sP_2$, etc. 
After the card showing index $\L$ (recall that $\sP_t\subseteq [1\!:\!\L]$), the next card starts with index $1$ again (in  Fig.~\ref{fig:cards}, the $6$th card shows $\L=6$ and goes to $\sP_2$ and the $7$th card shows $1$ and goes to $\sP_{3}$).
This scheme works as long as we %do not come to a situation where 
avoid assigning an index to a set $\sP_t$ to which that index was already assigned in a previous round.
\begin{figure}[tbp]
\centering
\begin{tikzpicture}[node distance = 6em]
\tikzset{card/.style={rectangle, draw, fill=blue!20,  
    text width=1.5em,  text badly centered, rounded corners, minimum height=1em,label={below:#1}}}
\tikzset{player/.style={rectangle, draw, 
    text width=8em, rounded corners, inner sep=1.2em,label={above:#1}}}
\node [card=\scriptsize{1st card}] (card1) {1};
\node [card=\scriptsize{2nd card}, right of=card1] (card2) {2};
\node [card=\scriptsize{3rd card}, right of=card2] (card3) {3};
\node [card=\scriptsize{4th card}, right of=card3] (card4) {4};
\node [card=\scriptsize{5th card}, below of=card1, node distance = 3em] (card5) {5};
\node [card=\scriptsize{6th card}, right of=card5] (card6) {6};
\node [card=\scriptsize{7th card}, right of=card6] (card7) {1};
\node [card=\scriptsize{8th card}, right of=card7] (card8) {2};
\node [card, below of=card5, fill=white, white, node distance = 4em] (card9b) {};
\node [card=\scriptsize{9th card}, below of=card5, node distance = 3em] (card9) {3};
\node [card=\scriptsize{10th card}, right of=card9] (card10) {4};
\node [card=\scriptsize{11th card}, right of=card10] (card11) {5};
\node [card, below of=card8, fill=white, white, node distance = 4em] (card12b) {};
\node [card=\scriptsize{12th card}, right of=card11] (card12) {6};
\node [card, below of=card10, fill=white, white, node distance = 4em] (card13b) {};
\node [card=\scriptsize{13th card}, below of=card10, node distance = 3em] (card13) {1};
\node [card, below of=card11, fill=white, white, node distance = 4em] (card14b) {};
\node [card=\scriptsize{14th card}, right of=card13] (card14) {2};
\node [player=$\sP_1$, fit=(card1)(card9b)] (player1) {};
\node [player=$\sP_2$, fit=(card2)(card13b)] (player2) {};
\node [player=$\sP_3$, fit=(card3)(card14b)] (player3) {};
\node [player=$\sP_4$, fit=(card4)(card12b)] (player4) {};
%\draw[rounded corners=2mm] 
%      (-5.1,-2.1) rectangle (5.1,2.1) node[above] {$[1\!:\!\L-\nreq]$};
%\draw[rounded corners=2mm] 
%      (-5,-2) rectangle (-4,2)
%      (-4,-2) rectangle (-3,2)
%      (-1,-2) rectangle (0,2);
%\draw[rounded corners=2mm] 
%      (0,-2) rectangle (1,2)
%      (1,-2) rectangle (2,2)
%      (4,-2) rectangle (5,2);
%\node at (-4.5,0) {$\sG_1$};
%\node at (-3.5,0) {$\sG_2$};
%\node at (-2,0) {$\cdots$};
%\node at (-0.5,0) {$\sG_{\Tprop}$};
%\node at (0.5,0) {$\sL_1$};
%\node at (1.5,0) {$\sL_2$};
%\node at (3,0) {$\cdots$};
%\node at (4.5,0) {$\sL_{\Tprop}$};
%\draw (-4.5,-1.5) circle (.35);
%\node at (-4.5,-1.5) {$g_1$};
%\draw (-3.5,-.8) circle (.35);
%\node at (-3.5,-.8) {$g_2$};
%\draw (-.5,1.5) circle (.35);
%\node at (-.47,1.5) {$g_{\Tprop}$};
\end{tikzpicture}
%\vspace{-3mm}
\caption{Construction of the sets $\sP_t$ for $\Tprop=4$, $\L=6$, and $\vartheta_{\R}=14$.}
\label{fig:cards}
%\vspace{-5mm}
\end{figure}
%%%%%%%%%%%
(In Fig.~\ref{fig:cards}, this would happen after the $12$th card. The $13$th card shows $1$ and the algorithm would set $1\in \sP_1$, which was already assigned to $\sP_1$ in the first round.)
%Here is an example where this situation occurs. 
%Let $\L=4$ and $\Tprop=2$. 
%According to the algorithm just described, we need to assign $1\in \sP_1$, $2\in \sP_2$, $3\in \sP_1$, $4\in \sP_2$. 
%If we want to assign another position, the algorithm would set $1\in \sP_1$, which was already assigned to $\sP_1$ in the first round. 
To avoid this issue, we introduce an offset and skip one set (resulting in the $13$th card going to $\sP_2$ in Fig.~\ref{fig:cards}) and proceed as before. 
The algorithm stops when $\vartheta_{\R}$ indices (cards) have been assigned to the sets (players) $\sP_t$.

We now present a mathematical formulation of the algorithm we just outlined.
%To construct the sets $\sP_t$ 
Let the function $\betav\colon [1\!:\!\Tprop\L]  \rightarrow [1\!:\!\Tprop]\times[1\!:\!\L]$ be defined as
\ba
& \betav(j)=
\begin{pmatrix}
\beta_1(j) \\[2mm]
\beta_2(j)
\end{pmatrix}
\triangleq
\begin{pmatrix}
\Big(j+\Big\lfloor \frac{j - 1}{\lcm(\Tprop,\L)}\Big\rfloor \Big) \mymod \Tprop \\[3mm]
j \mymod \L 
\end{pmatrix}, \notag \\[1mm]
& \rule{55mm}{0mm} j\in [1\!:\!\Tprop\L]\,.
\label{eq:beta}
\ea
Here $\lcm(\cdot,\cdot)$ denotes the least common multiple and  
\[
a \mymod b \triangleq a-b\bigg\lfloor\frac{a-1}{b} \bigg\rfloor
\]
 denotes the residuum of $a$ divided by $b$ in $[1\!:\!b]$ (and not in $[0\!:\!b-1]$ as commonly done). 
We use the function $\betav$ to assign up to $\Tprop\L$ elements (note that $\vartheta_{\R}\leq \Tprop\L$) to the sets $\sP_t$ as follows: 
for $j\in [1\!:\!\vartheta_{\R}]$, the function $\beta_1(j)$ specifies $t\in [1\!:\!\Tprop]$ (equivalently, one of the sets $\sP_t,$ $t\in [1\!:\!\Tprop]$), and the function $\beta_2(j)$ specifies the index $i \in [1\!:\!\L]$ that is assigned to $\sP_t$ 
(again invoking our card game metaphor, the $j$th card shows the index $\beta_2(j)$ and is assigned to player $\sP_{\beta_1(j)}$). 
Using $\beta_1(j)$ and $\beta_2(j)$, we can compactly describe each set $\sP_t$ as follows:\footnote{For a set $\sA\subseteq [1\!:\!\Tprop\L]$, we use the notation $\beta_2(\sA)$ to denote the image of the set $\sA$ under the function $\beta_2$, i.e., $\beta_2(\sA)=\{\beta_2(j):j\in \sA\}$.}
\be\label{eq:defpt}
\sP_t\triangleq \beta_2\big(\beta_1^{-1}(t)\cap [1\!:\!\vartheta_{\R}]\big), \quad t\in [1\!:\!\Tprop].
\ee
Here, the set $\beta_1^{-1}(t)$ consists of all values $j\in [1\!:\!\Tprop\L]$ that correspond to an assignment of an index $i$ to the set $\sP_t$. Since we only want to assign a total of $\vartheta_{\R}$ indices, we take the intersection with $[1\!:\!\vartheta_{\R}]$. For each $j\in \beta_1^{-1}(t)\cap [1\!:\!\vartheta_{\R}]$, the function $\beta_2$ now chooses an index $i \in [1\!:\!\L]$, and we obtain the definition~\eqref{eq:defpt}.

The sets $\sP_t$ in~\eqref{eq:defpt} satisfy the properties listed in the following lemma.
\begin{lemma}\label{LEMproppt}
Suppose that
%\footnote{Note that in Proposition~\ref{pro:prelim}, we required the weaker condition $\R\geq \Tprop$. 
%Because we will need Lemma~\ref{LEMproppt} only in the inductive step reducing the number of receive antennas, the assumption $\R \!>\! \Tprop$ turns out to be reasonable. 
%The other two inequalities are the constraints assumed throughout the proof of Lemma~\ref{LEMnotvanish}, cf.~\eqref{eq:assumption1} and~\eqref{eq:assumption2}.} 
$\R \geq \Tprop$, $\L>\Tprop\Q$, and $\R\leq \lceil\Tprop(\L-1)/(\L-\Tprop\Q)\rceil$. 
Let the sets $\{\sP_t\}_{t\in [1:\Tprop]}$ be defined  as in~\eqref{eq:defpt}.
Then the following properties hold:
\begin{enumerate} %[(i)]
\renewcommand{\theenumi}{\roman{enumi}}
\renewcommand{\labelenumi}{(\roman{enumi})}
\renewcommand{\theenumii}{-\alph{enumii}}
\renewcommand{\labelenumii}{\alph{enumii})}
\item $\sum_{t\in[1:\Tprop]}\abs{\sP_t} \,=\, \vartheta_{\R}$;\vspace{1mm}~  
\label{item:sizept}
\item $\abs{\sP_t}\leq \Tprop\Q$; \label{item:sizep}
\end{enumerate}
If $\R>\Tprop$, let $\{\sPt_t\}_{t\in [1:\Tprop]}$ be the corresponding sets for the case of $\R \!-\! 1$ receive antennas, i.e., 
\be\label{eq:defpttilde}
\sPt_t\triangleq \beta_2\big(\beta_1^{-1}(t)\cap [1\!:\!\vartheta_{\R-1}]\big)\,.
\ee
Furthermore, we set
\be\label{eq:deflt}
\sL_t\triangleq \sPt_t\setminus \sP_t
\ee
and
\be\label{eq:defsLt}
\sLt\triangleq\bigcup_{t\in [1:\Tprop]}\sL_t\,.
\ee 
Then the following properties hold:
\begin{enumerate}
\renewcommand{\theenumi}{\roman{enumi}}
\renewcommand{\labelenumi}{(\roman{enumi})}
\renewcommand{\theenumii}{-\alph{enumii}}
\renewcommand{\labelenumii}{\alph{enumii})}
\setcounter{enumi}{2}
\item $\sL_t\cap \sL_{t'}=\emptyset$ for 
\vspace{.5mm}
$t\neq t'$; \label{item:ldisjoint}
\item $\sL_t\subseteq [1\!:\!\L-\nreq]$, where $\nreq$ is defined in~\eqref{eq:defnreq};\label{item:lsubi}
\vspace{.5mm}
\item There exist sets $\sG_t\subseteq [1\!:\!\L-\nreq]$, $t\in [1\!:\!\Tprop]$ satisfying  \label{item:partitiong}
\begin{enumerate}
\item $\abs{\sG_t}=\Q$,\label{en:partitiong1}
\item $\sG_t\cap \sG_{t'}= \emptyset$ for $t\neq t'$,\label{en:partitiong2}
\item $\sG_t\cap \sP_t\neq \emptyset$,\label{en:partitiong3}
\item $\bigcup_{t\in [1:\Tprop]}\sG_t \ist=\ist\sG \ist\triangleq\ist  [1\!:\!\L-\nreq]\setminus\sLt$.\label{en:partitiong4}
\end{enumerate}
\vspace{1.5mm}
\end{enumerate}
\end{lemma}
\begin{IEEEproof}[\hspace{-1em}Proof]
See Appendix~\ref{app:proofproppt}.
%\vspace{1.5mm} 
\end{IEEEproof}
\begin{remark}
Property~\eqref{item:sizept} states that the sets $\sP_t$ have the correct size (see~\eqref{eq:sizep1}). 
Properties~\eqref{item:ldisjoint}, \eqref{item:lsubi}, and~\eqref{item:partitiong} state that we can partition the set $[1\!:\!\L-\nreq]$ into $2\Tprop$ disjoint sets $\sL_t$ and $\sG_{t'}$, $t, t'\in [1\!:\!\Tprop]$, 
i.e., $\sG_t\cap \sG_{t'}= \emptyset$  for $t\neq t'$ (see~\eqref{en:partitiong2}), $\sL_t\cap \sL_{t'}= \emptyset$ for $t\neq t'$ (see~\eqref{item:ldisjoint}), and $\sG_t\cap \sL_{t'}= \emptyset$  for $t, t'\in [1\!:\!\Tprop]$ (see~\eqref{en:partitiong4}). Furthermore, in each $\sG_{t}$ there is a point $g_{t}\in \sP_{t}$ (see~\eqref{en:partitiong3}). 
%This partitioning is sketched in Fig.~\ref{fig:gtlt}.
%\begin{figure}[bth]
%\centering
%\begin{tikzpicture}
%\draw[rounded corners=2mm] 
%      (-5.1,-2.1) rectangle (5.1,2.1) node[above] {$[1\!:\!\L-\nreq]$};
%\draw[rounded corners=2mm] 
%      (-5,-2) rectangle (-4,2)
%      (-4,-2) rectangle (-3,2)
%      (-1,-2) rectangle (0,2);
%\draw[rounded corners=2mm] 
%      (0,-2) rectangle (1,2)
%      (1,-2) rectangle (2,2)
%      (4,-2) rectangle (5,2);
%\node at (-4.5,0) {$\sG_1$};
%\node at (-3.5,0) {$\sG_2$};
%\node at (-2,0) {$\cdots$};
%\node at (-0.5,0) {$\sG_{\Tprop}$};
%\node at (0.5,0) {$\sL_1$};
%\node at (1.5,0) {$\sL_2$};
%\node at (3,0) {$\cdots$};
%\node at (4.5,0) {$\sL_{\Tprop}$};
%\draw (-4.5,-1.5) circle (.35);
%\node at (-4.5,-1.5) {$g_1$};
%\draw (-3.5,-.8) circle (.35);
%\node at (-3.5,-.8) {$g_2$};
%\draw (-.5,1.5) circle (.35);
%\node at (-.47,1.5) {$g_{\Tprop}$};
%\end{tikzpicture}
%\caption{Sketch of the sets $\sG_t$ and $\sL_t$}
%\label{fig:gtlt}
%\end{figure} 
\end{remark}
%
%\begin{example}
%For the simplified setting we have $\vartheta_{3}=\max\{2, 3\cdot 2\cdot 1-(3 - 2)\cdot4\}=2$. Hence, we get with Example~\ref{ex:defpt} that $\sP_1=\{1\}$ and $\sP_2 = \{2\}$. Furthermore, we have $\vartheta_{2}=\max\{2, 2\cdot 2\cdot 1-(2 -2)\cdot4\}=4$ and thus $\sPt_1 = \{1, 3\}$ and $\sPt_2=\{2, 4\}$.
%\end{example}
%
%\subsection{General Proof, Part III: Induction Argument} \label{sub:part3}
%\vspace{-2mm}

\subsubsection{Construction of $\Zm$, $\sv$, and $\xv$}
For the choice of $\{\sP_t\}_{t\in[1:\Tprop]}$  described above, it now remains to find a triple $(\Zm,\sv,\xv)$ for which
$p(\Zm, \sv, \xv) = \det\big(\Jm_{\phi_{[\xv]^{}_{\sP}}}\!(\sv,[\xv]^{}_{\sD})\big)$ is nonzero. 
This will be done by an induction argument over $\R\geq\Tprop$. 
For this purpose, it is convenient to define the \vspace{-.5mm}sets 
\be\label{eq:dtaspt}
\sD_t\triangleq [1\!:\!\L]\setminus\sP_t\,.
\ee
Note that by~\eqref{eq:pilots} and because $\sD = [1\!:\!\Tprop\L]\setminus \sP$, we have that
\ba
\sD & = [1\!:\!\Tprop\L]\setminus \sP \notag \\
& = [1\!:\!\Tprop\L]\setminus\{i+(t-1)\L: i\in \sP_t, t\in [1\!:\!\Tprop]\}  \notag \\
& =\{i+(t-1)\L: i\in \sD_t, t\in [1\!:\!\Tprop]\}
\label{eq:data}
\ea
i.e.,  $\sD_t \subseteq[1\!:\!\L]$  specifies the positions of the data symbols in the vector $\xv_t$, $t\in [1\!:\!\Tprop]$.
Furthermore, we will make repeated use of the next result,
which follows from
\cite[Sec.~0.8.5]{hojo85}. 
%\vspace{1.5mm}~

\begin{lemma} \label{LEMdet}
Let $\Mm\in \IC^{n\times n}\rmv$, and let $\sE, \sF\subseteq [1\!:\!n]$ 
with $|\sE|=|\sF|$. If ${[\Mm]}_{[1:n]\setminus \sE}^{\sF} \rmv=\rmv \0v$ or ${[\Mm]}^{[1:n]\setminus \sF}_{\sE} \!\!=\rmv \0v$, and if ${[\Mm]}_{\sE}^{\sF}$ is nonsingular, 
then $\det(\Mm) \not= 0$ 
%% does not vanish 
if and only if 
%\vspace{1.5mm}
$\det\rmv\Big( {[\Mm]}_{[1:n]\setminus \sE}^{[1:n]\setminus \sF} \Big) \not= 0$.
%%  does not %\vspace{1.5mm}vanish.
\end{lemma}

\begin{remark}
Lemma~\ref{LEMdet} is just an abstract way to describe a situation where given a matrix $\Mm$, one is able to perform row and column interchanges that yield a new matrix of the form
$
\Big(\begin{smallmatrix}
\Am & \Bm \\[1mm]
\0v & \Cm
\end{smallmatrix}\Big)
$,
where $\Am$ and $\Cm$ are square matrices.
In this case, a basic result in linear algebra states that the determinant of $\Mm$ equals the product of the determinants of $\Am$ and $\Cm$, and hence, assuming that $\Cm$ is nonsingular, $\det(\Mm)\neq 0$ if and only if $\det(\Am)\neq 0$.
\end{remark}

%% Recall that we have to choose $\xv$, $\sv$, and $\Qm$ such that $\det(\Jm_{\phi_{[\xv]^{}_{\sP}}}(\sv,[\xv]^{}_{\sD}))\neq 0$. 
%% We will conclude the proof of Lemma~\ref{LEMnotvanish} 
%Due to space limitations we will only sketch the proof and omit many details.
%Because $\xv$ does not depend on $\R$, we can specify it before we start with the induction argument and set $\xv=(1, \dots, 1)^{\trans}$.

We will now present the inductive construction of $\Zm$, $\sv$, and $\xv$.

\subsubsection*{Induction hypothesis}
For $\Tprop\leq \R\leq \lceil\Tprop(\L-1)/(\L-\Tprop\Q)\rceil$, $\Tprop\Q<\L$ (as assumed throughout the proof), and $\{\sP_t\}_{t\in[1:\Tprop]}$ as in~\eqref{eq:defpt},
%, and $\nreq$ as in~\eqref{eq:defnreq},
there exists a triple $(\Zm, \sv, \xv)$ with $\xv=(1  \cdots  1)^{\trans}$ such that $p(\Zm, \sv, \xv)=\det\big(\Jm_{\phi_{[\xv]^{}_{\sP}}}\!(\sv,[\xv]^{}_{\sD})\big)$ is nonzero.

\subsubsection*{Base case (proof for $\R \!=\! \Tprop$)}
%We have to show that the determinant of the matrix in~\eqref{eq:Jacobian1} is nonzero for 
%$\R \!=\! \Tprop$.
%\be\label{eq:JacobianReqT}
%%\Jm_{\phi_{[\xv]^{}_{\sP}}}(\sv,[\xv]^{}_{\sD}) = 
%\left(
%\tilde{\Xim}_1 \,\cdots \, \tilde{\Xim}_{\Tprop} \;
%\begin{matrix}
%\Am_{1,1} & \hspace*{-2mm}\cdots\hspace*{-2mm} & \Am_{1,\Tprop}  \\[-.8mm]
%\vdots    &     & \vdots     \\
%\Am_{\Tprop,1} & \hspace*{-2mm}\cdots\hspace*{-2mm} &\Am_{\Tprop,\Tprop}
%\end{matrix}
%\right),
%\ee
%where (cf.\~\eqref{eq:Jacobian3a}, noting that $\Xm_t=\Iv_{\L}$)
%\[
%\tilde{\Xim}_t\triangleq 
%\begin{pmatrix}
%\Qm_{1,t} \\[-.8mm]
%& \hspace*{-2mm}\ddots\hspace*{-2mm} \\
%&& \Qm_{\Tprop,t}
%\end{pmatrix}.
%\]
When $\R \!=\! \Tprop$,~\eqref{eq:sizep1} reduces to
$ %\label{eq:sizepcasert}
\sum_{t\in[1:\Tprop]}\abs{\sP_t}= {\Tprop}^2\Q$.
Using Property~\eqref{item:sizep} in Lemma~\ref{LEMproppt}, this implies that $\abs{\sP_t}= \Tprop\Q$. 
Furthermore, $\nreq =0$ (see~\eqref{eq:defnreq}), resulting in $\sI=[1\!:\!\R\L]$.
To establish the desired result, we first choose $\sv_{r,t}=\0v$ for $r\neq t$.
With this choice, the matrix $\Jm_{\phi_{[\xv]^{}_{\sP}}}\!(\sv,[\xv]^{}_{\sD})$ in~\eqref{eq:Jacobian1} looks as follows:
\begin{align}\label{eq:Jacobianbasecase}
\Jm_{\phi_{[\xv]^{}_{\sP}}}\!(\sv,[\xv]^{}_{\sD}) & =
\left[
\begin{pmatrix}
\Bm \hspace{-2.5mm}&\hspace{-2.5mm}
\left[
\begin{pmatrix}
\Am_{1,1} & \hspace*{-2.5mm}\cdots\hspace*{-2.5mm}  & \Am_{1,\Tprop}  \\[-.8mm]
 \vdots    &   \hspace*{-2.5mm}\hspace*{-2.5mm}     & \vdots     \\[-1mm]
\Am_{\Tprop,1} & \hspace*{-2.5mm}\cdots\hspace*{-2.5mm} & \Am_{\Tprop,\Tprop}
\end{pmatrix}
\right]^{\sD}
\end{pmatrix}
\right]^{}_{\sI} \notag \\
& 
%\hspace{-5mm}
%%%%pagebreak!!!!%%%%%%%
=
\begin{pmatrix}
\Bm_1 &&& \hspace{-2.5mm}[\Am_{1,1}]^{\sD_1}   \\[-1.5mm]
&\hspace{-2.5mm}\ddots \hspace{-2.5mm} &&   &  \hspace{-3.5mm} \ddots \hspace{-2.5mm}   \\[-1.5mm]
&& \hspace{-1.5mm} \Bm_{\Tprop} & &  &  \hspace{-1.5mm} [\Am_{\Tprop,\Tprop}]^{\sD_{\Tprop}}
\end{pmatrix} \notag \\[2mm]
& \rule{18mm}{0mm}\in \IC^{(\Tprop^2\Q+\abs{\sD}) \times (\Tprop^2\Q+\abs{\sD})}
\end{align}
where we used the sets $\{\sD_t\}_{t\in[1:\Tprop]}$ given in~\eqref{eq:dtaspt}, and where (cf.~\eqref{eq:ybardisc2})
\[
\Bm_r=(\Zm_{r,1} \cdots \Zm_{r,\Tprop}), \quad r\in[1\!:\!\Tprop] 
\]
and (cf.~\eqref{EQArt})
\be
\Am_{t,t} \ist=\ist
\diag ( \av_{t,t}), \; t\in[1\!:\!\Tprop], 
\; \text{with } \av_{t,t} \triangleq\ist \Zm_{t,t}\sv_{t,t}\,.  \label{eq:Jacobianbasecase2}
\ee
We choose%
\footnote{Note that so far we used the index $t$ for the sets $\sP_t$. 
Now we consider the matrix $[\Bm_r]^{}_{\sP_t}$ for $t=r$. Thus, it is convenient to use only the index $r$.} 
$[\Zm_{r,t}]^{}_{\sP_r} \in \IC^{\Tprop\Q\times \Q}$ such that the square matrices
$
[\Bm_r]^{}_{\sP_r} = \big[
\big(
\Qm_{r,1}\, \cdots \, \Qm_{r,\Tprop}
\big)
\big]^{}_{\sP_r} \in \IC^{\Tprop\Q\times \Tprop\Q}
$
are non\-singular.
Furthermore, we have that $[\Am_{t,t}]^{\sD_t}_{\sP_t}=\0v$ (by~\eqref{eq:Jacobianbasecase2}, $\Am_{t,t}$ is a diagonal matrix, and because $\sP_t\cap \sD_t\stackrel{\eqref{eq:dtaspt}}=\emptyset$, the matrix $[\Am_{t,t}]^{\sD_t}_{\sP_t}$ contains only off-diagonal entries). 
We will use Lemma~\ref{LEMdet} with 
$\Mm = \Jm_{\phi_{[\xv]^{}_{\sP}}}\!(\sv,[\xv]^{}_{\sD})$ given by~\eqref{eq:Jacobianbasecase}, 
$n={\Tprop}^2\Q+\abs{\sD}$,
%% being the matrix in 
$\sE=\sP$ (i.e., the rows where $[\Am_{t,t}]^{\sD_t}$ is zero),
 and $\sF=[1:{\Tprop}^2\Q]$ (i.e., the columns of all $\Bm_r$, $r\in [1\!:\!\Tprop]$).
This choice yields
$
{[\Mm]}_{\sE}^{\sF} = \diag\big(
[
\Bm_1 ]^{}_{\sP_1}, \dots, 
[
\Bm_{\Tprop} ]^{}_{\sP_{\Tprop}}
\big)
$, which is nonsingular because it is a block-diagonal matrix where each block on the diagonal, $[\Bm_r ]^{}_{\sP_r}$, was chosen nonsingular. 
Furthermore, we have that ${[\Mm]}_{\sE}^{[1:n]\setminus \sF} = \diag\big([\Am_{1,1}]^{\sD_1}_{\sP_1}, \dots, [\Am_{\Tprop,\Tprop}]^{\sD_{\Tprop}}_{\sP_{\Tprop}} \big)=\0v$.
Thus, the requirements of Lemma~\ref{LEMdet} are met and, hence, $\det(\Mm) =\det\big(\Jm_{\phi_{[\xv]^{}_{\sP}}}\!(\sv,[\xv]^{}_{\sD})\big) \not= 0$ if and only if the determinant of the following matrix is nonzero:
\be
\label{eq:M_A}
[\Mm]^{[1:n]\setminus \sF}_{[1:n]\setminus \sE} =
\begin{pmatrix}
{[\Am_{1,1}]}_{\sD_1}^{\sD_1} \hspace*{-2mm} &\hspace*{-4mm}&\hspace*{-2mm} \\[-1mm]
\hspace*{-2mm} & \hspace*{-2mm}\ddots\hspace*{-2mm} \\[-1.5mm]
\hspace*{-2mm} & \hspace*{-4mm} &  {[\Am_{\Tprop,\Tprop}]}_{\sD_{\Tprop}}^{\sD_{\Tprop}} 
\end{pmatrix}\,.
\ee
Because of~\eqref{eq:Jacobianbasecase2}, we have $[\Am_{t,t}]^{\sD_{t}}_{\sD_{t}} =
[\diag (\av_{t,t})]^{\sD_{t}}_{\sD_{t}}$. Hence, the matrix in~\eqref{eq:M_A} is a diagonal matrix and can be chosen to have nonzero diagonal entries by choosing 
${[\Zm_{t,t}]}_{\sD_t}$ and $\sv_{t,t}$ such that $[\av_{t,t}]^{}_{i}=[\Zm_{t,t}]^{}_{\{i\}}\sv_{t,t}\neq 0$ for all $i\in \sD_t$ (again see~\eqref{eq:Jacobianbasecase2}). 
Thus, 
%the matrix in~\eqref{eq:M_A} is
%a diagonal matrix with nonzero entries and hence 
its determinant is nonzero and, in turn, $\det(\Mm)\neq 0$.

%\begin{figure*}[!t]
%% ensure that we have normalsize text
%\normalsize
%% Store the current equation number.
%%\newcounter{MYtempeqncnt}
%\setcounter{MYtempeqncnt}{\value{equation}}
%% Set the equation number to one less than the one
%% desired for the first equation here.
%% The value here will have to changed if equations
%% are added or removed prior to the place these
%% equations are referenced in the main text.
%\setcounter{equation}{71}
%\be\label{eq:Jacobianreduced}
%%\Jm_{\phi_{[\xv]^{}_{\sP}}}(\sv,\xv_{\D}) = 
%\Km\triangleq[\Mm]^{[1:n]\setminus\sF}_{[1:n]\setminus\sE}=
%\begin{pmatrix}
%\Bm_1 &  & &
%[\Am_{1,1}]^{\sD_1} & \dots & [\Am_{1,\Tprop}]^{\sD_{\Tprop}} \\
%& \ddots &  & \vdots && \vdots \\
%&& \Bm_{\R-1}  & [\Am_{\R-1,1}]^{\sD_1} & \dots & [\Am_{\R-1,\Tprop}]^{\sD_{\Tprop}} \\[1.5mm]
%&\0v & & \hspace*{-2mm}[\Am_{\R,1}]^{\sD_1}_{\sLt} & \hspace*{-2mm}\cdots\hspace*{-2mm} & [\Am_{\R,\Tprop}]^{\sD_{\Tprop}}_{\sLt}
%\end{pmatrix}
%\ee
%% Restore the current equation number.
%\setcounter{equation}{\value{MYtempeqncnt}}
%% IEEE uses as a separator
%\hrulefill
%% The spacer can be tweaked to stop underfull vboxes.
%\vspace*{4pt}
%\end{figure*}

\subsubsection*{Inductive step (transition from $\R-1$ to $\R$)}
Assuming that $\Zm_{r,t}$ and $\sv_{r,t}$ for $t \!\in\! [1\!:\!\Tprop]$, $r \!\in\! [1\!:\!\R-1]$ have already been chosen such that the determinant of $\Jm_{\phi_{[\xv]^{}_{\sP}}}\!(\sv,[\xv]^{}_{\sD})$ is nonzero in the $\R-1$ setting,
we want to show that there exist $\Zm_{\R,t}$ and $\sv_{\R,t}$, $t \!\in\! [1\!:\!\Tprop]$ for which the determinant of the matrix $\Jm_{\phi_{[\xv]^{}_{\sP}}}\!(\sv,[\xv]^{}_{\sD})$ 
in~\eqref{eq:Jacobian1} is nonzero. 
To facilitate the exposition, we rewrite the matrices involved in a more convenient form. 
For the case of $\R$ receive antennas, denoted by the superscript $[\R]$, we rewrite the Jacobian matrix $\Jm_{\phi_{[\xv]^{}_{\sP}}}\!(\sv,[\xv]^{}_{\sD})$ in~\eqref{eq:Jacobian1} as
\ba\label{eq:JacobianR}
& \Jm_{\phi_{[\xv]^{}_{\sP}}}\!(\sv,[\xv]^{}_{\sD})^{[\R]} \rmv=\rmv
%\left[
%\begin{pmatrix}
\big(
[\Bm]^{}_{\sI} \;
[\Am]^{\sD}_{\sI}
\big)
%\end{pmatrix}
%\right] 
%\notag \\
%& \rule{30mm}{0mm}
\in \IC^{(\R\L-\nreq) \times (\R\Tprop\Q + \abs{\sD})}
\ea
with
\ba
[\Bm]^{}_{\sI} & = 
\begin{pmatrix}
\Bm_1  \\[-.8mm]
&  \hspace{-2.5mm} \ddots \hspace{-2.5mm} \\[-1mm]
&& \hspace{-1.5mm} \Bm_{\R-1} 
\\
& & & \hspace{-1.5mm}[\Bm_{\R}]^{}_{[1:\L-\nreq]} 
\end{pmatrix}
\notag
\ea
and
\ba
[\Am]^{\sD}_{\sI} & = 
\left[
\begin{pmatrix}
\Am_{1,1} & \hspace*{-2.5mm}\cdots\hspace*{-2.5mm}  & \Am_{1,\Tprop}  \\[-.8mm]
 \vdots    &   \hspace*{-2.5mm}\hspace*{-2.5mm}     & \vdots     \\[-1mm]
\Am_{\R,1} & \hspace*{-2.5mm}\cdots\hspace*{-2.5mm} & \Am_{\R,\Tprop}
\end{pmatrix}
\right]^{\sD}_{\sI} \notag \\
& = 
\begin{pmatrix}
[\Am_{1,1}]^{\sD_1} & \hspace{-2.5mm} \cdots \hspace{-2.5mm} & [\Am_{1,\Tprop}]^{\sD_{\Tprop}} \\[-.8mm]
 \vdots && \vdots \\[-1mm]
 [\Am_{\R-1,1}]^{\sD_1} & \hspace{-2.5mm} \cdots \hspace{-2.5mm} & [\Am_{\R-1,\Tprop}]^{\sD_{\Tprop}}
\\[2mm]
[\Am_{\R,1}]^{\sD_1}_{[1:\L-\nreq]} & \hspace{-2.5mm} \cdots \hspace{-2.5mm} & [\Am_{\R,\Tprop}]^{\sD_{\Tprop}}_{[1:\L-\nreq]}
\end{pmatrix}
\notag
\ea
where we used~\eqref{eq:data} and~\eqref{eq:defi}.
For the $\R-1$ case, the Jacobian matrix is given by
\begin{align}\label{eq:JacobianRminus1}
& \Jm_{\phi_{[\xv]^{}_{\sP}}}\!(\sv,[\xv]^{}_{\sD})^{[\R-1]} \notag \\
& \rule{10mm}{0mm} = 
\begin{pmatrix}
\Bm_1 &  &  &
[\Am_{1,1}]^{\sDt_1} & \hspace{-2.5mm} \cdots \hspace{-2.5mm} & \hspace{-2.5mm} [\Am_{1,\Tprop}]^{\sDt_{\Tprop}} \\[-1mm]
& \hspace{-2.5mm} \ddots \hspace{-2.5mm} &  &  \vdots && \hspace{-2.5mm} \vdots \\[-1mm]
&& \hspace{-1.5mm} \Bm_{\R-1} \hspace{-2.5mm} &  [\Am_{\R-1,1}]^{\sDt_1} \hspace{-2.5mm} & \hspace{-2.5mm} \cdots \hspace{-2.5mm} & \hspace{-2.5mm} [\Am_{\R-1,\Tprop}]^{\sDt_{\Tprop}}
\end{pmatrix}\notag \\
& \rule{25mm}{0mm} \in \IC^{(\R-1)\L \times \big((\R-1)\Tprop\Q+\sum_{t\in [1:\Tprop]}\abs{\sDt_t}\big)}
\end{align}
where 
\be\label{eq:defdtt}
\sDt_t\triangleq [1\!:\!\L]\setminus \sPt_t
\ee
 with the sets $\sPt_t$ introduced in Lemma~\ref{LEMproppt}.
Note that in~\eqref{eq:JacobianRminus1}, we do not need to truncate the matrix when selecting the rows in the set $\sI$ as required by~\eqref{eq:Jacobian1}. 
This follows because $\nreq=0$ for $\R-1\leq \Tprop(\L-1)/(\L-\Tprop\Q)$ (which holds because $\R\leq \lceil\Tprop(\L-1)/(\L-\Tprop\Q)\rceil$) and, hence, $\sI = [1\!:\!(\R-1)\L]$. 

Let $\sG$, $\sG_t$, and $\sL_t$ be defined as in Lemma~\ref{LEMproppt}. 
Set $[\Zm_{\R,t}]^{}_{\sG\setminus\sG_t}=\0v$ for all $t\in [1\!:\!\Tprop]$, 
and choose $[\Zm_{\R,t}]^{}_{\sG_t}\in \IC^{\Q\times \Q}$ nonsingular for all $t\in [1\!:\!\Tprop]$.
With these choices, and recalling that we set $\xv=(1  \cdots  1)^{\trans}$ in the induction hypothesis (whence $\Xm_t=\Iv_{\L}$),
%and our earlier choice $\xv=(1  \cdots  1)^{\trans}$ (whence $\Xm_t=\Iv_{\L}$), 
it follows from~\eqref{eq:ybardisc2}
%% Thus, we obtain 
that $[\Bm_{\R}]^{}_{\sG}=\big([\Zm_{\R,1}]^{}_{\sG}\,\cdots\, [\Zm_{\R,\Tprop}]^{}_{\sG}\big)$
is nonsingular. 
Next, for each $t\in [1\!:\!\Tprop]$, select an index $g_t$ in the set $\sG_t \rmv\cap\rmv \sP_t$ (note that this set is non-empty due to Property~\eqref{en:partitiong3} in Lemma~\ref{LEMproppt}). 
Furthermore, choose $\sv_{\R,t}$ to be 
orthogonal to the rows of $[\Zm_{\R,t}]^{}_{\sG_t\setminus \{g_t\}}\in \IC^{(\Q-1)\times \Q}$ and to satisfy $[\Zm_{\R,t}]^{}_{\{g_t\}}\sv_{\R,t}\neq 0$
(note that since $\sv_{r,t}\in \IC^{\Q}$, it is always possible to choose $\sv_{r,t}$ such that it is orthogonal to $\Q-1$ vectors  of a set of $\Q$ linearly independent vectors and not orthogonal to the last one). 
Recalling~\eqref{EQArt}, we have
\[
[\Am_{\R,t}]^{}_{\sG} = \big[\diag ( \av_{\R,t})\big]^{}_{\sG}\,, \quad
t\in[1\!:\!\Tprop]
\]
where $[\av_{\R,t}]^{}_{i} = [\Zm_{\R,t}]^{}_{\{i\}}\sv_{\R,t}=0$ for $i\in \sG\setminus \sG_t$ by our choice $[\Zm_{\R,t}]^{}_{\sG\setminus\sG_t}=\0v$, and for $i\in \sG_t\setminus \{g_t\}$ because we chose $\sv_{\R,t}$ to be orthogonal to the rows of $[\Zm_{\R,t}]^{}_{\sG_t\setminus \{g_t\}}$. 
Thus, $[\Am_{\R,t}]^{}_{\sG}$ has only one nonzero entry $[\av_{\R,t}]^{}_{g_t}$, which is in the $g_t$th column. 
But since $g_t\in \sP_t$ and $\sP_t\cap \sD_t=\emptyset$, taking only the columns indexed by $\sD_t$ results in $[\Am_{\R,t}]^{\sD_t}_{\sG}=\0v$.
We will use Lemma~\ref{LEMdet} with $\Mm=\Jm_{\phi_{[\xv]^{}_{\sP}}}\!(\sv,[\xv]^{[\R]}_{\sD})$ given in~\eqref{eq:JacobianR}, 
$n=\R\Tprop\Q+\abs{\sD}$,
$\sE=\{i+(\R-1)\L: i\in \sG\}$ (i.e., the rows of $[\Bm_{\R}]^{}_{[1:\L-\nreq]}$ specified by $\sG$),
and $\sF=[(\R-1)\Tprop\Q+1 : \R\Tprop\Q]$ (i.e.,  the columns of $[\Bm_{\R}]^{}_{[1:\L-\nreq]}$).
This choice yields
\[
[\Mm]^{\sF}_{\sE}=[\Bm_{\R}]^{}_{\sG}=\big([\Zm_{\R,1}]^{}_{\sG} \,\cdots \, [\Zm_{\R,\Tprop}]^{}_{\sG}\big)
\]
 which is nonsingular as noted above. 
Furthermore, we have
\[
[\Mm]^{[1:n]\setminus\sF}_{\sE}= \Big(\0v \;\; [\Am_{\R,1}]^{\sD_1}_{\sG} \, \cdots \, [\Am_{\R,\Tprop}]^{\sD_{\Tprop}}_{\sG}\Big)=\0v\,.
\]
Hence, the requirements of Lemma~\ref{LEMdet} are satisfied. We obtain that the determinant of $\Mm=\Jm_{\phi_{[\xv]^{}_{\sP}}}\!(\sv,[\xv]^{}_{\sD})^{[\R]}$ in~\eqref{eq:JacobianR} 
is nonzero if and only if the determinant of the following matrix 
%$\Km$ in \eqref{eq:Jacobianreduced} \addtocounter{equation}{1}at the top of this page 
is nonzero:
\ba\label{eq:Jacobianreduced}
%\Jm_{\phi_{[\xv]^{}_{\sP}}}(\sv,\xv_{\D}) = 
\Km & \triangleq[\Mm]^{[1:n]\setminus\sF}_{[1:n]\setminus\sE} \notag \\
& =
\begin{pmatrix}
\Bm_1 &  & &
[\Am_{1,1}]^{\sD_1} & \dots & [\Am_{1,\Tprop}]^{\sD_{\Tprop}} \\
& \hspace{-2.5mm} \ddots &  & \vdots && \vdots \\
&& \hspace{-2.5mm} \Bm_{\R-1}  & [\Am_{\R-1,1}]^{\sD_1} & \dots & [\Am_{\R-1,\Tprop}]^{\sD_{\Tprop}} \\[1.5mm]
&\0v & & \hspace*{-2mm}[\Am_{\R,1}]^{\sD_1}_{\sLt} & \hspace*{-2mm}\cdots\hspace*{-2mm} & [\Am_{\R,\Tprop}]^{\sD_{\Tprop}}_{\sLt}
\end{pmatrix}.
\ea
Here, we used Property~\eqref{en:partitiong4} in Lemma~\ref{LEMproppt}, i.e., that $[1:\L-\nreq]\setminus \sG =\sLt$.

So far, we specified only the rows $[\Zm_{\R,t}]^{}_{\sG}$.
Because $\sG\cap \sLt=\emptyset$ by Property~\eqref{en:partitiong4} in Lemma~\ref{LEMproppt}, we can still freely choose the remaining rows $[\Zm_{\R,t}]^{}_{\sLt}$. 
We first choose the rows indexed by $\sL_t$ such that $[\Zm_{\R,t}]^{}_{\sL_t}\sv_{\R,t}$ does not have zero entries (e.g., $[\Zm_{\R,t}]^{}_{\{i\}}=\sv_{\R,t}^{\operatorname{H}}$ for $i\in \sL_t$, resulting in $[\Zm_{\R,t}]^{}_{\{i\}}\sv_{\R,t}=\norm{\sv_{\R,t}}^2\neq 0$).
Next, we choose the remaining rows, indexed by $\sLt \setminus \sL_t$, to be zero, i.e., $[\Zm_{\R,t}]^{}_{\sLt \setminus \sL_t}=\0v$.
With these choices and using~\eqref{EQArt}, we obtain 
%% achieve that
$[\Am_{\R,t}]^{\sD_t}_{\sLt\setminus \sL_t} =\0v$
and  $\det\rmv\big({[\Am_{\R,t}]}_{\sL_t}^{\sL_t}\big) \rmv\neq\rmv 0$. 

We will next use another application of Lemma~\ref{LEMdet}
with $\Mm=\Km$ given in~\eqref{eq:Jacobianreduced},
$n=(\R-1)\Tprop\Q+\abs{\sD}$,
\[
\sE=[(\R-1)\L +1 : (\R-1)\Tprop\Q+\abs{\sD}]
\]
(i.e., all rows of $\Km$ below $\Bm_{\R-1}$), and
\[
\sF= \bigcup_{t\in [1:\Tprop]} \bigg\{i+(\R-1)\Tprop\Q + \sum_{t'\in [1:t-1]}\abs{\sD_{t'}}: i\in \sL_t \bigg\}
\]
 (i.e., the columns of $[\Am_{\R,t}]^{\sD_t}_{\sL_{t}}$ for all $t\in [1\!:\!\Tprop]$).
This choice results in 
\[
[\Mm]^{\sF}_{\sE}=\diag\big([\Am_{\R,1}]^{\sL_1}_{\sL_1}, \dots, [\Am_{\R,\Tprop}]^{\sL_{\Tprop}}_{\sL_{\Tprop}}\big)
\]
 which is nonsingular because $\det\rmv\big({[\Am_{\R,t}]}_{\sL_t}^{\sL_t}\big) \rmv\neq\rmv 0$.
Furthermore, we have
\bas
 [\Mm]^{[1:n]\setminus\sF}_{\sE} & =\Big(\0v \;\, [\Am_{\R,1}]^{\sD_1}_{\widetilde{\sL}\,\setminus \, \sL_1}\, \cdots \, [\Am_{\R,\Tprop}]^{\sD_{\Tprop}}_{\widetilde{\sL}\, \setminus \, \sL_{\Tprop}}
\Big)  =\0v\,.
\eas
%where $\widetilde{\sL}\triangleq \bigcup_{t'\in [1:\Tprop]}\sL_{t'}$.
Thus, the requirements of Lemma~\ref{LEMdet} are satisfied, and we obtain that the determinant of
$\Km$ in~\eqref{eq:Jacobianreduced} is nonzero if and only if the determinant of the following matrix is nonzero:
\ba\label{eq:JacobianRminus12}
%\Jm_{\phi_{[\xv]^{}_{\sP}}}(\sv,\xv_{\D}) = 
& [\Mm]^{[1:n]\setminus\sF}_{[1:n]\setminus\sE} \notag \\
& = \begin{pmatrix}
\Bm_1 &  & &
[\Am_{1,1}]^{\sD_1\setminus\sL_1} & \hspace{-2.5mm} \dots \hspace{-2.5mm} & \hspace{-2mm} [\Am_{1,\Tprop}]^{\sD_{\Tprop}\setminus\sL_{\Tprop}} \\
& \hspace{-2.5mm} \ddots \hspace{-2.5mm}&  & \vdots && \vdots \\[1mm]
&& \hspace{-2.5mm} \Bm_{\R-1}  & [\Am_{\R-1,1}]^{\sD_1\setminus\sL_1} & \hspace{-2.5mm} \dots \hspace{-2.5mm} & \hspace{-2mm} [\Am_{\R-1,\Tprop}]^{\sD_{\Tprop}\setminus\sL_{\Tprop}} 
\end{pmatrix}. %\notag \\[-5mm]
\ea
By the definitions $\sL_t=\sPt_t\setminus \sP_t$, $\sD_t=[1\!:\!\L]\setminus \sP_t$, and $\sDt_t=[1\!:\!\L]\setminus \sPt_t$ (see~\eqref{eq:deflt},~\eqref{eq:dtaspt}, and~\eqref{eq:defdtt}), we obtain 
\[
\sD_t\setminus \sL_t=([1\!:\!\L]\setminus \sP_t)\setminus (\sPt_t\setminus \sP_t)\stackrel{(a)}=[1\!:\!\L]\setminus \sPt_t=\sDt_t
\]
for all $t\in [1\!:\!\Tprop]$, where $(a)$ holds because $\sP_t\subseteq \sPt_t$. 
Thus, $[\Mm]^{[1:n]\setminus\sF}_{[1:n]\setminus\sE}$ in \eqref{eq:JacobianRminus12} is equal to
$\Jm_{\phi_{[\xv]^{}_{\sP}}}\!(\sv,[\xv]^{}_{\sD})^{[\R-1]}$ in \eqref{eq:JacobianRminus1}.
Altogether, we obtain that the determinant of $\Jm_{\phi_{[\xv]^{}_{\sP}}}\!(\sv,[\xv]^{}_{\sD})^{[\R]}$ in \eqref{eq:JacobianR} is nonzero if and only if
the determinant of $[\Mm]^{[1:n]\setminus\sF}_{[1:n]\setminus\sE}= \Jm_{\phi_{[\xv]^{}_{\sP}}}\!(\sv,[\xv]^{[\R-1]}_{\sD})$ in \eqref{eq:JacobianRminus1} is nonzero. 
But the determinant of $\Jm_{\phi_{[\xv]^{}_{\sP}}}\!(\sv,[\xv]^{}_{\sD})^{[\R-1]}$ is nonzero by the induction hypothesis.

%%\vspace{-1mm}

%%%%%%%%%%%%%%%%%%%%%%%%%%%%%%%
\section{Proof of Lemma~\ref{LEMchangeh}}\label{app:proofpartition}
%%%%%%%%%%%%%%%%%%%%%%%%%%%%%%%

%%\vspace{1mm}

%% \begin{IEEEproof}[\hspace{-1em}Proof]
\subsection{Proof of Part (I)}
To prove part (I) of Lemma~\ref{LEMchangeh}, i.e., that almost all of $\sM$ can be covered by the union of disjoint measurable subsets $\sU_k$, we will use the following lemma, which is an application of the result reported  in \cite[Cor.~3.2.4]{fed69}.%\vspace{1.5mm}~
\begin{lemma}\label{LEMrestrictinj}
Let $\sA\subseteq \IC^n$  be a Lebesgue measurable set and $\kappa\colon \IC^n \!\rightarrow\rmv \IC^n$ a continuously differentiable mapping (e.g., the mapping in Lemma~\ref{LEMchangeh}).
Then there exists a Lebesgue measurable set $\sB \!\subseteq\! \sA \cap \{\uv\in \IC^n : \abs{\Jm_{\kappa}(\uv)}\neq 0\}$ such that 
$\kappa\big|_\sB$ is one-to-one and $\kappa(\sA)\setminus \kappa(\sB) \!=\! \sN$, where $\sN$ is a set of Lebesgue measure  %\vspace{1.5mm} 
zero. 
%Furthermore, $\abs{\psi^{-1}(\vv)\cap (\sA\setminus \sB)}\leq m' \!-\! 1 < \infty$ for all %\vspace{1.5mm}$\vv\in \IC^n$.
\end{lemma}

We will use Lemma~\ref{LEMrestrictinj} repeatedly to construct the disjoint sets $\{\sU_j\}_{j\in [1:m]}$.%\vspace{1.5mm}~
\begin{lemma}\label{LEMreducemult}
Let $\kappa$ and $\sM$ be as in Lemma~\ref{LEMchangeh}, i.e.,
$\kappa\colon \IC^n \!\rightarrow \IC^n$ is a continuously differentiable mapping with Jacobian matrix $\Jm_{\kappa}$ such that $\Jm_{\kappa}(\uv)$ is nonsingular a.e.\ and
$\sM\triangleq \{\uv \rmv\in\rmv \IC^n \!: \absdet{\Jm_{\kappa}(\uv)}\neq 0\}$. 
Again as in Lemma~\ref{LEMchangeh}, assume that for all $\vv\in \IC^n$, the cardinality of the set $\kappa^{-1}(\vv)\cap \sM$ satisfies
$\abs{\kappa^{-1}(\vv)\cap \sM}\leq m < \infty$, for some 
%% constant 
$m\in \IN$ (i.e., $\kappa\big|_{\sM}$ is finite-to-one).
Then, for $k\in [1\!:\!m]$, there exist disjoint Lebesgue measurable sets $\{\sU_j\}_{j\in [1:k]}$ with $\sU_j\subseteq \sM$ such that $\kappa\big|_{\sU_j}$ is one-to-one for $j\in [1\!:\!k]$. Furthermore, there exists a set $\sN_k$ of Lebesgue measure zero such that 
\ba
& \bigabs{\kappa^{-1}(\vv)\cap \bigg(\sM\setminus \bigcup_{j\in [1:k]}\sU_j\bigg)} \leq m - k\,, \notag \\
& \rule{20mm}{0mm} \text{for all } \vv\in \kappa\bigg(\sM\setminus \bigcup_{j\in [1:k-1]}\sU_j\bigg)\setminus \sN_k \,.%\vspace{1.5mm}
\label{eq:reducemulthyp}
\ea
\end{lemma}
%Using Lemma~\ref{LEMrestrictinj} we construct the sets $\sU_j$, $j\in [1\!:\!m]$ as folows:
\begin{IEEEproof}[\hspace{-1em}Proof]
We prove Lemma~\ref{LEMreducemult} by induction over $k$.

\subsubsection*{Base case (proof for $k=1$)}  By Lemma~\ref{LEMrestrictinj} with $\sA=\sM$, we obtain a set $\sB\subseteq \sM$ (recall that $\sM = \{\uv \rmv\in \IC^n \!: \absdet{\Jm_{\kappa}(\uv)}\neq 0\}$ and thus $\sM\cap \{\uv \rmv\in\rmv \IC^n \!: \absdet{\Jm_{\kappa}(\uv)}\neq 0\}= \sM$) such that  
%$\abs{\kappa(\sM\setminus \sB)}\leq m-1$ and 
$\kappa\big|_{\sB}$ is one-to-one.
Furthermore, $\kappa(\sM)\setminus \kappa(\sB)=\sN_1$ for a set $\sN_1$ of Lebesgue measure zero. 
Because $\kappa(\sB)\subseteq\kappa(\sM)$, this implies $\kappa(\sM)\setminus\sN_1=\kappa(\sB)$.
Thus, for each $\vv\in \kappa(\sM)\setminus\sN_1$, there exists $\uv\in \sB$ such that $\kappa(\uv) = \vv$. Equivalently, $\kappa^{-1}(\vv)\cap\sB \neq \emptyset$. 
Hence, for $\vv\in \kappa(\sM)\setminus\sN_1$,
\begin{align}
\abs{\kappa^{-1}(\vv)\cap (\sM\setminus \sB)} & = \abs{(\kappa^{-1}(\vv)\cap\sM)\setminus (\kappa^{-1}(\vv)\cap\sB)} \notag \\
& \stackrel{\hidewidth (a) \hidewidth}= \abs{\kappa^{-1}(\vv)\cap\sM} - \abs{\kappa^{-1}(\vv)\cap\sB}  \notag \\
& \stackrel{\hidewidth (b) \hidewidth}= \abs{\kappa^{-1}(\vv)\cap \sM} - 1  \notag  \\
& \stackrel{\hidewidth (c) \hidewidth}\leq m \!-\! 1
\label{eq:reducemult}
\end{align}
 where  $(a)$ holds because $\kappa^{-1}(\vv)\cap\sB \subseteq \kappa^{-1}(\vv)\cap\sM$, $(b)$ holds because $\kappa^{-1}(\vv)\cap\sB$ is nonempty and contains at most one element since $\kappa\big|_{\sB}$ is one-to-one, and $(c)$ holds because we assumed that $\abs{\kappa^{-1}(\vv)\cap \sM} \leq m$.
We set $\sU_1\triangleq \sB$ and, by~\eqref{eq:reducemult}, the property~\eqref{eq:reducemulthyp} is satisfied for $k=1$.
Furthermore, $\kappa\big|_{\sU_1}=\kappa\big|_{\sB}$ is one-to-one, which concludes the proof for the base case.
%For $m=1$ this concludes the proof.
%We start by constructing the set $\sU_1\triangleq \sB$ by applying  Lemma~\ref{LEMrestrictinj} with $\psi=\kappa, \sA=\sM$, and $m'=m$. 

\subsubsection*{Inductive step (transition from $k$ to $k+1$)} 
Suppose we already constructed the $k$ disjoint measureable sets $\{\sU_j\}_{j\in [1:k]}$ and the set $\sN_k$ satisfying~\eqref{eq:reducemulthyp}.
To simplify notation, define
\[
\sU^{[k]}\triangleq \bigcup_{j\in [1:k]}\sU_j\,.
\]
Note that~\eqref{eq:reducemulthyp} can now be written as 
\ba
&\big\lvert\kappa^{-1}(\vv)\cap \big(\sM\setminus \sU^{[k]}\big)\big\rvert \leq m - k\,, \notag \\
&\rule{25mm}{0mm} \text{for all }\vv\in \kappa\big(\sM\setminus \sU^{[k-1]}\big)\setminus \sN_k\,.
\label{eq:reducemulthyp2}
\ea
%For $k$ disjoint measureable sets $\{\sU_j\}_{j\in [1:k]}$,  
By Lemma~\ref{LEMrestrictinj} with  $\sA=\sM\setminus\sU^{[k]}$, 
we obtain 
a set $\sB$ such that $\kappa\big|_{\sB}$ is one-to-one and
\be\label{eq:bsubsetuk}
\sB\subseteq \sM\setminus\sU^{[k]}\,.
\ee
%$\Big|\kappa\Big(\big(\sM\setminus \bigcup_{j\in [1:k]}\sU_j\big) \setminus \sB\Big)\Big|\leq m-k-1$ and 
Furthermore,
$\kappa\big(\sM\setminus \sU^{[k]}\big) \setminus \kappa(\sB) =\widetilde{\sN}_{k+1}$ for a set $\widetilde{\sN}_{k+1}$ of Lebesgue measure zero. 
Because $\kappa(\sB)\subseteq \kappa\big(\sM\setminus \sU^{[k]}\big)$,  this implies $\kappa\big(\sM\setminus \sU^{[k]}\big)\setminus\widetilde{\sN}_{k+1}=\kappa(\sB)$.
Hence, for $\vv\in \kappa\big(\sM\setminus  \sU^{[k]}\big)\setminus\widetilde{\sN}_{k+1}$, there exists $\uv\in \sB$ such that $\kappa(\uv) = \vv$, or equivalently, $\kappa^{-1}(\vv)\cap\sB \neq \emptyset$.
Thus, similarly to~\eqref{eq:reducemult}, we obtain for $\vv\in \kappa\big(\sM\setminus \sU^{[k]}\big)\setminus\big(\widetilde{\sN}_{k+1} \cup \sN_{k}\big)$ 
%(here $\sN_k$ denotes the subset of $\kappa\big(\sM\setminus \sU^{[k]}\big)$ of measure zero where~\eqref{eq:reducemulthyp} does not hold)
\begin{align}
& \big\lvert\kappa^{-1}(\vv)\cap \big(\big(\sM\setminus \sU^{[k]}\big)\setminus \sB\big)\big\rvert \notag \\
& \rule{15mm}{0mm}= \big\lvert\big(\kappa^{-1}(\vv)\cap\big(\sM\setminus \sU^{[k]}\big)\big)\setminus (\kappa^{-1}(\vv)\cap\sB)\big\rvert \notag \\
& \rule{15mm}{0mm} \stackrel{\hidewidth (a) \hidewidth}= \big\lvert\kappa^{-1}(\vv)\cap\big(\sM\setminus \sU^{[k]}\big)\big\rvert - \abs{\kappa^{-1}(\vv)\cap\sB}  \notag \\
& \rule{15mm}{0mm} \stackrel{\hidewidth (b) \hidewidth}= \big\lvert\kappa^{-1}(\vv)\cap \big(\sM\setminus \sU^{[k]}\big)\big\rvert - 1  \notag  \\
& \rule{15mm}{0mm} \stackrel{\hidewidth (c) \hidewidth}\leq m - k - 1
\label{eq:reducemult2}
\end{align}
 where $(a)$ holds because $\kappa^{-1}(\vv)\cap\sB \subseteq \kappa^{-1}(\vv)\cap\big(\sM\setminus \sU^{[k]}\big)$, $(b)$ holds because $\kappa^{-1}(\vv)\cap\sB$ is nonempty and contains at most one element since $\kappa\big|_{\sB}$ is one-to-one, and $(c)$ holds because of our induction hypothesis~\eqref{eq:reducemulthyp2}.
Setting $\sU_{k+1}\triangleq \sB$, the left-hand side in~\eqref{eq:reducemult2} is equal to $\big\lvert\kappa^{-1}(\vv)\cap \big(\big(\sM\setminus \sU^{[k]}\big)\setminus \sU_{k+1}\big)\big\rvert =\big\lvert\kappa^{-1}(\vv)\cap \big(\sM\setminus \sU^{[k+1]}\big)\big\rvert$,
so that~\eqref{eq:reducemult2} becomes $\big\lvert\kappa^{-1}(\vv)\cap \big(\sM\setminus \sU^{[k+1]}\big)\big\rvert\leq m - k - 1$ for all $\vv\in \kappa\big(\sM\setminus \sU^{[k]}\big)\setminus\big(\widetilde{\sN}_{k+1} \cup \sN_{k}\big)$.
This is exactly the property~\eqref{eq:reducemulthyp2} with $k$ replaced by $k+1$ and $\sN_k$ replaced by $\sN_{k+1}\triangleq\widetilde{\sN}_{k+1} \cup \sN_{k}$.
Furthermore, we have by~\eqref{eq:bsubsetuk} that $\sU_{k+1} =\sB \subseteq \sM\setminus \sU^{[k]}$ and thus 
$\sU_{k+1}\cap \sU_j=\emptyset$ for $j\in [1\!:\!k]$. 
Finally, $\kappa\big|_{\sU_{k+1}}=\kappa\big|_{\sB}$ is one-to-one, which concludes the proof. %\vspace{1.5mm} 
\end{IEEEproof}

%The next step is to show that~\eqref{eq:reducemultfin} holds for all $\vv\in \kappa(\sM) \setminus \big(\bigcup_{j\in [1:m]}\sN_{j}\big)$.
%Thus, it remains to show~\eqref{eq:reducemultfin} for 
%$\vv\in \Big(\kappa(\sM) \setminus \big(\bigcup_{j\in [1:m]}\sN_{j}\big)\Big)
%\setminus 
%\Big(\kappa\big(\sM\setminus \bigcup_{j\in [1:m-1]}\sU_j\big) \setminus \big(\bigcup_{j\in [1:m]}\sN_{j}\big)\Big)$
%For these $\vv$ we have that $\vv\in \kappa(\sM)\setminus \kappa\big(\sM\setminus \bigcup_{j\in [1:m-1]}\sU_j\big)$ and hence $\vv\notin \kappa\big(\sM\setminus \bigcup_{j\in [1:m-1]}\sU_j\big)$.
%This is equivalent to $\kappa^{-1}(\vv)\cap\big(\sM\setminus \bigcup_{j\in [1:m-1]}\sU_j\big)=\emptyset$ from which~\eqref{eq:reducemultfin} follows.
The sets $\{\sU_j\}_{j\in [1:m]}$ constructed in Lemma~\ref{LEMreducemult} are disjoint and $\kappa\big|_{\sU_{j}}$ is one-to-one for all $j\in [1\!:\!m]$.
It remains to be shown that $\sU^{[m]}=\bigcup_{j\in [1:m]}\sU_j$ covers almost all of $\sM$.
To this end, we first show that $\kappa\big(\sM\setminus  \sU^{[m]}\big)\setminus \sN_m$ is empty.
Assume by contradiction that $\vv\in \kappa\big(\sM\setminus  \sU^{[m]}\big)\setminus \sN_m$. 
By~\eqref{eq:reducemulthyp} with $k=m$, we have that for all $\vv\in \kappa\big(\sM\setminus  \sU^{[m-1]}\big)\setminus \sN_m$
\[%\label{eq:reducemultfin}
\big\lvert\kappa^{-1}(\vv)\cap\big(\sM\setminus \sU^{[m]}\big)\big\rvert\leq m-m = 0
\]
i.e., there exists no $\uv\in \sM\setminus \sU^{[m]}$ such that $\kappa(\uv)=\vv$. 
This is a contradiction to the assumption $\vv\in \kappa\big(\sM\setminus \sU^{[m]}\big)\setminus \sN_m$, and thus we conclude that there is no $\vv\in \kappa\big(\sM\setminus  \sU^{[m]}\big)\setminus \sN_m$, i.e.,
$\kappa\big(\sM\setminus \sU^{[m]}\big)\setminus \sN_{m}=\emptyset$.
Hence,  we have
\be\label{eq:onlynull}
\kappa\big(\sM\setminus \sU^{[m]}\big)\subseteq \sN_{m}\,.
\ee
%By construction the sets $\{\sU_j\}_{j\in [1:m]}$ are disjoint and $\kappa\big|_{\sU_j}$ is one-to-one. 
%It remains to show that they cover almost all of $\sM$.
We next use the integral transformation reported in~\cite[Th.~3.2.3]{fed69} to obtain
\bas
\int_{\sM \setminus \sU^{[m]}} \abs{\Jm_{\kappa}(\uv)}^2 \, d\uv & \leq 
m \int_{\kappa(\sM \setminus \sU^{[m]})} \, d\vv \\
& \stackrel{\eqref{eq:onlynull}} \leq m \int_{\sN_{m}} \, d\vv \\
& = 0\,.
\eas
Because the function $\abs{\Jm_{\kappa}(\uv)}$ is positive on $\sM$, it follows that the Lebesgue measure of the set $\sM \setminus \sU^{[m]}$ has to be zero, i.e., $\sU^{[m]}$ covers almost all of $\sM$.
%and  $\bigcup_{j\in [1:m]}\sU_j = \sM$.
This concludes the proof of part (I). 

\subsection{Proof of Part (II)}
To establish part (II), i.e., the bound~\eqref{eq:bounddiffe}, we first note that
%\vspace{-.5mm}
\be\label{eq:rvvdelta}
h(\rvv) \ist\geq\ist h(\rvv \ist |\ist \randk) \ist=\rmv \sum_{k\in [1:m]} \!\rmv h(\rvv\ist |\ist \randk \!=\! k) \, p_k
%\vspace{-.5mm}
\ee
where $\randk$ is the discrete random variable that takes on the value $k$ when $\ruv\in \sU_k$, and $p_k \triangleq\ist \operatorname{Pr}\{\ruv\in \sU_k\} = \int_{\sU_k} \rmv f_{\ruv}(\uv)\ist d\uv$.
We assume without loss of generality\footnote{%
If $p_k \!=\! 0$ for some $k$, we simply omit
%% eliminate 
the corresponding term 
%% from the sum 
in~\eqref{eq:rvvdelta}.} that $p_k \!\neq\! 0$, $k \!\in\! [1\!:\!m]$. 
Since $\kappa\big|_{\sU_k}\!$ is one-to-one, 
%% the differential entropy 
 we can use the transformation rule 
for one-to-one mappings  \cite[Lemma 3]{moridu13} to relate $h(\rvv\ist|\ist\randk \!=\! k)$ to $h(\ruv\ist|\ist\randk \!=\! k)$:
\ba\label{eq:rvvdeltak}
& h(\rvv\ist|\ist\randk \!=\! k) = h(\ruv\ist|\ist\randk \!=\! k) +\rmv \int_{\IC^n} \!f_{\ruv|\randk=k}(\uv)\log (\absdet{\Jm_{\kappa}(\uv)}^2) \, d\uv \ist.
\ea
The conditional probability density function of $\ruv$ given $\randk \!=\! k$ is $f_{\ruv|\randk=k}(\uv)=\mathbbmss{1}_{\sU_k}\rmv(\uv) \ist f_{\ruv}(\uv)/p_k$. 
Thus,
$h(\ruv\ist|\ist\randk \!=\! k) = - \int_{\ist \sU_k} \!\rmv \big(f_{\ruv}(\uv)/p_k\big) \log\rmv\big(f_{\ruv}(\uv)/p_k \big) \ist d\uv$, and~\eqref{eq:rvvdeltak} becomes
\begin{align*}
 h(\rvv\ist|\ist\randk \!=\! k) %\\
%& \rule{9mm}{0mm}
& \ist=\ist \frac{1}{p_k} \bigg[ \rmv-\! \int_{\sU_k} \!f_{\ruv}(\uv)\log\rmv\bigg( \rmv\frac{f_{\ruv}(\uv)}{p_k}\rmv \bigg) \ist d\uv \\
& \rule{20mm}{0mm} +\int_{\sU_k} \!f_{\ruv}(\uv)\log (\absdet{\Jm_{\kappa}(\uv)}^2) \,d\uv \bigg] \notag\\
& 
%\rule{9mm}{0mm} 
\ist=\ist \frac{1}{p_k} \bigg[ \rmv-\! \int_{\sU_k} \!f_{\ruv}(\uv)\log\rmv\big(f_{\ruv}(\uv) \big) \ist d\uv \\
& \rule{6mm}{0mm} +\int_{\sU_k} \!f_{\ruv}(\uv)\log (\absdet{\Jm_{\kappa}(\uv)}^2) \,d\uv  + p_k \log (p_k)  \bigg] \,.
%% \label{eq:rvvdeltak2}
\end{align*}
Inserting this expression into~\eqref{eq:rvvdelta}, and recalling that the sets $\sU_k$ are disjoint, that $\sU^{[m]}=\bigcup_{k\in[1:m]}\sU_k$ covers almost all of $\sM$,
and that $\IC^{n}\setminus \sM$ has Lebesgue measure zero,
%% and~\eqref{eq:rvvdeltak2} 
%and using the trivial inequality $p_k\leq 1$,
we 
%\vspace{-.5mm}
obtain
\begin{align*}
&  h(\rvv) \geq
\sum_{k\in[1:m]} \bigg[ \rmv-\! \int_{\sU_k} \!f_{\ruv}(\uv)\log\rmv\big(f_{\ruv}(\uv) \big) \ist d\uv \notag\\
& \rule{22mm}{0mm} +\int_{\sU_k} \!f_{\ruv}(\uv)\log (\absdet{\Jm_{\kappa}(\uv)}^2) \,d\uv   + p_k \log (p_k) \bigg] \\
%h(\rvv\ist|\ist\randk) \notag\\
& = -\! \int_{\sU^{[m]}}\! f_{\ruv}(\uv) \log\rmv\big(f_{\ruv}(\uv)\big) \ist d\uv 
  \notag\\%[1mm]
& \rule{10mm}{0mm} + \int_{\sU^{[m]}}\! f_{\ruv}(\uv)\log (\absdet{\Jm_{\kappa}(\uv)}^2) \ist d\uv 
%\\ & \rule{40mm}{0mm} 
 + \underbrace{\sum_{k\in [1:m]} \!p_k \log (p_k)}_{-H(\randk)} \notag \\[-3.5mm]
& = -\! \int_{\IC^n}\! f_{\ruv}(\uv) \log\rmv\big(f_{\ruv}(\uv)\big) \ist d\uv 
  \notag\\
& \rule{25mm}{0mm} + \int_{\IC^n}\! f_{\ruv}(\uv)\log (\absdet{\Jm_{\kappa}(\uv)}^2) \ist d\uv \ist-\ist H(\randk) \notag \\[.5mm]
& \ist=\ist h(\ruv) \ist+\rmv \int_{\IC^n}\! f_{\ruv}(\uv)\log (\absdet{\Jm_{\kappa}(\uv)}^2) \ist d\uv \ist-\ist H(\randk)\,.
\end{align*}

%%%%%%%%%%%%%%%%%%%%%%%%%%%%%%%
\section{Proof of Lemma~\ref{LEMboundanalytic}} \label{app:proofboundanalytic}
%%%%%%%%%%%%%%%%%%%%%%%%%%%%%%%

%\vspace{1mm}

Since $f$ is not identically zero, there exists a $\xiv_0 \!\in\rmv \IC^{n}\rmv$ such that $f(\xiv_0)\neq 0$. 
%% Suppose that $f(\xiv_0)\neq 0$ for some $\xiv_0\in \IC^{N}$.
The function $g(\xiv)\triangleq f(\xiv+\xiv_0)$ is an analytic function that satisfies $g(\0v)\neq 0$.
By performing the change of variables $\xiv \mapsto \xiv +\xiv_0$, we can rewrite $I_1$ in~\eqref{eq:expec} in the following more convenient form:
\[
%% \label{eq:bound1a}
I_1 \ist=\int_{\IC^n} \! \exp(-\norm{\xiv+\xiv_0}^2)\log(\abs{g(\xiv)})\,d\xiv\,.
%\vspace{-1mm}
\]
We have
\begin{align}
\norm{\xiv+\xiv_0}^2 & \ist \leq \,
 (\norm{\xiv}+\norm{\xiv_0})^2 \notag\\
& \ist=\, \norm{\xiv}^2+2\norm{\xiv}\norm{\xiv_0}+\norm{\xiv_0}^2 \notag\\
& \ist\leq\, \norm{\xiv}^2+2\max \{\norm{\xiv}^2, \norm{\xiv_0}^2\}+\norm{\xiv_0}^2 \notag \\
& \ist\leq\, 3\norm{\xiv}^2+ 3\norm{\xiv_0}^2\,. \label{eq:boundxi}
\end{align}
Using~\eqref{eq:boundxi}, we lower-bound $I_1$ as follows:
\be\label{eq:bound1}
I_1 \ist\geq\,
c \rmv \int_{\IC^n} \!\exp(-3\norm{\xiv}^2)\log(\abs{g(\xiv)})\,d\xiv \,\triangleq\, I_2
%\vspace{-.5mm}
\ee
where $c\triangleq \exp(-3\norm{\xiv_0}^2)$.
%By \cite[Example 2.6.1.3]{Azarin09} the function $\log(\abs{g(\xiv)})$ is subharmonic 
%as a function of all $2N$ real and imaginary parts of $\xiv$. 
We next define
%% With 
the mapping $\varphi \colon \IR^{2n} \!\!\rightarrow\rmv \IC^n$;
$\xv \mapsto \big([\xv]^{}_{[1:n]}+i[\xv]^{}_{[n+1:2n]}\big)$,
%% \[
%% \varphi\colon \begin{cases}
%% \IR^{2N} &\hspace*{-3mm}\rightarrow \IC^N \\
%% \xv &\hspace*{-3mm} \mapsto \big(\xv_{[1:N]}+i\xv_{[N+1:2N]}\big)\,,
%% \end{cases}
%% \]
and rewrite $I_2$ in~\eqref{eq:bound1} as
\be\label{eq:subharm}
I_2 \,=\, c \rmv \int_{\IR^{2n}} \!\exp(-3\norm{\xv}^2)\,u(\xv)\,d\xv
\ee
with $u(\xv)\triangleq \log(\abs{g(\varphi(\xv))})$. Since $g(\0v) \!\neq\! 0$, we have that $u(\0v)>-\infty$.
By \cite[Example 2.6.1.3]{Azarin09}, $u(\xv)$ is a \emph{subharmonic function}. 
We shall use the following  property of subharmonic 
functions, which is a special case of the more general result reported 
%\vspace{1.5mm} 
in
\cite[Th.~2.6.2.1]{Azarin09}.

\begin{lemma} \label{lem:subharmonic}
Let $u$ be a subharmonic function on $\sW\subseteq \IR^{2n}\rmv$.
%, and let $\xv \!\in\! \IR^{2n}\rmv$.
If $\{\xv \rmv\in\rmv \IR^{2n} \!: \norm{\xv}\leq r\} \!\subseteq\! \sW$ for some $r \!>\! 0$,
then 
\[
u(\0v) \,\leq\, \frac{1}{\sigma_{2n}\, r^{2n-1}} \rmv \int_{\sS_{r}}\!\rmv u(\xv) \,ds(\xv)
\]
where $\sS_{r}\triangleq \{\xv \rmv\in\rmv \IR^{2n} \!: \norm{\xv} \rmv=\rmv r\}$, the constant $\sigma_{2n}$ denotes the area of the unit sphere in $\IR^{2n}\rmv$, 
and $ds$ denotes integration with respect to the $(2n \!-\! 1)$-dimensional Hausdorff measure (cf.\ \cite[Sec.~2.10.2]{fed69}).
%\vspace{1.5mm}
\end{lemma}

Using a well-known measure-theoretic result (see, e.g., \cite[Th.~3.2.12]{fed69}), we have for $u(\xv)=\log(\abs{g(\varphi(\xv))})$
\ba\label{eq:help26}
& \int_{\IR^{2n}} \!\exp(-3\norm{\xv}^2) \,u(\xv)\,d\xv \notag \\
& \rule{15mm}{0mm} =
\int_{0}^{\infty} \bigg[ \int_{\sS_{r}} \! u(\xv)\,ds(\xv)\bigg] \exp(-3 r^2)\,dr \,.
\ea
Inserting~\eqref{eq:help26} in~\eqref{eq:subharm}, we 
obtain
\begin{align*}
I_2 
& = c \rmv \int_{0}^{\infty} \bigg[ \int_{\sS_{r}} u(\xv)\,ds(\xv) \bigg] \exp(-3 r^2) \,dr \\[.3mm]
& \stackrel{\hidewidth (a) \hidewidth}\geq c\, \sigma_{2n}\, u(\0v)\int_{0}^{\infty} \exp(-3 r^2) \, r^{2n-1} \ist dr \\[-.3mm]
& \stackrel{\hidewidth (b) \hidewidth}> -\infty\,.
\end{align*}
Here, $(a)$  is due to Lemma~\ref{lem:subharmonic} and $(b)$ holds because $u(\0v) \!>\! -\infty$ and $0<\int_{0}^{\infty}  \exp(-3 r^2) \, r^{2n-1} \ist dr < \infty$. Using~\eqref{eq:bound1}, we conclude that %\vspace{1mm} 
$I_1>-\infty$.

%%%%%%%%%%%%%%%%%%%%%%%%%%%%%%%
\section{Proof of Lemma~\ref{LEMproppt}} \label{app:proofproppt}
%%%%%%%%%%%%%%%%%%%%%%%%%%%%%%%

\subsection{Bijectivity of $\betav$}

In order to prove Lemma~\ref{LEMproppt}, we will use the following property of the function $\betav$ in~\eqref{eq:beta}.
%\vspace{1.5mm}~
\begin{lemma}\label{LEMbetabij}
The function $\betav$ defined in~\eqref{eq:beta} is  
%\vspace{1.5mm} 
bijective.
\end{lemma}
\begin{IEEEproof}[\hspace{-1em}Proof]
To facilitate the exposition, we introduce the notation
\[
\shortlcm\triangleq\lcm(\Tprop,\L)\,.
\]
Recall that $\betav(j)=\big(\beta_1(j)\; \beta_2(j)\big)^{\trans}$ with $\beta_1(j)=\big(j+\lfloor (j - 1)/\shortlcm\rfloor \big) \mymod \Tprop\in [1\!:\!\Tprop]$ and $\beta_2(j)=j \mymod \L \in [1\!:\!\L]$, for $j\in[1\!:\!\Tprop\L]$.
We start by proving that $\betav$ is one-to-one. Assume that there exist $j_1,j_2 \in [1\!:\!\Tprop\L]$ with $j_1\leq j_2$ such that $\betav(j_1)=\betav(j_2)$. 
From $\beta_2(j_1)=\beta_2(j_2)$, it follows that $j_1\mymod \L = j_2 \mymod \L$ and, hence,\footnote{%
Recall that we defined $a \mymod b \triangleq a-b\lfloor(a-1)/b \rfloor$ to be the residuum of $a$ divided by $b$ in $[1\!:\!b]$ (and not in $[0\!:\!b-1]$ as commonly done).} 
$j_2=j_1+n\L$ for some $n\in [0\!:\!\Tprop-1]$. 
Similarly, $\beta_1(j_1)=\beta_1(j_2)$ implies that
\[
j_1+\bigg\lfloor \frac{j_1 \!-\! 1}{\shortlcm}\bigg\rfloor = j_2+\bigg\lfloor \frac{j_2 \!-\! 1}{\shortlcm}\bigg\rfloor-m\Tprop
\] 
for some $m\in \IN$,
and thus
\[
j_1+\bigg\lfloor \frac{j_1 \!-\! 1}{\shortlcm}\bigg\rfloor = j_1+n\L+\bigg\lfloor \frac{j_1+n\L \!-\! 1}{\shortlcm}\bigg\rfloor -m\Tprop
\]
or, equivalently,
\be\label{eq:necj1}
m\Tprop- n\L = \bigg\lfloor \frac{j_1+n\L \!-\! 1}{\shortlcm}\bigg\rfloor - \bigg\lfloor \frac{j_1 \!-\! 1}{\shortlcm}\bigg\rfloor\,.
\ee
We can write  $j_1=k\shortlcm+\tilde{\jmath}_1$ with some $k\in \IN$ and $\tilde{\jmath}_1\in [1\!:\!\shortlcm]$ and simplify~\eqref{eq:necj1} as follows:
\begin{align}
m\Tprop -  n\L 
& = \bigg\lfloor \frac{k\shortlcm+\tilde{\jmath}_1+n\L \!-\! 1}{\shortlcm}\bigg\rfloor - \bigg\lfloor \frac{k\shortlcm+\tilde{\jmath}_1 \!-\! 1}{\shortlcm}\bigg\rfloor  \notag \\
& = k + \bigg\lfloor \frac{\tilde{\jmath}_1+n\L \!-\! 1}{\shortlcm}\bigg\rfloor - k - \bigg\lfloor \frac{\tilde{\jmath}_1 \!-\! 1}{\shortlcm}\bigg\rfloor  \notag \\
& \stackrel{\hidewidth (a) \hidewidth}= \bigg\lfloor \frac{\tilde{\jmath}_1 +n\L \!-\! 1}{\shortlcm}\bigg\rfloor\,.
\label{eq:gcdteilt}
\end{align}
Here, $(a)$ holds because $\tilde{\jmath}_1 \!-\! 1 < \shortlcm$ and thus $\lfloor (\tilde{\jmath}_1 \!-\! 1)/\shortlcm\rfloor=0$.
We will next show that the right-hand side of~\eqref{eq:gcdteilt} is zero,
by establishing the following chain of inequalities:
%\be
%0 
%\leq \bigg\lfloor \frac{\tilde{\jmath}_1 +n\L \!-\! 1}{\shortlcm}\bigg\rfloor 
%\stackrel{(a)}\leq \bigg\lfloor \frac{j_1 +n\L \!-\! 1}{\shortlcm}\bigg\rfloor 
%\stackrel{(b)}\leq\bigg\lfloor \frac{\Tprop\L \!-\! 1}{\shortlcm}\bigg\rfloor
%\stackrel{(c)}= \bigg\lfloor \gcd(\Tprop,\L)- \frac{1}{\shortlcm}\bigg\rfloor
%= \gcd(\Tprop,\L) - 1\,. \label{eq:lessgcd}
%\ee
\begin{align}
0 & \leq \bigg\lfloor \frac{\tilde{\jmath}_1 +n\L \!-\! 1}{\shortlcm}\bigg\rfloor \notag \\
& \stackrel{\hidewidth (a) \hidewidth}\leq \bigg\lfloor \frac{j_1 +n\L \!-\! 1}{\shortlcm}\bigg\rfloor \notag \\
& \stackrel{\hidewidth (b) \hidewidth}\leq\bigg\lfloor \frac{\Tprop\L \!-\! 1}{\shortlcm}\bigg\rfloor \notag \\
& \stackrel{\hidewidth (c) \hidewidth}= \bigg\lfloor \gcd(\Tprop,\L)- \frac{1}{\shortlcm}\bigg\rfloor\notag \\
& = \gcd(\Tprop,\L) - 1\,. \label{eq:lessgcd}
\end{align}
Here, $(a)$ holds because $\tilde{\jmath}_1\leq j_1$, $(b)$ holds because $j_1+n\L =j_2\leq\Tprop\L,$ and $(c)$ holds because $\Tprop\L=\gcd(\Tprop,\L)\shortlcm$ \cite[Th.~52]{hawr1975} (here, $\gcd(\cdot,\cdot)$ denotes the greatest common divisor).
Note now that $\gcd(\Tprop,\L)$ divides the left-hand side of~\eqref{eq:gcdteilt} and, hence, also the right-hand side. 
But by~\eqref{eq:lessgcd}, the right-hand side of~\eqref{eq:gcdteilt} is an element of $[0\!:\!\gcd(\Tprop,\L)-1]$. % where only $0$ is divisible by $\gcd(\Tprop,\L)$. 
Hence, it must be zero, and thus~\eqref{eq:gcdteilt} becomes
\be\label{eq:mteqnn}
m\Tprop -  n\L= \bigg\lfloor \frac{\tilde{\jmath}_1 +n\L \!-\! 1}{\shortlcm}\bigg\rfloor = 0\,.
\ee
Therefore, $\tilde{\jmath}_1 +n\L \!-\! 1 < \shortlcm$. 
Since $n\L\leq \tilde{\jmath}_1 +n\L \!-\! 1$, we obtain $n\L<\shortlcm$. 
Furthermore, by~\eqref{eq:mteqnn}, we have that $m\Tprop=n\L$.
Thus, $n\L$ is a common multiple of $\Tprop$ and $\L$ that is less than the least (positive) common multiple.
Therefore, $n=0$ and, hence, $j_1=j_1+n\L = j_2$.
We have thus shown that $\betav(j_1)=\betav(j_2)$ implies $j_1=j_2$, which means that $\betav$ is one-to-one. 
 Since  the domain of $\betav$, $[1\!:\!\Tprop\L]$, and its codomain, $[1\!:\!\Tprop] \times [1\!:\!\L]$, are finite and of the same cardinality (namely, $\Tprop\L$), we conclude that $\betav$ is also bijective. %\vspace{1.5mm} 
\end{IEEEproof}

We will now prove the individual properties stated in Lemma~\ref{LEMproppt}.

\subsection{Proof of  Property~(\ref{item:sizept})}\label{sec:proofprop1}
We first show that $\beta_2\big|_{\beta_1^{-1}(t)}$ is one-to-one, i.e., if  $\beta_2(j_1)=\beta_2(j_2)$ for $j_1, j_2\in \beta_1^{-1}(t)$ then $j_1=j_2$. 
To this end, let $j_1, j_2\in \beta_1^{-1}(t)$ (i.e., $\beta_1(j_1)=\beta_1(j_2)=t$) and assume that $\beta_2(j_1)=\beta_2(j_2)=i$. 
Then $\betav(j_1)=\betav(j_2)=(t\;i)^{\trans}$. 
Since $\betav$ is one-to-one by Lemma~\ref{LEMbetabij}, we conclude that $j_1=j_2$. 
Hence, $\beta_2\big|_{\beta_1^{-1}(t)}$ is one-to-one. 
Furthermore, since $\beta_1^{-1}(t)\cap [1\!:\!\vartheta_{\R}] \subseteq \beta_1^{-1}(t)$, we have (cf.~\eqref{eq:defpt}) 
\be \label{eq:abspt}
\abs{\sP_t} = \big|\beta_2\big(\beta_1^{-1}(t)\cap [1\!:\!\vartheta_{\R}]\big)\big|
= \big|\beta_1^{-1}(t)\cap [1\!:\!\vartheta_{\R}]\big|
\ee
for $t\in [1\!:\!\Tprop]$. 
To conclude the proof, we will use the following basic 
%\vspace{1.5mm} 
lemma.
\begin{lemma}\label{partbeta1}
The sets $\{\beta_1^{-1}(t)\}_{t\in [1:\Tprop]}$ form a partition of the domain $[1\!:\!\Tprop\L]$ of $\beta_1$, i.e.,
\be\label{eq:partdisjoint}
\beta_1^{-1}(t)\cap \beta_1^{-1}(t')=\emptyset, \quad \text{for } t,t'\in [1\!:\!\Tprop] \text{ with } t\neq t'
\ee
and
\be\label{eq:partunion}
\bigcup_{t\in[1:\Tprop]}\beta_1^{-1}(t)=[1\!:\!\Tprop\L]\,.%\vspace{1.5mm} 
\ee
%%\vspace{-2mm}
\end{lemma}
\begin{IEEEproof}[\hspace{-1em}Proof]
This lemma follows from the definition of a function, i.e., the fact that $\beta_1$ maps every element in the domain to exactly one element in the codomain.
%Assume that $j\in \beta_1^{-1}(t)\cap \beta_1^{-1}(t')$ for $t\neq t'$. 
%Then by the definition of the preimage, $\beta_1(j)=t$ and $\beta_1(j)=t'$. 
%We thus have $t=t'$, which contradicts the assumption $t\neq t'$ and, hence, we obtain~\eqref{eq:partdisjoint}. 
%To prove~\eqref{eq:partunion}, we first note that $\beta_1^{-1}([1\!:\!\Tprop])=[1\!:\!\Tprop\L]$ because $\beta_1\colon [1\!:\!\Tprop\L]\to [1\!:\!\Tprop]$.
%Furthermore, $\beta_1^{-1}(t)\subseteq\beta_1^{-1}([1\!:\!\Tprop])=[1\!:\!\Tprop\L]$, for $t\in [1\!:\!\Tprop]$. 
%Thus, we have
%\be\label{eq:partunionsub}
%\bigcup_{t\in[1:\Tprop]}\beta_1^{-1}(t)\subseteq[1\!:\!\Tprop\L]\,.
%\ee
%To prove the inclusion in the other direction, let $j\in[1\!:\!\Tprop\L]= \beta_1^{-1}([1\!:\!\Tprop])$. 
%This implies $\beta_1(j)\in [1\!:\!\Tprop]$ and, hence, $\beta_1(j)=t'$ for a specific $t'\in [1\!:\!\Tprop]$. Thus, $j\in\beta_1^{-1}(t') \subseteq \bigcup_{t\in[1:\Tprop]}\beta_1^{-1}(t)$. Since $j\in[1\!:\!\Tprop\L]$, we conclude that $[1\!:\!\Tprop\L]\subseteq \bigcup_{t\in[1:\Tprop]}\beta_1^{-1}(t)$. Combining with~\eqref{eq:partunionsub} yields~\eqref{eq:partunion}.
%\vspace{1.5mm} 
\end{IEEEproof}

By Lemma~\ref{partbeta1}, we obtain
\begin{align}
\sum_{t\in[1:\Tprop]} \abs{\sP_t} \,& \stackrel{\hidewidth \eqref{eq:abspt} \hidewidth}=
\sum_{t\in[1:\Tprop]} \big|\beta_1^{-1}(t)\cap [1\!:\!\vartheta_{\R}]\big|  \notag \\
& \stackrel{\hidewidth \eqref{eq:partdisjoint} \hidewidth}= \,\Bigg| \Bigg(\bigcup_{t\in[1:\Tprop]}\beta_1^{-1}(t)\Bigg)\cap [1\!:\!\vartheta_{\R}]\Bigg| \notag \\
& \stackrel{\hidewidth \eqref{eq:partunion} \hidewidth}= \,\big\lvert[1\!:\!\Tprop\L]\cap [1\!:\!\vartheta_{\R}]\big\rvert \notag \\
& = \,\min\{\Tprop\L, \vartheta_{\R}\}\,. \label{eq:absbeta1}
\end{align}
Since $\L>\Tprop\Q$, we have that $\vartheta_{\R} = \max\{\Tprop, \R\Tprop\Q-(\R \rmv-\rmv \Tprop)\L\} = \max\{\Tprop, \Tprop\L - \R(\L-\Tprop\Q)\} < \Tprop\L$. 
Combining this with~\eqref{eq:absbeta1}, we conclude that
\begin{align*}
\sum_{t\in[1:\Tprop]}\abs{\sP_t} \,=\, \vartheta_{\R}\,.
\end{align*}

\subsection{Proof of  Property~(\ref{item:sizep})}
We will make use of the following  
%\vspace{1.5mm} 
lemma.
\begin{lemma}\label{LEMmultiples}
Let $p, q \in \IN$ with $p<q$. 
Then 
\[
\big|\big\{j\in [p+1\!:\!q]: (j+a) \mymod b =c \big\}\big|\leq \bigg\lceil\frac{q-p}{b}\bigg\rceil
\]
for all $a,b,c \in \IN$ with $b\geq 2$, $c\geq 1$, and $c\leq b$.
 %\vspace{1.5mm}
\end{lemma}
\begin{IEEEproof}[\hspace{-1em}Proof]
We prove Lemma~\ref{LEMmultiples} by contradiction. 
Assume 
\[
\big|\big\{j\in [p+1\!:\!q]: (j+a) \mymod b =c \big\}\big|> \bigg\lceil\frac{q-p}{b}\bigg\rceil\triangleq d\,.
\]
Thus, the set $\big\{j\in [p+1\!:\!q]: (j+a) \mymod b =c \big\}$ contains at least $d+1$ elements $\{j_i\}_{i\in [1:d+1]}$, i.e., there exist at least $d+1$ distinct elements $j_i\in [p+1\!:\!q]$ satisfying $(j_i+a) \mymod b = c$. 
Hence, there exist distinct $k_i\in \IN$, $i\in [1\!:\!d+1]$ such that 
\be\label{eq:jiki}
j_i + a = c + k_i b \in [p+1\!:\!q] \,.
\ee
Assume, without loss of generality, that $k_i<k_{i+1}$ for $i\in [1\!:\!d]$. 
Because $k_i\in \IN$, we obtain 
$k_i\leq k_{i+1}-1$ and thus, iteratively, $k_1\leq k_2-1\leq k_3-2\leq \cdots$, and finally 
\be\label{eq:distki}
k_1 \leq k_{d +1}-d\,.
\ee
Hence,
\bas
j_{d +1} - j_1 & \stackrel{\eqref{eq:jiki}}=
k_{d +1} b - k_1  b \\
& =
(k_{d +1} - k_1) b  \\
& \stackrel{ \eqref{eq:distki} }\geq
d\, b \\
& = \bigg\lceil\frac{q-p}{b}\bigg\rceil b \\
& \geq q-p
\eas
which contradicts $j_1, j_{d +1} \in [p+1\!:\!q]$.
%\vspace{1.5mm} 
\end{IEEEproof}
To prove Property~\eqref{item:sizep}, we first establish an upper bound on $\vartheta_{\R}$. 
We have that
\begin{align*}
\R\Tprop\Q -(\R-\Tprop)\L & =
(\R-\Tprop)\Tprop\Q- (\R-\Tprop)\L +{\Tprop}^2\Q \\
& = \underbrace{(\R-\Tprop)}_{\geq 0}\underbrace{(\Tprop\Q-\L)}_{<0} +\,{\Tprop}^2\Q  \\
& \leq {\Tprop}^2\Q
\end{align*}
and, hence,
\be\label{eq:boundvartheta}
\vartheta_{\R} = \max\{\Tprop, \R\Tprop\Q-(\R \rmv-\rmv \Tprop)\L\} \leq {\Tprop}^2\Q\,.
\ee
To bound the size of the sets $\sP_t$, we use~\eqref{eq:abspt} and the definition of $\beta_1$ to conclude that
\ba\label{eq:abspt2}
\hspace{-1mm}\abs{\sP_t} 
& = \big|\{j\in [1\!:\!\vartheta_{\R}]: \beta_1(j)=t\}\big| \notag \\
& = \bigabs{\bigg\{j\in [1\!:\!\vartheta_{\R}]: \!\bigg(j+\bigg\lfloor \frac{j - 1}{\shortlcm}\bigg\rfloor\bigg) \mymod \Tprop=t\bigg\}}\,.
\ea
Choose $m\in \IN$ such that $(m-1)\shortlcm<\vartheta_{\R}\leq m\shortlcm$. We can partition the set $[1\!:\!\vartheta_{\R}]$ as follows:
\ba\label{eq:partitionthetar}
[1\!:\!\vartheta_{\R}] & = \Bigg(\bigcup_{n\in [0:m-2]}\big[n\shortlcm+1:(n+1)\shortlcm\big]\Bigg) \notag \\
& \rule{30mm}{0mm} \cup \big[(m-1)\shortlcm+1:\vartheta_{\R}\big]\,.
\ea
Note that the intervals $\big[n\shortlcm+1:(n+1)\shortlcm\big]$, $n\in [0\!:\!m-2]$ and $\big[(m-1)\shortlcm+1:\vartheta_{\R}\big]$ in~\eqref{eq:partitionthetar} are disjoint and satisfy
\be\label{eq:casespt}
\bigg\lfloor \frac{j - 1}{\shortlcm}\bigg\rfloor = 
\begin{cases}
n, & \text{for } j\in   \big[n\shortlcm+1:(n+1)\shortlcm\big] \\
m-1, & \text{for } j\in   \big[(m-1)\shortlcm+1:\vartheta_{\R}\big]\,.
\end{cases}
\ee
Thus, using~\eqref{eq:partitionthetar} and~\eqref{eq:casespt} in~\eqref{eq:abspt2}, we obtain
\begin{align}
& \abs{\sP_t} = \sum_{n\in [0:m-2]}\big|\big\{j\in \big[n\shortlcm+1:(n+1)\shortlcm\big]: \notag \\[-4mm]
& \rule{45mm}{0mm}(j+n) \mymod \Tprop=t\big\}\big| \notag \\
& \rule{15mm}{0mm} + \big|\big\{j\in \big[(m-1)\shortlcm+1:\vartheta_{\R}\big]: \notag \\
& \rule{35mm}{0mm}(j+m-1) \mymod \Tprop=t\big\}\big|\,.  
\label{eq:sepabspt}
\end{align}
By Lemma~\ref{LEMmultiples}, we have
\ba\label{eq:boundint1}
& \big|\big\{j\in \big[n\shortlcm+1:(n+1)\shortlcm\big]: 
(j+n) \mymod \Tprop=t\big\}\big| \notag \\
& \rule{40mm}{0mm}\leq \bigg\lceil\frac{\shortlcm}{\Tprop}\bigg\rceil
=\frac{\shortlcm}{\Tprop}
\ea
and
\ba\label{eq:boundint2}
& \big|\big\{j\in \big[(m-1)\shortlcm+1:\vartheta_{\R}\big]\!:\! (j+m-1) \mymod \Tprop=t\big\}\big| \notag \\[2mm]
& \rule{40mm}{0mm} \leq \bigg\lceil \frac{\vartheta_{\R}-(m-1)\shortlcm}{\Tprop}\bigg\rceil\,.
\ea
Thus, inserting~\eqref{eq:boundint1} and~\eqref{eq:boundint2} into~\eqref{eq:sepabspt}, we obtain
\begin{align*}
\abs{\sP_t}
& \leq \, (m-1)\frac{\shortlcm}{\Tprop}+ \bigg\lceil \frac{\vartheta_{\R}-(m-1)\shortlcm}{\Tprop}\bigg\rceil \\[1mm]
& \stackrel{\hidewidth (a) \hidewidth}= \,(m-1)\frac{\shortlcm}{\Tprop}+ \bigg\lceil \frac{\vartheta_{\R}}{\Tprop}\bigg\rceil - (m-1)\frac{\shortlcm}{\Tprop}\\
& = \, \bigg\lceil \frac{\vartheta_{\R}}{\Tprop}\bigg\rceil \\[1mm]
& \stackrel{\hidewidth \eqref{eq:boundvartheta} \hidewidth}\leq \, \bigg\lceil \frac{{\Tprop}^2\Q}{\Tprop}\bigg\rceil \\[1mm]
& = \, \Tprop\Q
\end{align*}
where $(a)$ holds because $\shortlcm/\Tprop\in \IN$ (recall that $\shortlcm=\lcm(\Tprop, \L)$).

\subsection{Proof of Property~(\ref{item:ldisjoint})}
To prove Properties~\eqref{item:ldisjoint}--\eqref{item:partitiong}, we calculate the difference $\vartheta_{\R-1}-\vartheta_{\R}$. 
Because we assumed that $\R\leq \lceil\Tprop(\L-1)/(\L-\Tprop\Q)\rceil$, we have $\R-1< \Tprop(\L-1)/(\L-\Tprop\Q)$. 
This is easily verified to be equivalent to $(\R-1)\Tprop\Q-(\R-1-\Tprop)\L>\Tprop$.
Hence, using~\eqref{eq:sizep1}, 
\ba\label{eq:thetarminus1}
\vartheta_{\R-1} & =\max\{\Tprop, (\R-1)\Tprop\Q-(\R-1 \rmv-\rmv \Tprop)\L\} \notag \\
& =(\R-1)\Tprop\Q-(\R-1-\Tprop)\L\,.
\ea
Thus, we have
\begin{align}
& \vartheta_{\R-1}-\vartheta_{\R} \notag \\
& \rule{10mm}{0mm} = (\R-1)\Tprop\Q-(\R-1-\Tprop)\L \notag \\
& \rule{39mm}{0mm}- \max\{\Tprop, \R\Tprop\Q-(\R-\Tprop)\L\} \notag  \\
& \rule{10mm}{0mm}= \R\Tprop\Q-(\R-\Tprop)\L + \L -\Tprop\Q \notag \\
& \rule{39mm}{0mm} - \max\{\Tprop, \R\Tprop\Q-(\R-\Tprop)\L\} \notag  \\
& \rule{10mm}{0mm}= \L - \Tprop\Q  - \max\big\{\Tprop-\big(\R\Tprop\Q-(\R-\Tprop)\L\big), 0\big\}  \notag \\
& \rule{10mm}{0mm}= \L-\Tprop\Q-\nreq \label{eq:diffvartheta}
\end{align}
where $\nreq$ was defined in~\eqref{eq:defnreq}. 
Furthermore, by~\eqref{eq:boundnreq}, $\nreq<\L-\Tprop\Q$ and thus~\eqref{eq:diffvartheta} implies
\be\label{eq:varthetapos}
\vartheta_{\R-1}-\vartheta_{\R}>0\,.
\ee
%We shall now use~\eqref{eq:diffvartheta} to establish Properties~\eqref{item:ldisjoint}--\eqref{item:partitiong}.
We are now ready to prove Property~\eqref{item:ldisjoint}.
%\begin{IEEEproof}[\hspace*{-1em}Proof of Property~\eqref{item:ldisjoint}]
From the definitions $\sP_t\triangleq\beta_2\big(\beta_1^{-1}(t)\cap [1\!:\!\vartheta_{\R}]\big)$ in~\eqref{eq:defpt} and
 $\sPt_t\triangleq\beta_2\big(\beta_1^{-1}(t)\cap {[1\!:\!\vartheta_{\R-1}]}\big)$ in~\eqref{eq:defpttilde},
it follows that $\sL_t=\sPt_t\setminus \sP_t$ (recall~\eqref{eq:deflt}) can be written as
\begin{align}\label{eq:charlt}
\sL_t 
& =\beta_2\big(\beta_1^{-1}(t)\cap {[1\!:\!\vartheta_{\R-1}]}\big) \setminus \beta_2\big(\beta_1^{-1}(t)\cap [1\!:\!\vartheta_{\R}]\big) \notag \\
& \stackrel{\hidewidth (a) \hidewidth}=\beta_2\big((\beta_1^{-1}(t)\cap {[1\!:\!\vartheta_{\R-1}]})\setminus (\beta_1^{-1}(t)\cap [1\!:\!\vartheta_{\R}])\big) \notag \\
& = \beta_2\big(\beta_1^{-1}(t)\cap [\vartheta_{\R}+1\!:\!\vartheta_{\R-1}]\big)
\end{align}
where $(a)$ holds because $\beta_2\big|_{\beta_1^{-1}(t)}$ is one-to-one (see Section~\ref{sec:proofprop1}).
Since $\beta_2(j)=j\mymod \L$, the function $\beta_2$ is one-to-one on every set consisting of up to $\L$ consecutive integers. 
In particular,~\eqref{eq:varthetapos} and~\eqref{eq:diffvartheta} imply that
$\big|[\vartheta_{\R}+1\!:\!\vartheta_{\R-1}]\big| = \vartheta_{\R-1}-\vartheta_{\R}= \L-\Tprop\Q-\nreq$ and hence
$\beta_2\big|_{[\vartheta_{\R}+1:\vartheta_{\R-1}]}$ is one-to-one. 
Because  by Lemma~\ref{partbeta1} the sets $\beta_1^{-1}(t)$, $t\in [1\!:\!\Tprop]$ are pairwise disjoint, we conclude that the sets 
$\beta_1^{-1}(t)\cap [\vartheta_{\R}+1\!:\!\vartheta_{\R-1}]$, $t\in [1\!:\!\Tprop]$ are pairwise disjoint too. 
Hence, by~\eqref{eq:charlt} and because $\beta_2\big|_{[\vartheta_{\R}+1:\vartheta_{\R-1}]}$ is one-to-one, the sets $\sL_t$ are pairwise disjoint.

\subsection{Proof of Property~(\ref{item:lsubi})}
By~\eqref{eq:charlt}, we have
\be\label{eq:ltsubset}
\sL_t= \beta_2\big(\beta_1^{-1}(t)\cap [\vartheta_{\R}+1\!:\!\vartheta_{\R-1}]\big) \subseteq \beta_2([1\!:\!\vartheta_{\R-1}])\,.
\ee
Hence, it remains to prove that
\be\label{eq:beta2image}
\beta_2([1\!:\!\vartheta_{\R-1}])\subseteq [1\!:\!\L-\nreq]\,.
\ee
%from which the result follows almost immediately.
Recall that we assumed $\R \leq \lceil\Tprop(\L-1)/(\L-\Tprop\Q)\rceil$.
If $\R < \lceil\Tprop(\L-1)/(\L-\Tprop\Q)\rceil$, then $\R < \Tprop(\L-1)/(\L-\Tprop\Q)$ (because $\R\in \IN$), which implies $\R\L-(\R\Tprop\Q+\Tprop\L-\Tprop)< 0$;
hence, it follows from the definition of $\nreq$ in~\eqref{eq:defnreq} that $\nreq=0$.
In this case, it follows from the definition of $\beta_2$ in~\eqref{eq:beta}, i.e., $\beta_2(j)=j\mymod \L$ for $j\in [1\!:\!\Tprop\L]$, that~\eqref{eq:beta2image} is trivially true. 
For the complementary case $\R=\lceil\Tprop(\L-1)/(\L-\Tprop\Q)\rceil$,
%For $\R\geq k+2$ we obtain
%\[
%\R>\frac{\Tprop\L-\Tprop}{\L-\Tprop\Q}+1\,.
%\]
%Simple manipulations show that this is equivalent to 
%\[
%\Tprop>\Tprop\Q(\R-1)-(\R-1-\Tprop)\L
%\]
%which implies $\vartheta_{\R-1}=\Tprop$. This yields $\sL_t=\emptyset\subseteq \sI_{\R}$. It remains to prove~\eqref{item:lsubi} for $\R=k+1$. In this case we have $\sI_{\R}=[1:\Tprop\Q+\ell]$ where $\ell = \Tprop\L-\Tprop-(\L-\Tprop\Q)k$. 
we note that $\R\L-(\R\Tprop\Q+\Tprop\L-\Tprop)\geq 0$ and hence, using the definition of $\nreq$ in~\eqref{eq:defnreq},
\begin{align*}
\L-\nreq & = \L-(\R\L-\R\Tprop\Q-\Tprop\L+\Tprop) \\
& = \R\Tprop\Q-(\R-1-\Tprop)\L -\Tprop \\
& \geq (\R-1)\Tprop\Q-(\R-1-\Tprop)\L \\
& \stackrel{\hidewidth \eqref{eq:thetarminus1} \hidewidth}= \,\,\vartheta_{\R-1}\,.
\end{align*}
Thus, $[1\!:\!\vartheta_{\R-1}]\subseteq[1\!:\!\L-\nreq]$ and, further, $\beta_2([1\!:\!\vartheta_{\R-1}])\subseteq \beta_2([1\!:\!\L-\nreq])=[1\!:\!\L-\nreq]$, i.e.,~\eqref{eq:beta2image} is again true. 
Combining~\eqref{eq:ltsubset} and~\eqref{eq:beta2image} concludes the proof that $\sL_t\subseteq [1\!:\!\L-\nreq]$.

\subsection{Proof of Property~(\ref{item:partitiong})}
We have
\begin{align}\label{eq:charalllt}
\sLt \,
& \stackrel{\hidewidth \eqref{eq:defsLt} \hidewidth}= \bigcup_{t\in [1:\Tprop]}\sL_t \notag \\
& \stackrel{\hidewidth \eqref{eq:charlt} \hidewidth}= \bigcup_{t\in [1:\Tprop]} \beta_2\big(\beta_1^{-1}(t)\cap[\vartheta_{\R}+1\!:\!\vartheta_{\R-1}]\big) \notag \\
& \stackrel{\hidewidth (a) \hidewidth }= \, \beta_2\Bigg(\bigcup_{t\in [1:\Tprop]}\Big(\beta_1^{-1}(t)\cap[\vartheta_{\R}+1\!:\!\vartheta_{\R-1}]\Big)\Bigg) \notag \\
& = \, \beta_2\Bigg(\Bigg(\bigcup_{t\in [1:\Tprop]}\beta_1^{-1}(t)\Bigg)\cap[\vartheta_{\R}+1\!:\!\vartheta_{\R-1}]\Bigg) \notag \\
& \stackrel{\hidewidth \eqref{eq:partunion} \hidewidth}= \, \beta_2([\vartheta_{\R}+1\!:\!\vartheta_{\R-1}])
\end{align}
where $(a)$ holds because $\beta_2$ is one-to-one on every set consisting of up to $\L$ consecutive integers. % and $(b)$ holds because,   by~\eqref{eq:partunion}, we have $\bigcup_{t\in[1:\Tprop]}\beta_1^{-1}(t)=[1\!:\!\Tprop\L]$.
Thus, $\big\lvert\sLt\big\rvert = \big\lvert\beta_2([\vartheta_{\R}+1\!:\!\vartheta_{\R-1}])\big\rvert =
\vartheta_{\R-1} -\vartheta_{\R} \stackrel{\eqref{eq:diffvartheta}}=
\L-\Tprop\Q-\nreq$.
Furthermore, Property~\eqref{item:lsubi} implies that the set $\sLt$ is a subset of $[1\!:\!\L-\nreq]$, and hence  we obtain for the size of $\sG=[1\!:\!\L-\nreq]\setminus\sLt$ 
\bas
\abs{\sG} & =\big\lvert[1\!:\!\L-\nreq]\setminus\sLt \big\rvert \\
& =\L-\nreq- (\L-\Tprop\Q-\nreq) \\
& =\Tprop\Q\,.
\eas
Thus, we can partition $\sG$ as $\sG=\bigcup_{t\in[1:\Tprop]}\sG_t$, with disjoint $\sG_t$ of size $\Q$ each. 
We have thus shown the existence of sets $\sG_t$ satisfying~\eqref{en:partitiong1}, \eqref{en:partitiong2}, and~\eqref{en:partitiong4}.

It remains to show~\eqref{en:partitiong3}, i.e., that we can choose $\{\sG_t\}_{t\in[1:\Tprop]}$ such that each $\sG_t$ has a nonempty intersection with $\sP_t$. 
Because $\beta_2$  is one-to-one on sets of up to $\L$ consecutive integers and 
\bas
\vartheta_{\R-1}-(\vartheta_{\R}-\Tprop) 
& \,\stackrel{\hidewidth \eqref{eq:diffvartheta} \hidewidth }= \,\L-\Tprop\Q-\nreq +\Tprop \\
& \, =\L-\nreq -\Tprop(\Q-1) \\
& \,\leq \L-\nreq
\eas
we obtain that $\beta_2\big|_{[\vartheta_{\R}-\Tprop+1:\vartheta_{\R-1}]}$ is one-to-one.
Thus,
\ba\label{eq:beta2empty}
& \beta_2([\vartheta_{\R}-\Tprop+1\!:\!\vartheta_{\R}]) \cap \beta_2([\vartheta_{\R}+1\!:\!\vartheta_{\R-1}]) \notag  \\
& \rule{15mm}{0mm}=
\beta_2([\vartheta_{\R}-\Tprop+1\!:\!\vartheta_{\R}] \cap [\vartheta_{\R}+1\!:\!\vartheta_{\R-1}])\notag  \\
& \rule{15mm}{0mm}
= \beta_2(\emptyset)\notag  \\
& \rule{15mm}{0mm}
= \emptyset\,.
\ea
Inserting~\eqref{eq:charalllt} into~\eqref{eq:beta2empty}, we obtain
% we have that $\bigcup_{t\in [1:\Tprop]}\sL_t\subseteq \beta_2([\vartheta_{\R}+1\!:\!\vartheta_{\R-1}])$ and, hence,
\be\label{eq:beta2cuplt}
\beta_2([\vartheta_{\R}-\Tprop+1\!:\!\vartheta_{\R}])\cap\sLt
=\emptyset\,.
\ee
By the fact that $[\vartheta_{\R}-\Tprop+1\!:\!\vartheta_{\R}]\subseteq [1\!:\!\vartheta_{\R-1}]$ and~\eqref{eq:beta2image}, we have that $\beta_2([\vartheta_{\R}-\Tprop+1\!:\!\vartheta_{\R}])\subseteq
\beta_2([1\!:\!\vartheta_{\R-1}])\subseteq 
[1\!:\!\L-\nreq]$. 
Hence, \eqref{eq:beta2cuplt} implies that
\[%\label{eq:gtpt}
\beta_2([\vartheta_{\R}-\Tprop+1\!:\!\vartheta_{\R}])\subseteq [1\!:\!\L-\nreq]\setminus\sLt =\sG\,.
\]
Thus, we identified $\Tprop$ elements $\beta_2(\vartheta_{\R}-\Tprop+1), \beta_2(\vartheta_{\R}-\Tprop+2), \dots, \beta_2(\vartheta_{\R})$ in the set $\sG$, which will now be used to construct the sets $\sG_t$.
We will show that we can assign a different index $t\in [1\!:\!\Tprop]$ to each of these $\Tprop$ elements such that the element with index $t$ belongs to $\sP_t$, i.e.,
\ba\label{chargt}
& \beta_2([\vartheta_{\R}-\Tprop+1\!:\!\vartheta_{\R}])=\{g_1, \dots, g_{\Tprop}\}, \notag \\
& \rule{30mm}{0mm} \text{ with } g_t\in \sP_t, t\in [1\!:\!\Tprop]\,.
\ea
The desired sets $\sG_t$ are then obtained by assigning  $g_t$ to $\sG_t$, for $t\in [1\!:\!\Tprop]$.
Thus, recalling that $\abs{\sG_t}=\Q$,  $\sG_t$ consists of $g_t\in \sP_t$ and $\Q-1$ additional elements taken from the set 
$\sG\setminus\beta_2([\vartheta_{\R}-\Tprop+1\!:\!\vartheta_{\R}])$.

In order to prove~\eqref{chargt}, we distinguish two cases.
%\vspace{-2mm}
%\begin{itemize}
\subsection*{Case %$\nreq= 0$ and 
$n\shortlcm\notin [\vartheta_{\R}-\Tprop+1\!:\!\vartheta_{\R}-1]$ for All $n\in \IN$}
In this case, there exists $m\in \IN$ such that
$m\shortlcm\leq\vartheta_{\R}-\Tprop$ and $(m+1)\shortlcm\geq\vartheta_{\R}$. 
Thus, for all $j\in [\vartheta_{\R}-\Tprop+1\!:\!\vartheta_{\R}]$, we have
\be\label{eq:boundoffset1}
\bigg\lfloor \frac{j - 1}{\shortlcm}\bigg\rfloor \geq  
\bigg\lfloor \frac{\vartheta_{\R}-\Tprop}{\shortlcm}\bigg\rfloor \geq
\bigg\lfloor \frac{m\shortlcm}{\shortlcm}\bigg\rfloor = m
\ee
and
\be\label{eq:boundoffset2}
\bigg\lfloor \frac{j - 1}{\shortlcm}\bigg\rfloor <  
\bigg\lfloor \frac{\vartheta_{\R}}{\shortlcm}\bigg\rfloor \leq  
\bigg\lfloor \frac{(m+1)\shortlcm}{\shortlcm}\bigg\rfloor = m+1\,.
\ee
Combining~\eqref{eq:boundoffset1} and~\eqref{eq:boundoffset2}, we obtain that the offset in~\eqref{eq:beta} satisfies $\lfloor (j - 1)/\shortlcm\rfloor = m$ 
for all $j\in [\vartheta_{\R}-\Tprop+1\!:\!\vartheta_{\R}]$.
Thus, we have $\beta_1\big|_{[\vartheta_{\R}-\Tprop+1:\vartheta_{\R}]}(j)=(j+m)\mymod \Tprop$,
which implies that $\beta_1([\vartheta_{\R}-\Tprop+1\!:\!\vartheta_{\R}])=[1\!:\!\Tprop]$.
%$$ since $\beta_1$ takes $\Tprop$ consecutive numbers modulo $\Tprop$. 
Hence, we can write
\bas
& [\vartheta_{\R}-\Tprop+1\!:\!\vartheta_{\R}]=\{\tilde{\jmath}_1, \dots, \tilde{\jmath}_{\Tprop}\}, \notag \\
& \rule{30mm}{0mm}
\text{where } \tilde{\jmath}_t\in \beta_1^{-1}(t)
\text{ for } t\in [1\!:\!\Tprop]\,.
\eas
We then obtain
\[
\beta_2([\vartheta_{\R}-\Tprop+1\!:\!\vartheta_{\R}])=\{\beta_2(\tilde{\jmath}_1), \dots, \beta_2(\tilde{\jmath}_{\Tprop})\}
\]
and assign the indices $t\in [1\!:\!\Tprop]$ according to $g_t = \beta_2(\tilde{\jmath}_t)$.
By construction, we have both $g_t = \beta_2(\tilde{\jmath}_t)\in \beta_2(\beta_1^{-1}(t))$ and $g_t = \beta_2(\tilde{\jmath}_t)\in\beta_2([\vartheta_{\R}-\Tprop+1\!:\!\vartheta_{\R}])\subseteq \beta_2([1\!:\!\vartheta_{\R}])$, so that we also have
\[
g_t \in \beta_2(\beta_1^{-1}(t)\cap [1\!:\!\vartheta_{\R}]) = \sP_t
\]
(recall~\eqref{eq:defpt}).
Thus, our choice of the $g_t$ satisfies~\eqref{chargt}.
%Since $\tilde{\jmath}_t\in [\vartheta_{\R}-\Tprop+1\!:\!\vartheta_{\R}]\subseteq [1\!:\!\vartheta_{\R}]$ we choose
%$g_t=\beta_2(\tilde{\jmath}_t)\in \beta_2(\beta_1^{-1}(t)\cap [1\!:\!\vartheta_{\R}])=\sP_t$ and obtainc
%Hence, for each $t\in [1\!:\!\Tprop]$ there exists a different $j\in [\vartheta_{\R}-\Tprop+1\!:\!\vartheta_{\R}]$ with $\beta_2(j)\in \sP_t\cap\big([1\!:\!\L-\nreq]\setminus\bigcup_{t\in [1:\Tprop]}\sL_t\big)$.
%%
%%
%\vspace{-2mm}
\subsection*{Case %$\nreq= 0$ and 
$n\shortlcm\in [\vartheta_{\R}-\Tprop+1\!:\!\vartheta_{\R}-1]$ for Some $n\in \IN$} 
We first note that
\begin{align}
& \beta_2([\vartheta_{\R}-\Tprop+1\!:\!\vartheta_{\R}]) \notag \\
& \rule{5mm}{0mm} =\,\beta_2([\vartheta_{\R}-\Tprop+1\!:\!n\shortlcm]) \cup \beta_2([n\shortlcm+1\!:\vartheta_{\R}])  \notag \\
& \rule{5mm}{0mm} \stackrel{\hidewidth (a) \hidewidth}=\,\beta_2([\vartheta_{\R}-\Tprop+1\!:\!n\shortlcm]) 
%\notag \\ & \rule{23mm}{0mm} 
\cup \beta_2([n\shortlcm-\shortlcm+1\!:\vartheta_{\R}-\shortlcm]) \notag \\
&\rule{5mm}{0mm} =\,\beta_2([\vartheta_{\R}-\Tprop+1\!:\!n\shortlcm]) 
%\notag \\ & \rule{23mm}{0mm} 
\cup \beta_2([(n-1)\shortlcm+1\!:\vartheta_{\R}-\shortlcm])
\label{eq:imagebeta2}
\end{align}
where $(a)$ holds because 
(recall that $\shortlcm=\lcm(\Tprop, \L)$ is a multiple of $\L$)
\[
\beta_2(j) = j\mymod \L = (j-\shortlcm) \mymod \L = \beta_2(j-\shortlcm)
\]
 for $j> \shortlcm$.
 We will next calculate the offset $\lfloor (j - 1)/\shortlcm\rfloor$ in~\eqref{eq:beta} for $j$ belonging to either of the intervals in the arguments in~\eqref{eq:imagebeta2}, i.e., $j\in [\vartheta_{\R}-\Tprop+1\!:\!n\shortlcm]$ or $j\in [(n-1)\shortlcm+1\!:\vartheta_{\R}-\shortlcm]$.
Note that
\be\label{eq:nlcmin}
n\shortlcm\in [\vartheta_{\R}-\Tprop+1\!:\!\vartheta_{\R}-1]
\ee
and 
\be\label{eq:boundlcm}
\shortlcm \geq \Tprop \,.
\ee
Thus, we have
\be\label{eq:thetalcm}
(n-1)\shortlcm = n\shortlcm -\shortlcm \stackrel{\eqref{eq:nlcmin}}< \vartheta_{\R}-\shortlcm \stackrel{\eqref{eq:boundlcm}}\leq \vartheta_{\R}-\Tprop 
\ee
and
\be\label{eq:thetalcm2}
\vartheta_{\R} \stackrel{\eqref{eq:nlcmin}}< n\shortlcm +\Tprop \stackrel{\eqref{eq:boundlcm}}\leq (n+1)\shortlcm\,.
\ee
For $j\in [\vartheta_{\R}-\Tprop+1\!:\!n\shortlcm]$, we obtain that $j-1\geq \vartheta_{\R}-\Tprop \stackrel{\eqref{eq:thetalcm}}> (n-1)\shortlcm$ and $j-1\leq n\shortlcm-1$. 
Hence, $n-1<(j-1)/\shortlcm <n$ and further
\be\label{eq:jspecial1}
\bigg\lfloor \frac{j-1}{\shortlcm}\bigg\rfloor  = n-1, \quad \text{for } j\in [\vartheta_{\R}-\Tprop+1\!:\!n\shortlcm]\,.
\ee
Similarly, for  $j\in [(n-1)\shortlcm+1\!:\vartheta_{\R}-\shortlcm]$, we obtain
$j-1 \leq \vartheta_{\R}-\shortlcm-1 \stackrel{\eqref{eq:thetalcm2}}< (n+1)\shortlcm - \shortlcm-1 = n \shortlcm-1$ and $j-1\geq (n-1)\shortlcm$. 
Thus,  $n-1\leq(j-1)/\shortlcm <n$ and further
\be\label{eq:jspecial2}
\bigg\lfloor \frac{j-1}{\shortlcm}\bigg\rfloor  = n-1, \quad \text{for } j\in [(n-1)\shortlcm+1\!:\vartheta_{\R}-\shortlcm]\,.
\ee
Combining~\eqref{eq:jspecial1} and~\eqref{eq:jspecial2}, we conclude that the offset 
%$\lfloor (j - 1)/\shortlcm\rfloor$  
in~\eqref{eq:beta} satisfies
\begin{align}
&\bigg\lfloor \frac{j-1}{\shortlcm}\bigg\rfloor  = n-1\,, \;
 \text{for } j\in [\vartheta_{\R}-\Tprop+1\!:\!n\shortlcm] \notag \\
& \rule{35mm}{0mm} \cup [(n-1)\shortlcm+1\!:\vartheta_{\R}-\shortlcm]\,.
\label{eq:offsetnmin1}
\end{align}
%same for all elements in the set $[\vartheta_{\R}-\Tprop+1\!:\!n\shortlcm] \cup  [n\shortlcm-\shortlcm+1\!:\vartheta_{\R}-\shortlcm]$.
Let us next consider $\beta_1$ on the sets $[\vartheta_{\R}-\Tprop+1\!:\!n\shortlcm]$ and $[(n-1)\shortlcm+1\!:\vartheta_{\R}-\shortlcm]$.
We obtain
\begin{align}
&\beta_1([\vartheta_{\R}-\Tprop+1\!:\!n\shortlcm]) 
 \notag \\
& \rule{5mm}{0mm}= \, \big\{k= \beta_1(j)=\big(j + \lfloor (j - 1)/\shortlcm\rfloor \big) \mymod \Tprop : \notag \\
& \rule{49mm}{0mm} j\in [\vartheta_{\R}-\Tprop+1\!:\!n\shortlcm]\big\}  \notag \\
&  \rule{5mm}{0mm}\stackrel{\hidewidth \eqref{eq:offsetnmin1} \hidewidth}= \, \big\{k= \beta_1(j)=(j + n-1) \mymod \Tprop :   \notag \\
& \rule{49mm}{0mm}  j\in [\vartheta_{\R}-\Tprop+1\!:\!n\shortlcm]\big\}  \notag \\
&  \rule{5mm}{0mm}=\,  \big\{k= j  \mymod \Tprop\rmv:\rmv j\in [\vartheta_{\R}\rmv-\rmv\Tprop+n\!:\!n\shortlcm+n-1] \big\}  
\notag \\ 
& \rule{5mm}{0mm}\stackrel{\hidewidth (a) \hidewidth}= \big\{k \in [1\!:\!\Tprop]: \exists \, m\in \IN \text{ such that } \notag \\
& \rule{17mm}{0mm} k+m\Tprop\in [\vartheta_{\R}-\Tprop+n\!:\!n\shortlcm+n-1]\big\}
 \label{eq_firstpartimage}
\end{align}
%\begin{align}
%\beta_1([\vartheta_{\R}-\Tprop+1\!:\!n\shortlcm]) 
%& = \{k \in [1\!:\!\Tprop]: \exists \, j\in [\vartheta_{\R}-\Tprop+1\!:\!n\shortlcm]  \text{ such that } k = \beta_1(j)=\big(j + \lfloor (j - 1)/\shortlcm\rfloor \big) \mymod \Tprop\}  \notag \\
%&  \stackrel{\hidewidth \eqref{eq:offsetnmin1} \hidewidth}= \{k \in [1\!:\!\Tprop]: \exists \, j\in [\vartheta_{\R}-\Tprop+1\!:\!n\shortlcm] \text{ such that } k = \beta_1(j)=(j + n-1) \mymod \Tprop\}  \notag \\
%&  = \{k \in [1\!:\!\Tprop]: \exists \, j\in [\vartheta_{\R}-\Tprop+n\!:\!n\shortlcm+n-1] 
%%\notag \\ & \rule{35mm}{0mm} 
%\text{ such that } k = j  \mymod \Tprop\}  
%\notag \\ 
%&  \stackrel{\hidewidth (a) \hidewidth}= \{k \in [1\!:\!\Tprop]: \exists \, m\in \IN \text{ such that } k+m\Tprop\in [\vartheta_{\R}-\Tprop+n\!:\!n\shortlcm+n-1]\}
% \label{eq_firstpartimage}
%\end{align}
where $(a)$ holds because $k = j  \mymod \Tprop$  is equivalent to $j= k+m\Tprop$ for some $m\in \IN$.
Similarly, 
\begin{align}
& \beta_1([(n-1)\shortlcm+1\!:\vartheta_{\R}-\shortlcm])  \notag \\
& \rule{5mm}{0mm} = \, \big\{k = \beta_1(j)=\big(j + \lfloor (j - 1)/\shortlcm\rfloor \big) \mymod \Tprop : \notag \\
& \rule{40mm}{0mm} j\in [(n-1)\shortlcm+1\!:\vartheta_{\R}-\shortlcm]\big\}  \notag \\
& \rule{5mm}{0mm} \stackrel{\hidewidth \eqref{eq:offsetnmin1} \hidewidth}= \,\big\{ k = \beta_1(j)=(j + n-1) \mymod \Tprop : \notag \\
& \rule{40mm}{0mm} j\in [(n-1)\shortlcm+1\!:\vartheta_{\R}-\shortlcm]\big\}  \notag \\
& \rule{5mm}{0mm} = \, \big\{k = j  \mymod \Tprop : \notag \\
& \rule{25mm}{0mm} j\in [(n-1)\shortlcm+n\!:\vartheta_{\R}-\shortlcm+n-1]\big\} 
\notag \\ 
& \rule{5mm}{0mm} = \,\big\{k \in [1\!:\!\Tprop]: \exists \, m\in \IN 
\text{ such that } \notag \\
& \rule{15mm}{0mm} k+m\Tprop\in [(n-1)\shortlcm+n\!:\vartheta_{\R}-\shortlcm+n-1]\big\} \notag \\ 
& \rule{5mm}{0mm} \stackrel{\hidewidth (a) \hidewidth}= \,\big\{k \in [1\!:\!\Tprop]: \exists \, m\in \IN 
%\notag \\ & \rule{35mm}{0mm} 
\text{ such that }  \notag \\
& \rule{20mm}{0mm} k+m\Tprop\in [n\shortlcm+n\!:\vartheta_{\R}+n-1]\big\}
\label{eq_secondpartimage}
\end{align}
where $(a)$ holds because a shift of the interval by $\shortlcm$ (which is a multiple of $\Tprop$) can be compensated by choosing a different $m\in \IN$.
Combining~\eqref{eq_firstpartimage} and~\eqref{eq_secondpartimage}, we 
obtain
\begin{align}
&\beta_1\big([\vartheta_{\R}-\Tprop+1\!:\!n\shortlcm] \cup [(n-1)\shortlcm+1\!:\vartheta_{\R}-\shortlcm]\big) \notag \\
& \rule{5mm}{0mm}= \big\{k \in [1\!:\!\Tprop]: \exists \, m\in \IN \text{ such that }  \notag \\
& \rule{20mm}{0mm} k+m\Tprop\in [\vartheta_{\R}-\Tprop+n\!:\!n\shortlcm+n-1] \notag \\
& \rule{45mm}{0mm} \cup [n\shortlcm+n\!:\vartheta_{\R}+n-1]\big\} \notag \\
& \rule{5mm}{0mm}= \big\{k \in [1\!:\!\Tprop]: \exists \, m\in \IN \text{ such that } \notag \\
& \rule{25mm}{0mm}  k+m\Tprop\in [\vartheta_{\R}-\Tprop+n\!:\vartheta_{\R}+n-1]\big\} \notag \\
& \rule{5mm}{0mm} \stackrel{\hidewidth (a) \hidewidth}= [1\!:\!\Tprop]
\label{eq:imagebeta1}
\end{align}
where $(a)$ holds because $[\vartheta_{\R}-\Tprop+n\!:\vartheta_{\R}+n-1]$ is an interval of length $\Tprop$ and thus for every $k\in [1\!:\!\Tprop]$ we can find an $m\in \IN$ such that $k+m\Tprop\in [\vartheta_{\R}-\Tprop+n\!:\vartheta_{\R}+n-1]$.
Similarly to the previous case,~\eqref{eq:imagebeta1} allows us to write
\bas
&[\vartheta_{\R}-\Tprop+1\!:\!n\shortlcm] \cup [(n-1)\shortlcm+1\!:\vartheta_{\R}-\shortlcm]
%\notag \\ & \rule{50mm}{0mm} 
=\{\tilde{\jmath}_1, \dots, \tilde{\jmath}_{\Tprop}\}
\eas
where $\tilde{\jmath}_t\in \beta_1^{-1}(t)$ for $t\in [1\!:\!\Tprop]$.
By \eqref{eq:imagebeta2}, we then obtain
\bas
& \beta_2([\vartheta_{\R}-\Tprop+1\!:\!\vartheta_{\R}]) \\
& \rule{5mm}{0mm} =
\beta_2([\vartheta_{\R}-\Tprop+1\!:\!n\shortlcm] \cup [(n-1)\shortlcm+1\!:\vartheta_{\R}-\shortlcm]) \\
& \rule{5mm}{0mm} 
=\{\beta_2(\tilde{\jmath}_1), \dots, \beta_2(\tilde{\jmath}_{\Tprop})\}\,.
\eas
By the same arguments as in the previous case, we find that assigning $g_t= \beta_2(\tilde{\jmath}_t)$ satisfies~\eqref{chargt}.

%%%%%%%%%%%%%%%%%%%%%%%%%%%%%%%
\section*{Acknowledgment} 
%%%%%%%%%%%%%%%%%%%%%%%%%%%%%%%

%\vspace{1mm}

The authors would like to thank Dr. Shaowei Lin for pointing them to the weak version of 
B\'ezout's theorem.
Furthermore, they would like to thank the associate editor and the anonymous reviewers,
whose insightful comments helped them improve the presentation of their results.

%% %\vspace{1mm}

%\renewcommand{\baselinestretch}{1.07}\small\normalsize

\bibliography{references}
\bibliographystyle{IEEEtran}

\begin{IEEEbiographynophoto}{G\"unther Koliander}(S'13) received the Master degree in Technical Mathematics (with distinction) from Vienna University of Technology, Austria, in 2011. Since 2011 he has been with the Institute of Telecommunications, Vienna University of Technology, Austria, where he is currently working towards his PhD. 
He twice held visiting researcher positions at Chalmers University of Technology, Gothenburg, Sweden.
His research interests are in the areas of noncoherent communications and information theory.
\end{IEEEbiographynophoto}

\begin{IEEEbiographynophoto}{Erwin Riegler} (M'07) received the Dipl-Ing. degree in Technical Physics (with distinction) in 2001 and the Dr. techn. degree in Technical Physics (with distinction) in 2004 from Vienna University of Technology. From 2005 to 2006, he was a post-doctoral researcher at the Institute for Analysis and Scientific Computing, Vienna University of Technology. From 2007 to 2010, he was a senior researcher at the Telecommunications Research Center Vienna (FTW). From 2010 to 2014, he was a post-doctoral researcher at the Institute of Telecommunications, Vienna University of Technology. Since 2014, he has been a senior researcher with the Communication Theory Group at ETH Zurich, Switzerland. 

Dr. Riegler was a visiting researcher at the Max Planck Institute for Mathematics in the Sciences in Leipzig, Germany (Sep. 2004 to Feb. 2005),  
the Communication Theory Group at ETH Zurich, Switzerland (Sep. 2010 to Feb. 2011 and June 2012 to Nov. 2012),  the Department of Electrical and Computer Engineering at The Ohio State University in Columbus, Ohio (Mar. 2012), and the Department of Signals and Systems at Chalmers University of Technology in Gothenburg, Sweden (Nov. 2013). He is a co-author of a paper that won a Student Paper Award at the 2012 International Symposium on Information Theory. 

His research interests include noncoherent communications, machine learning, interference management, large system analysis, and transceiver design.
\end{IEEEbiographynophoto}

\begin{IEEEbiographynophoto}{Giuseppe Durisi}(S'02--M'06--SM'12) received the Laurea degree summa cum laude and the Doctor degree both from Politecnico di Torino, Italy, in 2001 and 2006, respectively. From 2002 to 2006, he was with Istituto Superiore Mario Boella, Torino, Italy. From 2006 to 2010 he was a postdoctoral researcher at ETH Zurich, Switzerland. Since 2010 he has been with Chalmers University of Technology, Gothenburg, Sweden, where he is now an associate professor. He held visiting researcher positions at IMST, Germany, University of Pisa, Italy, ETH Zurich, Switzerland, and Vienna University of Technology, Austria.

Dr. Durisi is a senior member of the IEEE. He is the recipient of the 2013 IEEE ComSoc Best Young Researcher Award for the Europe, Middle East, and Africa Region, and is co-author of a paper that won a Student Paper Award at the 2012 International Symposium on Information Theory, and of a paper that won the 2013 IEEE Sweden VT-COM-IT joint chapter best student conference paper award. He served as TPC member in several IEEE conferences, and is currently publications editor of the {\sc IEEE Transactions on Information Theory}. His research interests are in the areas of communication and information theory.
\end{IEEEbiographynophoto}

\begin{IEEEbiographynophoto}{Franz Hlawatsch}(S'85--M'88--SM'00--F'12) received the Diplom-Ingenieur, Dr. techn., and Univ.-Dozent (habilitation) degrees in electrical engineering/signal processing from Vienna University of Technology, Vienna, Austria in 1983, 1988, and 1996, respectively.
Since 1983, he has been with the Institute of Telecommunications, Vienna University of Technology, where he is currently an Associate Professor. During 1991--1992, as a recipient of an Erwin Schr\"odinger Fellowship, he spent a sabbatical year with the Department of Electrical Engineering, University of Rhode Island, Kingston, RI, USA. In 1999, 2000, and 2001, he held one-month Visiting Professor positions with INP/ENSEEIHT, Toulouse, France and IRCCyN, Nantes, France. He (co)authored a book, three review papers that appeared in the {\sc IEEE Signal Processing Magazine}, about 200 refereed scientific papers and book chapters, and three patents. He coedited three books. His research interests include 
%% signal processing for 
wireless communications, sensor networks, and statistical and compressive signal processing.

Prof. Hlawatsch was Technical Program Co-Chair of EUSIPCO 2004 and served on the technical committees of numerous IEEE conferences. He was an Associate Editor for the {\sc IEEE Transactions on Signal Processing} from 2003 to 2007 and for the {\sc IEEE Transactions on Information Theory} from 2008 to 2011. From 2004 to 2009, he was a member of the IEEE SPCOM Technical Committee. He is coauthor of papers that won an IEEE Signal Processing Society Young Author Best Paper Award and a Best Student Paper Award at IEEE ICASSP 2011.
\end{IEEEbiographynophoto}

\end{document}